\documentclass[1p,times,sort&compress]{elsarticle}

\usepackage{amsmath}
\usepackage{amssymb}
\usepackage{lineno}
\usepackage{bbold}
\usepackage{dsfont}
\usepackage{rotating}
\usepackage{color}

\journal{Progress in Particle and Nuclear Physics}

\newcommand{\be}{\begin{equation}}
\newcommand{\ee}{\end{equation}}
\newcommand{\ba}{\begin{eqnarray}}
\newcommand{\ea}{\end{eqnarray}}

\newcommand{\bi}{\begin{itemize}}
\newcommand{\ei}{\end{itemize}}
\newcommand{\<}{\langle}
\renewcommand{\>}{\rangle}
\newcommand{\eq}[1]{Eq.\,(\ref{#1})}
\newcommand{\eqs}{Eqs.~}
\newcommand{\fig}[1]{Figure\,\ref{#1}}
\newcommand{\tab}[1]{Table\,\ref{#1}}
\newcommand{\la}{\label}

\newcommand{\ahvp}{a_\mu^{\rm hvp}}

\newcommand{\amuhlbl}{a_\mu^{\rm hlbl}}
\newcommand{\ahlbl}{a_\mu^{\rm hlbl}}
\newcommand{\Pihat}{\hat{\Pi}}

\newcommand{\mev}{{\rm{MeV}}}
\newcommand{\gev}{{\rm{GeV}}}
\newcommand{\fm}{{\rm{fm}}}
\newcommand{\rmO}{{\rm{O}}}
\newcommand{\rme}{{\rm{e}}}
\newcommand{\psibar}{\overline{\psi}}
\newcommand{\zv}{Z_{\rm V}}
\newcommand{\Nf}{N_{\rm f}}

\newcommand{\dZ}{{\mathds{Z}}}

\newcommand{\rb}[1]{\raisebox{1.5ex}[-1.5ex]{#1}}
\newcommand{\xcut}{x_0^{\rm{cut}}}
\renewcommand{\vec}[1]{\boldsymbol{#1}}

\begin{document}

\begin{frontmatter}

\title{Lattice QCD and the anomalous magnetic moment of the muon}

\author{Harvey B.\ Meyer}
\author{Hartmut Wittig}
\address{PRISMA Cluster of Excellence,  Institut f\"ur Kernphysik \& Helmholtz-Institute Mainz,
Johannes Gutenberg-Universit\"at Mainz, 55099 Mainz, Germany}

\begin{abstract} 
The anomalous magnetic moment of the muon, $a_\mu$, has been measured
with an overall precision of 540\,ppb by the E821 experiment at
BNL. Since the publication of this result in 2004 there has been a
persistent tension of $3.5$ standard deviations with the theoretical
prediction of $a_\mu$ based on the Standard Model. The uncertainty of
the latter is dominated by the effects of the strong interaction,
notably the hadronic vacuum polarisation (HVP) and the hadronic
light-by-light (HLbL) scattering contributions, which are commonly
evaluated using a data-driven approach and hadronic models,
respectively. Given that the discrepancy between theory and experiment
is currently one of the most intriguing hints for a possible failure
of the Standard Model, it is of paramount importance to determine both
the HVP and HLbL contributions from first principles. In this review
we present the status of lattice QCD calculations of the leading-order
HVP and the HLbL scattering contributions, $\ahvp$ and $\ahlbl$. After
describing the formalism to express $\ahvp$ and $\ahlbl$ in terms of
Euclidean correlation functions that can be computed on the lattice,
we focus on the systematic effects that must be controlled to achieve
a first-principles determination of the dominant strong interaction
contributions to $a_\mu$ with the desired level of precision. We also
present an overview of current lattice QCD results for $\ahvp$ and
$\ahlbl$, as well as related quantities such as the transition form
factor for $\pi^0\to\gamma^\ast\gamma^\ast$. While the total error of
current lattice QCD estimates of $\ahvp$ has reached the few-percent
level, it must be further reduced by a factor $\sim5$ to be
competitive with the data-driven dispersive approach. At the same
time, there has been good progress towards the determination of
$\ahlbl$ with an uncertainty at the $10-15$\%-level.

\end{abstract}


\end{frontmatter}



\section{Introduction\la{sec:intro}}

The anomalous magnetic moment of the muon, $a_\mu$ is one of the most
precisely measured quantities in particle physics. It is defined as
the deviation of the $g$-factor, which determines the strength of the
muon's magnetic moment, from the value $g=2$ predicted by the Dirac
equation, i.e.
\be
  g_\mu=2(1+a_\mu), \quad a_\mu=\frac{1}{2}(g-2)_\mu.
\ee
The deviation, caused by quantum loop corrections, is a characteristic
property of the particle. Both $a_\mu$ and the corresponding anomalous
magnetic moment of the electron, $a_e$, have been measured
experimentally with very high precision
\cite{Bennett:2006fi,Hanneke:2010au},
\ba
  & & a_\mu^{\rm exp} = (116\,592\,089\pm 63)\cdot 10^{-11} \quad
 (540\,\rm ppb) \label{eq:amuexp} \\
  & & a_e^{\rm exp} = (115\,965\,218\,073 \pm 28)\cdot 10^{-14} \quad
 (0.24\,\rm ppb)
\ea
The particular interest in $a_\mu$ comes from the high sensitivity to
effects from physics beyond the Standard Model. The anomalous magnetic
moment of a generic lepton, $a_\ell$, receives a contribution from
quantum fluctuations induced by heavy particles proportional to
\be
   \delta a_\ell = m_\ell^2/M^2,
\ee
where $m_\ell$ is the lepton mass, and $M$ denotes either the mass of
a particle which is not part of the Standard Model (SM) or the energy
scale beyond which the SM loses its validity. This implies that the
sensitivity of $a_\mu$ to ``new physics'' is increased by a factor
$(m_\mu/m_e)^2\approx 4\cdot10^4$ relative to $a_e$. Against this
backdrop it is intriguing that there has been a persistent discrepancy
between the measured value of $a_\mu$ and its prediction based on the
SM, $a_\mu^{\rm exp}-a_\mu^{\rm SM}=(266\pm76)\cdot10^{-11}$, which
amounts to $\sim3.5$ standard deviations (see
\tab{tab:amustatus}).\footnote{It is interesting to note that a recent
  improved determination of the fine structure constant
  \cite{Parker:2018sci} has resulted in a similar but less significant
  deviation between the experimental and SM estimates of the electron
  anomalous magnetic moment, i.e. $a_e^{\rm exp}-a_e^{\rm
    SM}=(-87\pm36)\cdot10^{-14}$, which corresponds to $-2.4\sigma$.}

Within the SM, the anomalous magnetic moment of the muon receives
contributions from QED, the electroweak sector, and the strong
interaction:
\be
   a_\mu^{\rm SM} = a_\mu^{\rm QED}+a_\mu^{\rm EW}+a_\mu^{\rm had},
\ee
where the superscript ``had'' indicates that the effects of the strong
interaction must be quantified at typical hadronic scales. An overview
which specifies the contributions from electromagnetism, the weak and
the strong interactions to $a_\mu$ is provided in
Table\,\ref{tab:amustatus}. Extensive reviews of the subject, which
detail the various contributions, can be found in
Refs.\,\cite{Jegerlehner:2009ry,Blum:2013xva,Jegerlehner:2017gek}.

\begin{table}[t]
\begin{center}
\begin{tabular}{l r@{.}l r@{.}l c l }
\hline\hline
 & \multicolumn{2}{c}{Value}  & \multicolumn{2}{c}{Error}
 & $a_\mu^{\rm exp}-a_\mu^{\rm SM}$ & \\
\hline
QED        & 11\,658\,471&895 & 0&008 & & 10$^{\rm th}$ order \cite{Aoyama:2012wk} \\
EW         &           15&36  & 0&11  & & Two loop \cite{Czarnecki:2002nt,Gnendiger:2013pva} \\
HVP, LO    &          693&1   & 3&4   & & DHMZ\,17 \cite{Davier:2017zfy} \\
HVP, NLO   &         $-9$&84  & 0&07  & & HMNT \cite{Kurz:2014wya} \\
HLBL       &           10&5   & 2&6   & & PdeRV \cite{Prades:2009tw} \\
\hline
Total SM   & 11\,659\,182&3   & 4&3   & $3.5\sigma$ & DHMZ\,17 \\
\hline
Experiment & 11\,659\,208&9   & 6&3   & & BNL E821 \cite{Bennett:2006fi} \\
\hline\hline
\end{tabular}
\caption{Contributions to the SM prediction for $a_\mu$ from QED, the
  electroweak (EW) and hadronic sectors, in units of
  $10^{-10}$. \la{tab:amustatus}}
\end{center}
\end{table}

\begin{figure}
\begin{center}
\includegraphics[width=10cm]{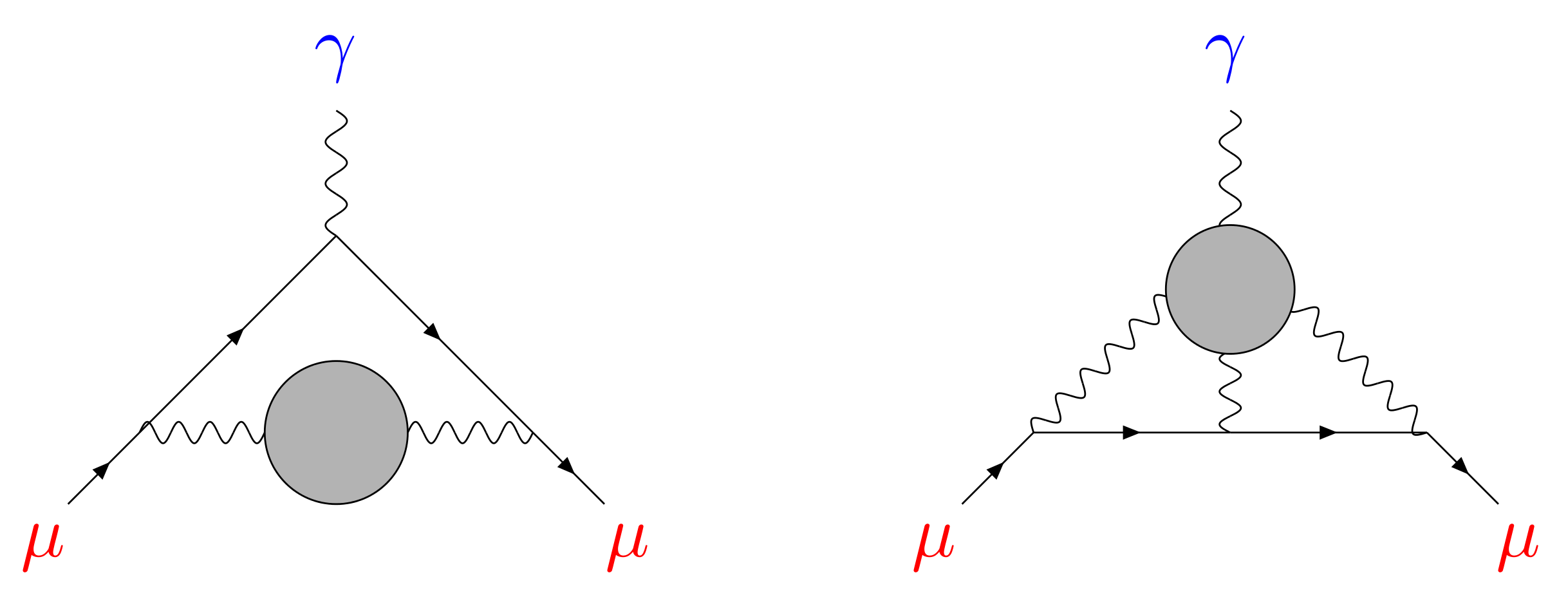}
\caption{The diagrams representing the leading hadronic vacuum
  polarisation (left) and light-by-light scattering
  contributions. Grey circles denote the hadronic loops.
  \la{fig:hadronic}}
\end{center}
\end{figure}

The overall precision of the SM prediction is limited by hadronic
contributions, as is evidenced by \tab{tab:amustatus}. In particular,
the uncertainties ascribed to the leading hadronic vacuum polarisation
(HVP) and hadronic light-by-light (HLbL) scattering contributions (see
Fig\,\ref{fig:hadronic}) dominate the total error of $a_\mu^{\rm
  SM}$. Efforts have therefore been concentrated on corroborating the
actual estimates and reducing the associated uncertainties.

The leading (i.e. $\rmO(\alpha^2)$) HVP contribution, $\ahvp$, which
enters the SM estimate has been determined via dispersion
relations. In the conventions and notation of
\cite{Jegerlehner:2009ry} the relevant expression reads
\be\la{eq:dispersion}
   \ahvp = \left(\frac{\alpha m_\mu}{3\pi}\right)^2
   \int_{m_{\pi^0}^2}^\infty ds\,\frac{\hat{K}(s)}{s^2}\,R(s),
\ee
where $\alpha$ is the fine-structure constant, $\hat{K}(s)$ is a known
QED kernel function \cite{Brodsky:1967sr}, and $R(s)$ denotes the
cross section for $e^{+}e^{-}\to\hbox{hadrons}$ normalised by
$\sigma(e^{+}e^{-}\to\mu^{+}\mu^{-})$ at tree level in the limit $s\gg
m_\mu^2$:
\be\la{eq:Rratio}
  R(s)=\frac{\sigma(e^{+}e^{-}\to\hbox{hadrons})}{4\pi\alpha^2/(3s)}
\ee
A high energies the ratio can be approximated with sufficient accuracy
in perturbative QCD. However, at low energies, where the dispersion
integral is dominated by the $\rho$-resonance, one has to resort to
experimental data for $R(s)$. In practice one splits the integration
into two intervals:
\be
   R(s)\longrightarrow \left\{ \begin{array}{ll}
        R(s)^{\rm data}, & m_{\pi^0}^2 \leq s < E_{\rm cut}^2 \\
        R(s)^{\rm pQCD}, & s > E_{\rm cut}^2 \end{array} \right. .
\ee
The resulting estimates for $\ahvp$ from several independent analyses
\cite{Davier:2010nc, Jegerlehner:2011ti, Hagiwara:2011af,
  Davier:2017zfy, Jegerlehner:2017lbd, Keshavarzi:2018mgv} based on
the combined data for $e^{+}e^{-}\to\hbox{hadrons}$ are listed in
Table\,\ref{tab:HVPdisp}. Currently, several issues are still being
debated: The first concerns the consistency of the experimental data
in the $\pi^{+}\pi^{-}$ channel determined using the ISR (initial
state radiation) method \cite{Ambrosino:2008aa, Ambrosino:2010bv,
  Babusci:2012rp, Lees:2012cj, Ablikim:2015orh}, as well as the
treatment of this particular contribution in the evaluation of the
dispersion integral. The second issue concerns the question whether a
more precise result for $\ahvp$ can be obtained by including data from
hadronic $\tau$ decays in order to estimate the spectral function
\cite{Alemany:1997tn,Davier:2010nc,Jegerlehner:2011ti}. Progress has
been achieved on both of these issues, and some of the most recent
analyses of the SM contribution to $a_\mu$ report a slightly increased
discrepancy of about $4\,\sigma$ with the direct measurement (see
\tab{tab:HVPdisp}). While the dispersive approach produces estimates
for $\ahvp$ with a total error at the sub-percent level, it is clear
that the resulting SM estimate is subject to experimental
uncertainties. This is one of the main motivations for working towards
a result based on a first-principles approach such as lattice QCD.

\begin{table}[t]
\begin{center}
\begin{tabular}{l r@{.}l r@{.}l c l }
\hline\hline
Author(s) & \multicolumn{2}{c}{$\ahvp\cdot10^{-10}$}  & \multicolumn{2}{c}{Error}
 & $a_\mu^{\rm exp}-a_\mu^{\rm SM}$ & Comment \\
\hline
DHMZ\,11 \cite{Davier:2010nc}    & 692&3  & 4&2  & $3.6\sigma$ & $e^+ e^-$ data \\
                                 & 701&5  & 4&7  & $2.4\sigma$ & $\tau$ data \\[0.8ex]
FS\,11 \cite{Jegerlehner:2011ti} & 690&75 & 4&72 &             & $e^+e^-$ data \\
                                 & 690&96 & 4&65 & $3.3\sigma$ & $e^+e^-$ and $\tau$ data \\[0.8ex]
HLMNT\,11 \cite{Hagiwara:2011af} & 694&9  & 4&3  & $3.2\sigma$ & $e^+ e^-$ data \\[0.8ex]
DHMZ\,17 \cite{Davier:2017zfy}   & 693&1  & 3&4  & $3.5\sigma$ & $e^+e^-$ data \\[0.8ex]
Jegerlehner\,17 \cite{Jegerlehner:2017lbd}
                                 & 688&07 & 4&14 & $4.0\sigma$ & $e^+e^-$ data \\
                                 & 688&77 & 3&38 & $4.1\sigma$ & $e^+e^-$ and $\tau$ data \\[0.8ex]
KNT\,18 \cite{Keshavarzi:2018mgv}& 693&27 & 2&46 & $3.7\sigma$ & $e^+ e^-$ data \\
\hline\hline
\end{tabular}
\caption{Compilation of results for the leading order (O($\alpha^2$))
  hadronic vacuum polarisation contribution $\ahvp$ from the
  data-driven dispersive approach.\la{tab:HVPdisp}}
\end{center}
\end{table}

\begin{figure}
\begin{center}
\includegraphics[width=0.8\textwidth]{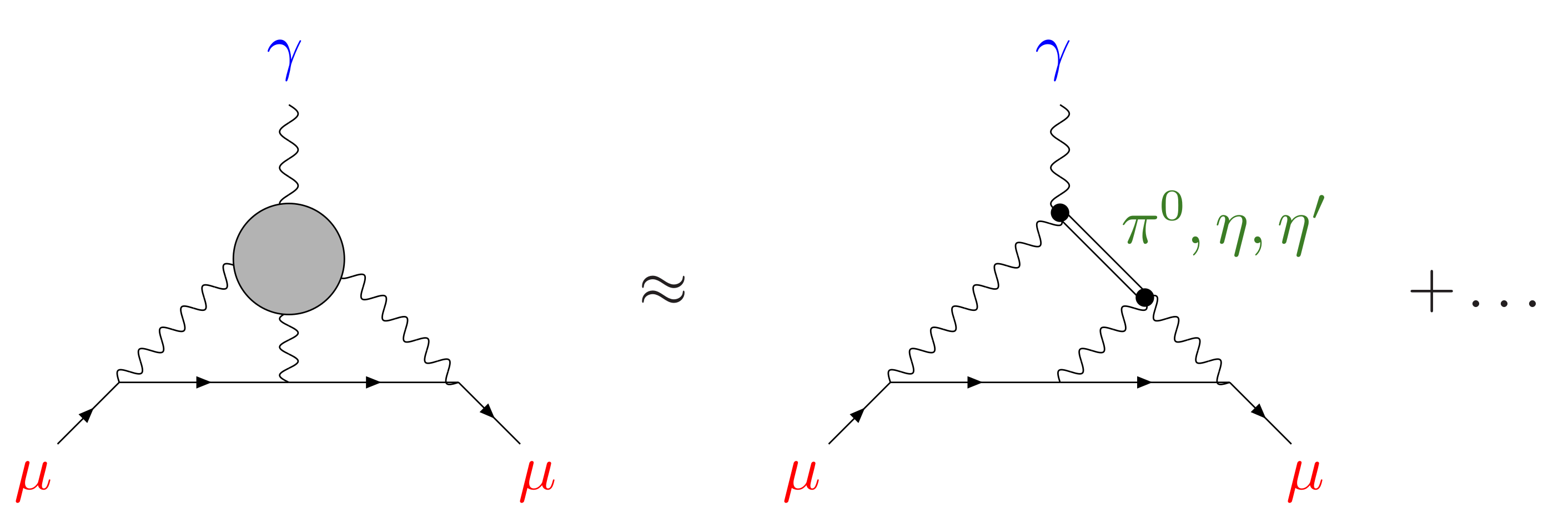}
\caption{The expected dominant contributions to the HLbL scattering
  amplitude. \la{fig:PShlbl}}
\end{center}
\end{figure}

The hadronic light-by-light scattering contribution, $\amuhlbl$, has
so far been determined via model estimates (see \cite{Hayakawa:1996ki,
  Hayakawa:1997rq, Bijnens:1995cc, Bijnens:1995xf, Bijnens:2001cq,
  Knecht:2001qf, Melnikov:2003xd, Nyffeler:2009tw,
  Jegerlehner:2009ry,Prades:2009tw,Blum:2013xva,Bijnens:2015jqa}),
though recent efforts have focussed on developing a dispersive
framework \cite{Colangelo:2014dfa, Colangelo:2014pva,
  Colangelo:2015ama, Colangelo:2017qdm, Colangelo:2017fiz} and other
data-driven approaches \cite{Pascalutsa:2010sj, Pascalutsa:2012pr,
  Pauk:2014jza, Pauk:2014rfa, Danilkin:2016hnh, Nyffeler:2016gnb}. The
most widely used model estimate that enters the current SM estimate is
known as the ``Glasgow consensus'' \cite{Prades:2009tw},
$\amuhlbl=(10.5\pm2.6)\cdot 10^{-10}$. An alternative, but compatible
estimate of $\amuhlbl=(11.6\pm3.9)\cdot 10^{-10}$ is quoted
in\,\cite{Jegerlehner:2009ry,Nyffeler:2009tw}, while a recent
update\,\cite{Jegerlehner:2017lbd} finds $\amuhlbl=(10.3\pm2.9)\cdot
10^{-10}$.

Since a comprehensive treatment of the full hadronic light-by-light
scattering tensor $\Pi_{\mu\nu\lambda\rho}$ is a rather complex task,
it is useful to focus on particular subprocesses, even though this
introduces a dependence on hadronic models. The value of $\ahlbl$ is
expected to be dominated by the pion pole, with additional corrections
provided by the $\eta$ and $\eta^\prime$ \cite{Hayakawa:1997rq,
  Bijnens:1995cc, Bijnens:1995xf, Bijnens:2001cq, Knecht:2001qf} (see
Figure\,\ref{fig:PShlbl}). In order to quantify the pion pole
contribution, it is then necessary to constrain the off-shell
pion-photon-photon transition form factor
${\cal{F}}_{\pi^0\gamma^\ast\gamma^\ast}$, which is usually done using
hadronic models \cite{Nyffeler:2016gnb}, lattice QCD
\cite{Gerardin:2016cqj} and a data-driven phenomenological approach
\cite{Hoferichter:2018dmo}.

The need to obtain more precise results for the HVP and HLbL
contributions is underlined by the fact that the sensitivity of future
experimental measurements of $a_\mu$ will exceed the uncertainties
associated with HVP and HLbL. Two new experiments with very different
setups are expected to improve the precision of the experimental
determination by a factor four: The E989 experiment at Fermilab
\cite{Kaspar:2015jwa,Fertl:2016nij} uses the original storage ring
of the older BNL experiment. A number of technical improvements will
provide a much cleaner muon sample, better magnetic field calibration
and more efficient detectors to record the muon decay. The goal is a
measurement of the the anomalous precession frequency of the muon spin
with a precision of 70\,ppb, with statistical and other systematic
uncertainties expected at the level of 100 and 70\,ppb,
respectively. Combining all projected uncertainties in quadrature
yields the target precision of 140\,ppb for the new measurement of
$a_\mu$. First results are expected in 2019.

The E34 experiment at J-PARC \cite{Otani:2015lra} is based on a very
different setup, designed to determine both $a_\mu$ and the muon's
electric dipole moment. This is made possible by working without an
electric field, $\vec{E}=0$. The technical challenge then consists in
producing an accurately collimated muon beam without any focussing
that is normally provided by the electric field. A beam of ultracold
muons with low emittance is produced via resonant laser ionisation of
muonium. The muons are subsequently re-accelerated to reduce their
transverse dispersion to a level of $10^{-5}$. Eventually they are
injected into the storage magnet equipped with detectors to measure
the anomalous precession frequency of the muon spin. The electric
dipole moment can be extracted from the amplitude of the
oscillation. The goal for the first phase of the experiment is the
determination of $a_\mu$ at the level of 370\,ppb. In the long term
one aims for a total precision of 100\,ppb.

From these considerations it is clear that the precision of the SM
estimate must keep pace with the expected error reduction provided by
the forthcoming direct measurements. In order to avoid any dependence
on experimental input in the dispersive approach to HVP and to
eliminate the model dependence in the current estimates of $\ahlbl$, a
first-principles approach to quantifying the main hadronic
contributions to $a_\mu$ is warranted. Lattice QCD has produced
precise results for a wide range of hadronic observables, including
not only hadron masses, decay constants, form factors and mixing
parameters characterising weak decay amplitudes, but also SM
parameters such as quark masses and the running coupling
\cite{Aoki:2016frl}.

The objective of this review is to provide an overview of recent
attempts to determine both the leading hadronic vacuum polarisation
and light-by-light scattering contributions to the muon $g-2$ using
lattice QCD. In order to test the significance of the tension between
the SM prediction and the direct measurement, lattice QCD must be able
to determine $\ahvp$ with an overall precision far below the percent
level. By contrast, a model-independent estimate of $\ahlbl$ with a
total uncertainty of $\rmO(10\%)$ would be a major achievement. As
will become apparent, both objectives present considerable challenges
to lattice calculations.

This article is organised as follows: Section~\ref{sec:HVP} is
focussed on the determination of the hadronic vacuum polarisation
contribution, $\ahvp$. We discuss various representations of $\ahvp$
that are amenable to lattice calculations and describe the particular
challenges that must be confronted in order to determine $\ahvp$ with
the desired precision. Section~\ref{sec:results} contains a
compilation of results for $\ahvp$ and a critical assessment of the
current status of lattice calculations. Section~\ref{sec:HLbL}
describes the efforts to gain information on $\ahlbl$ from first
principles. We introduce the general formalism that allows for the
calculation of $\ahlbl$ on the lattice with manageable numerical
effort. The crucial ingredient is the efficient treatment of the QED
kernel, which can be achieved either via stochastic sampling or by
performing a semi-analytic calculation. First results for $\ahlbl$ are
discussed in Section~\ref{sec:HLbL_latresu}, followed by a discussion
of related quantities that can be used in conjunction with
phenomenological models, including forward light-by-light scattering
amplitudes and the transition form factor for
$\pi^0\to\gamma^\ast\gamma^\ast$. We end the review with some
concluding remarks in Section~\ref{sec:concl}. A self-containted
introduction to the basic concepts of lattice QCD, including a
discussion of vector currents and correlators, is relegated to the
appendix.

\section{The hadronic vacuum polarisation \la{sec:HVP}}

A concrete proposal for determining the hadronic vacuum polarisation
contribution $\ahvp$ in lattice QCD was published in 2002
\cite{Blum:2002ii}. While early calculations in the quenched
approximation \cite{Blum:2002ii,Gockeler:2003cw} produced results that
were much smaller than the phenomenological value, the overall
feasibility of the lattice approach could be demonstrated. First
attempts to compute $\ahvp$ in full QCD were published in 2008
\cite{Aubin:2006xv}, and in the following years several studies
appeared
\cite{Feng:2011zk,Boyle:2011hu,DellaMorte:2011aa,Burger:2013jya},
employing a range of different discretisations of the quark action,
which were mostly aimed at investigating systematic effects. The most
recent calculations are focussed on reducing the overall uncertainties
to a level similar to that of the dispersive approach
\cite{Blum:2015you,Blum:2016xpd,Chakraborty:2016mwy,
  Borsanyi:2016lpl,DellaMorte:2017dyu,Giusti:2017jof,
  Chakraborty:2017tqp,Borsanyi:2017zdw,Blum:2018mom}. Here we
introduce the lattice approach for determining the hadronic vacuum
polarisation contribution. In particular, we present a detailed
discussion of systematic effects and give an overview of recent
results.

\subsection{Lattice approach to hadronic vacuum polarisation}

The relevant quantity for the determination of $\ahvp$ in lattice
QCD is the polarisation tensor
\be\la{eq:PolTens}
  \Pi_{\mu\nu}(Q) \equiv \int d^4x \, \rme^{iQ\cdot x} \<J_\mu(x) J_\nu(0)\>,
\ee
where $J_\mu(x)$ is the hadronic contribution to the electromagnetic
current, i.e.
\be\la{eq:emcurrent}
   J_\mu = {\textstyle \frac{2}{3}\overline{u}\gamma_\mu u -
     \frac{1}{3}\overline{d}\gamma_\mu d -
     \frac{1}{3}\overline{s}\gamma_\mu s +\ldots.}
\ee
Current conservation and O(4) invariance (which replaces Lorentz
invariance in the Euclidean formulation) imply the tensor structure
\be\la{eq:PimunuQ}
  \Pi_{\mu\nu}(Q) = \big(Q_\mu Q_\nu -\delta_{\mu\nu}Q^2\big) \Pi(Q^2).
\ee
Since the vacuum polarisation $\Pi(Q^2)$ still contains a logarithmic
divergence, one has to perform a subtraction in order to obtain a finite
quantity, which is defined as
\be\la{eq:Pihat}
 {\Pihat(Q^2)}\equiv {4\pi^2}\,[\Pi(Q^2)-\Pi(0)].
\ee
With these definitions, the leading hadronic contribution to
$(g-2)_\mu$ can be expressed in terms of a convolution integral over
Euclidean momenta $Q$ \cite{Lautrup:1971jf,Blum:2002ii}, i.e.
\be \la{eq:amublum2}
\ahvp = \left(\frac{\alpha}{\pi}\right)^2 \int_0^\infty {d Q^2}
\,f(Q^2)\, \Pihat(Q^2).
\ee
The QED kernel function $f$ which appears in this expression is
given by
\be
  \label{eq:kerK}
f(Q^2) = \frac{\hat s Z(\hat s)^3}{m_\mu^2} \cdot
\frac{1 - \hat s Z(\hat s)}{1 + \hat s Z(\hat s)^2}\,, \quad
Z(\hat s) = - \frac{\hat s - \sqrt{\hat s^2 + 4 \hat s}}{2  \hat s},
\ee
where $\hat s \equiv {Q^2}/{m_\mu^2}$.

Using a suitable transcription of the electromagnetic current and the
vacuum polarisation tensor for a Euclidean space-time lattice (details
are provided in \ref{app:vector} and~\ref{app:PolTens}), it is
straightforward to compute $\Pi(Q^2)$ via \eq{eq:PimunuQ} and
determine $\ahvp$ in lattice QCD. However, this procedure entails a
number of technical difficulties that limit the accuracy of the
result. First, the structure of the kernel function $f(Q^2)$ implies
that the convolution integral receives its dominant contribution from
the region near $Q^2\lesssim m_\mu^2\approx0.01\,\gev^2$. On a finite
hypercubic lattice the momentum is quantised in units of the inverse
box length, and hence the smallest non-zero value of $Q^2$ that can be
realised for spatial lengths of $L\approx 6\,\fm$ is $4-5$ times
larger than $m_\mu^2$. Furthermore, the statistical accuracy of
$\Pi(Q^2)$ deteriorates quickly in the small-momentum region. Thus,
any lattice calculation of $\ahvp$ must address the lack of
statistically precise data in the regime that provides the bulk of the
contribution.

\begin{figure}
\begin{center}
\includegraphics[width=0.7\textwidth]{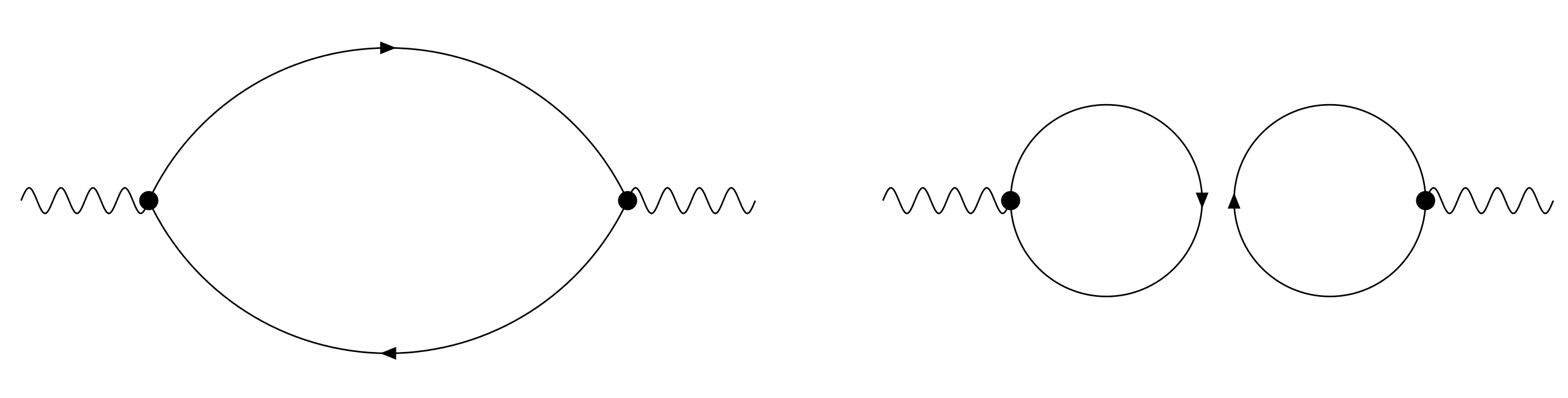}
\caption{The quark-connected and quark-disconnected diagrams that
  contribute to the correlator of the electromagnetic
  current Gluons lines are not shown. \la{fig:conndisc}} 
\end{center}
\end{figure}

In order to challenge or even surpass the accuracy of the estimates of
$\ahvp$ obtained using dispersion theory listed in \tab{tab:HVPdisp},
lattice calculations must control all sources of statistical and
systematic uncertainties at the sub-percent level. This includes the
inherent systematic effects that are common to all lattice
calculations, i.e. lattice artefacts, finite-volume effects and the
dependence on the light quark mass that are discussed
in\,\ref{sec:systematics}. Since simulations at or very near the
physical pion mass are the state of the art, the systematic error
associated with the chiral extrapolation is under increasingly good
control. Discretisation effects are potentially large for heavy
quarks, and since the charm quark makes a small but significant
contribution to $\ahvp$, the extrapolation to the continuum limit must
be sufficiently well controlled. Many quantities computed in lattice
QCD, such as hadron masses and decay constants do not receive large
finite-volume corrections relative to the typical statistical error,
provided that $m_\pi L\gtrsim 4$. However, this is only an empirical
statement derived from a finite set of quantities, and it is uncertain
whether this rule of thumb applies to $\ahvp$. At the sub-percent
level, isospin breaking effects arising from the mass splitting among
the up and down quarks, as well as from their different electric
charges cannot be neglected. This represents a major complication,
since calculations for $m_u\neq m_d$ are technically more involved and
because QED effects must be incorporated as well \cite{Blum:2010ym,
  Blum:2014oka, Boyle:2017gzv, Blum:2018mom, deDivitiis:2011eh,
  deDivitiis:2013xla, Carrasco:2015xwa, Lubicz:2016xro,
  Giusti:2017dmp, Borsanyi:2013lga, Borsanyi:2014jba, Fodor:2015pna,
  Chakraborty:2017tqp} (see also the recent review
\cite{Patella:2017fgk}). Finally, there is the issue of
quark-disconnected diagrams: after performing the Wick contractions
over the quark fields in the vector correlator of \eq{eq:PolTens} one
recovers the two types of diagrams shown in \fig{fig:conndisc}. Due to
the large inherent level of statistical fluctuations, special noise
reduction techniques must be applied in order to determine the
contributions from quark-disconnected diagrams with sufficient
accuracy. Isospin symmetry implies that, in the low-energy regime, the
disconnected contribution to $\hat\Pi(Q^2)$ amounts to $-1/10$ of the
connected one \cite{DellaMorte:2010aq,Francis:2013fzp}. While this
estimate for the ratio is essentially confirmed in chiral effective
theory at two loops \cite{Bijnens:2016ndo}, it is necessary to
evaluate disconnected contributions directly using actual simulation
data if the overall target precision is set below 1\%. We postpone a
detailed discussion of quark-disconnected diagrams to Section
\ref{sec:disc}.

\subsection{The infrared regime of $\Pi(Q^2)$}

In this subsection we will discuss the strategies that are employed to
determine the subtracted vacuum polarisation $\Pihat(Q^2)\equiv
4\pi^2(\Pi(Q^2)-\Pi(0))$ with sufficient accuracy in the low-momentum
region. We will focus, in particular, on the determination of the
additive renormalisation $\Pi(0)$. Recalling the relation between the
vacuum polarisation tensor and $\Pi(Q^2)$ in \eq{eq:PimunuQ} one
easily sees that the statistical accuracy of $\Pi(Q^2)$ deteriorates
near $Q^2=0$, which makes an accurate determination of $\Pi(0)$ quite
difficult. In early calculations of $\ahvp$ the value of $\Pi(0)$ was
determined by performing fits to $\Pi(Q^2)$ over the entire accessible
range in $Q^2$, using some {\it ansatz} for the momentum
dependence. The disadvantage of such a procedure lies in the fact that
the higher statistical accuracy of the data points at larger values of
$Q^2$ may lead to a systematic bias in the shape of $\Pi(Q^2)$ in the
momentum range from which the convolution integral in \eq{eq:amublum2}
receives its dominant contribution. This issue bears some resemblance
to the determination of the proton charge radius from $ep$ scattering
data \cite{Carlson:2015jba,Hill:2017wzi}.

\subsubsection{The ``hybrid method''}

\begin{figure}[t]
\begin{center}
\includegraphics[width=0.6\textwidth]{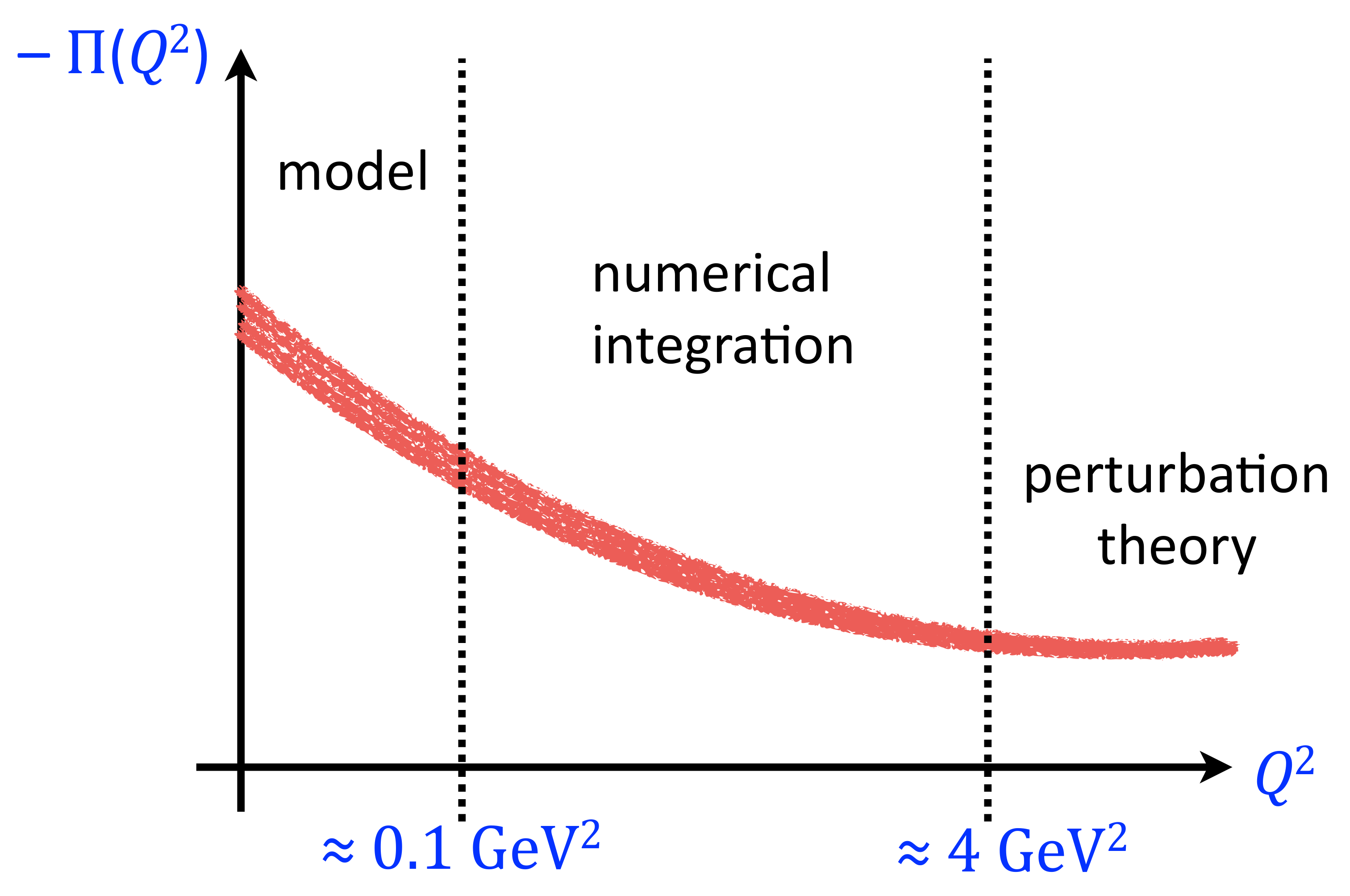}
\caption{Sketch of the hybrid method introduced in
  Ref.\,\cite{Golterman:2014ksa}. The red band denotes the
  unsubtracted vacuum polarisation $\Pi(Q^2)$. The model {\it ansatz}
  for fitting lattice data for $\Pi(Q^2)$ in the low-momentum regime
  are based on Pad\'e approximants or conformal polynomials.
  \la{fig:hybrid}}
\end{center}
\end{figure}

In Ref.\,\cite{Golterman:2014ksa} the so-called hybrid method was
proposed. Here the accessible $Q^2$-interval is divided into three
parts, as shown schematically in Figure\,\ref{fig:hybrid}. Fits to the
unsubtracted vacuum polarisation $\Pi(Q^2)$ are restricted to the
immediate vicinity of $Q^2=0$, i.e. to the interval $0\leq Q^2\leq
Q_{\rm low}^2$. Ideally, the scale $Q_{\rm low}$ should be chosen much
smaller than the mass of the lowest vector meson, $m_\rho$, in order
to avoid any bias arising from the parameterisation of $\Pi(Q^2)$. A
possible {\it ansatz} for the $Q^2$-behaviour in this regime is
provided by the Pad\'e approximant of order $[N,M]$:
\be\la{eq:Pade}
   \Pi_{[N,M]}(Q^2) = \Pi(0)+ \frac{a_1 Q^2+a_2 Q^4\ldots+a_N
     Q^{2N}}{1+b_1 Q^2+b_2 Q^4+\ldots+b_M Q^{2M}}.
\ee
One expects that Pad\'e approximants of increasingly higher degree
eventually converge towards a model-independent description of the
data \cite{Aubin:2012me}, so that a determination of the intercept
$\Pi(0)$ and the shape of $\Pi(Q^2)$ in the low-momentum region is
obtained which is free of any bias from data points at larger $Q^2$.
Alternatively, one may use conformal polynomials \cite{Hill:2010yb} in
the interval $0\leq Q^2\leq Q_{\rm low}^2$, i.e.
\be
  \Pi(Q^2)=\Pi(0)+\sum_{n=1}^{\infty}\,p_n w^n,\quad
  w=\frac{1-\sqrt{1+z}}{\sqrt{1+z}}, \quad z=Q^2/4m_\pi^2.
\ee
Given an estimate for $\Pi(0)$ one can determine $\Pihat(Q^2)$ over
the entire momentum range and evaluate the convolution integral. In
the intermediate momentum range, i.e. in the interval $Q_{\rm low}^2
\leq Q^2 \leq Q_{\rm high}^2$ the integration of $f(Q^2)\Pihat(Q^2)$
can be performed numerically using, for instance, the trapezoidal
rule. Typically $Q_{\rm high}^2$ is as large as a few $\gev^2$, and
hence one can use perturbation theory to continue the integration
above $Q_{\rm high}^2$.

Obviously, the success of the hybrid method depends on the
availability of statistically accurate data for $Q^2 \leq Q_{\rm
  low}^2$. In addition, a number of strategies for increasing the
number of data points in the low-momentum region have been
proposed. These include the use of twisted boundary conditions in
Ref.\,\cite{DellaMorte:2011aa} that allow for the realisation of
momenta which differ from the usual integer multiples of $2\pi/L$. To
this end one imposes spatial periodic boundary conditions on the quark
fields up to a phase factor
\cite{Bedaque:2004kc,deDivitiis:2004kq,Sachrajda:2004mi}
\be
  \psi(x+L\hat{k})=\rme^{i\theta_k}\psi(x).
\ee
This is equivalent to boosting the spatial momenta in the quark
propagator by $\theta_k/L$. By a suitable tuning of the phase angle
$\theta_k$ one can thus access much smaller values of $Q^2$ than those
which can be realised by the usual Fourier momenta. A potential
drawback of this procedure is the modification of the Ward identities
of the vacuum polarisation tensor due to twisting
\cite{Aubin:2013daa}, yet a recent investigation showed that
the effect is numerically insignificant \cite{Horch:2013lla}.

\subsubsection{Time moments}

\begin{figure}[t]
\begin{center}
\includegraphics[width=0.7\textwidth]{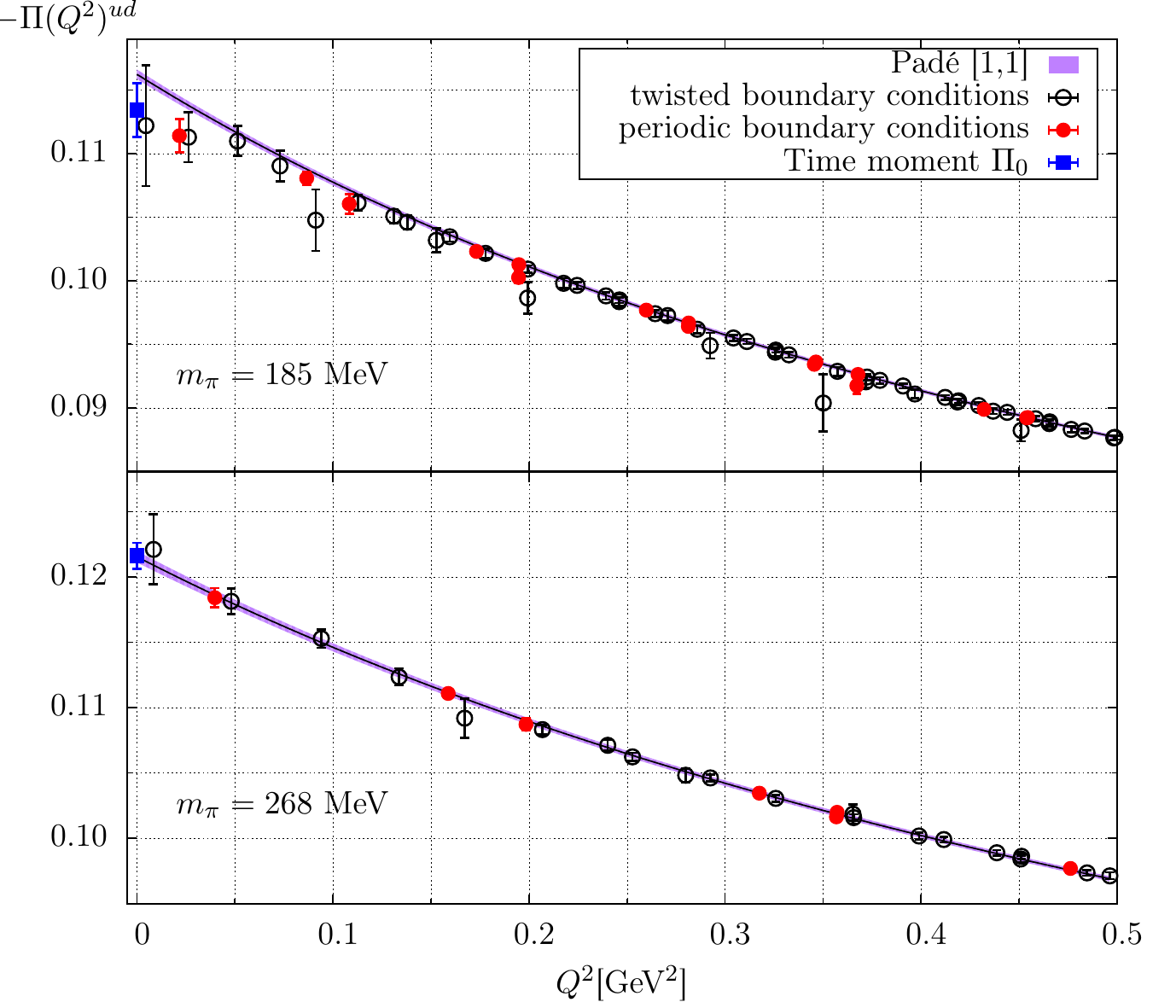}
\caption{The low-momentum representation of the $u, d$ contributions
  to $\Pi(Q^2)$ in terms of a [1,1]-Pad\'e approximation for pion
  masses of 190 (top) and 270\,MeV (bottom), taken from
  Ref.\,\cite{DellaMorte:2017dyu}. The curves represent fits to the
  data in the interval $0 \leq Q^2 \leq 0.5\,\gev^2$.  Blue filled
  squares indicate the value of $\Pi(0)$ determined from the second
  time moment. \la{fig:moments}}
\end{center}
\end{figure}

Another method, proposed in \cite{Chakraborty:2014mwa}, is based on
constructing the Pad\'e representation of $\Pi(Q^2)$ in the interval
$0 \leq Q^2 \leq Q_{\rm low}^2$ from the time moments of the vector
correlator. The starting point is the Taylor expansion
\be
  \Pi(Q^2)=\Pi_0+\sum_{j=1}^{\infty} \Pi_j\, Q^{2j},
\ee
with coefficients $\Pi_0, \Pi_1,\ldots$. Choosing $Q=(\omega,\vec{0})$
one finds that the non-vanishing components of the vacuum polarisation
tensor are given by (see \eq{eq:PimunuQ})
\be
   \Pi_{kk}(\omega)=\int_{-\infty}^{\infty}dx_0\,\rme^{i\omega x_0}\,
   \int d^3x\left\< J_k(x)J_k(0) \right\>.
\ee
If $G(x_0)$ denotes the spatially summed vector correlator defined by
\be\la{eq:Gx0def}
   G(x_0)\delta_{kl} = -\int d^3x\left\< J_k(x)J_l(0) \right\>,
\ee
it is easy to see that the expansion coefficients $\Pi_j$ can be
expressed in terms of the time moments $G_{2n}$ of $G(x_0)$, i.e.
\be\la{eq:momentsdef}
  G_{2n} \equiv \int_{-\infty}^{\infty}dx_0\,x_0^{2n}\,G(x_0)
  = (-1)^n \frac{\partial^{2n}}{\partial\omega^{2n}}
  \left\{ \omega^2\Pi(\omega^2)\right\}_{\omega^2=0}\,.
\ee
The expansion coefficients are then recovered as
\be\la{eq:taylorcoeffs}
   \Pi_j=(-1)^{j+1} \, \frac{G_{2j+2}}{(2j+2)!}.
\ee
In particular, the additive renormalisation $\Pi(0)$ is given by the
second moment, i.e.
\be
   \Pi(0)\equiv\Pi_0=-\frac{1}{2}G_2.
\ee
The Taylor coefficients can then be used to construct the Pad\'e
approximation of $\Pi(Q^2)$. For instance, the coefficients $a_n$ and
$b_m$ of the two lowest order Pad\'e approximations (see \eq{eq:Pade})
are related to the time moments via
\ba
  \Pi_{[1,1]}: && a_1=\Pi_1,\quad b_1=-\Pi_2/\Pi_1 \nonumber \\
  \Pi_{[2,1]}: && a_1=\Pi_1,\quad a_2=(\Pi_2^2-\Pi_1\Pi_3)/\Pi_2,
  \quad b_1=-\Pi_3/\Pi_1, 
\ea
while $\Pi(0)=\Pi_0$. Figure \ref{fig:moments} shows a comparison of
the [1,1]-Pad\'e approximation of $\Pi(Q^2)$ constructed from a
fit to the data for $\Pi(Q^2)$ in the interval $0 \leq Q^2 \leq
0.5\,\gev^2$. The value of the additive renormalisation, $\Pi(0)$,
determined from the intercept agrees well with the estimate from the
second time moment, $\Pi_0=-G_2/2$. Note that the results in the
figure have been obtained by restricting the electromagnetic current
to the quark-connected contributions from up and down quarks, only.

Although the use of time moments avoids the calculation of $\Pi(Q^2)$
at specific values of $Q^2$ as well as the subsequent fit to some {\it
  ansatz}, there are modelling issues that must still be addressed:
The fact that the Pad\'e representation is constructed from time
moments implies that the same considerations regarding any bias must
be applied as in the case where the Pad\'e is determined from fits to
$\Pi(Q^2)$. Secondly, while the moments are obtained by integrating
$G(x_0)$ up to infinitely large Euclidean time separations (see
\eq{eq:momentsdef}), the vector correlator is only accessible for a
finite number of time slices, due to the finite temporal extent of the
lattice and the rapidly decreasing signal-to-noise ratio. Therefore,
some degree of modelling is necessary to extrapolate $G(x_0)$ to
infinity. In fact, this issue becomes even more important for the
higher moments since the large-$|x_0|$ behaviour of the vector
correlator is enhanced by increasing powers of $x_0^2$.

\subsubsection{The time-momentum representation\label{sec:TMR}}

As was first shown in \cite{Bernecker:2011gh} the subtracted vacuum
polarisation function admits an integral representation in terms of
the spatially summed vector correlator, i.e.
\be\la{eq:TMR}
  \Pi(Q^2)-\Pi(0)=\frac{1}{Q^2}\int_0^{\infty}dx_0\, G(x_0)\,\left[
    Q^2x_0^2-4\sin^2\left({\textstyle\frac{1}{2}}Qx_0\right) \right].
\ee
When inserted into the convolution integral, \eq{eq:amublum2}, one can
re-arrange the order of the integrations, leading to the expression
\be\la{eq:TMRamu}
   \ahvp = \left(\frac{\alpha}{\pi}\right)^2\int_0^{\infty}\,
   dx_0\,w(x_0)G(x_0), 
\ee
where the kernel function $w(x_0)$ is given by
\be\la{eq:TMRkernel}
   w(x_0)=4\pi^2\int_0^{\infty}\frac{dQ^2}{Q^2}\,f(Q^2) \left[
    Q^2x_0^2-4\sin^2\left({\textstyle\frac{1}{2}}Qx_0\right) \right],
\ee
and $f(Q^2)$ denotes the momentum-space kernel of \eq{eq:kerK}.

The time-momentum representation is closely related to the expression
for $\Pihat(Q^2)$ in terms of time moments. By expanding the kernel
$\left\{Q^2x_0^2-4\sin^2\left({\textstyle\frac{1}{2}}Qx_0\right)\right\}$
in a Taylor series in $Q^2$ one recovers the expression for the
subtracted vacuum polarisation function in powers of $Q^2$ as
\be
   \Pi(Q^2)-\Pi(0) = \sum_{k=1}^{\infty}
     \left\{\frac{(-1)^{k+1}}{(2k+2)!} \int_{-\infty}^{\infty}
     dx_0\,x_0^{2k+2}\,G(x_0)\right\}Q^{2k}. 
\ee
Here the expression in curly brackets reproduces the time moment
$\Pi_k$, as can be seen from \eqs(\ref{eq:momentsdef})
and~(\ref{eq:taylorcoeffs}). Thus, the time-momentum representation is
equivalent to the exact Taylor series of $\Pihat(Q^2)$.

In both methods, the vector correlator must be integrated up to
infinite Euclidean time. On a finite lattice with temporal dimension
$T$ and periodic boundary conditions the maximum time extension that
can be achieved is $T/2$. More importantly, however, the relative
statistical precision of the vector correlator declines sharply
\cite{Parisi:1983ae,Lepage:1989hd} so that the computed data for
$G(x_0)$ provide only an increasingly inaccurate constraint on the
long-distance part of the integrand in \eq{eq:TMRamu}. It is customary
to split the vector correlator according to
\be
   G(x_0) = \left\{\begin{array}{ll} G(x_0)_{\rm data}, & x_0\leq
   \xcut \\ G(x_0)_{\rm ext}, & x_0> \xcut \end{array} \right.,
\ee
where $\xcut\gtrsim 1.5-2$\,fm, and the subscript ``ext'' indicates
that the correlator is being extended by a continuous function in
$x_0$. 

For the following discussion it is useful to consider the
decomposition of the electromagnetic current into an iso-vector
($I=1$) and an iso-scalar ($I=0$) part, according to
\ba
  & & J_\mu(x) = J_\mu^\rho(x)+J_\mu^{I=0}(x), \nonumber \\
  & & J_\mu^\rho = {\textstyle\frac{1}{2}}(\bar{u}\gamma_\mu u -
  \bar{d}\gamma_\mu d), \quad J_\mu^{I=0}=
      {\textstyle\frac{1}{6}}(\bar{u}\gamma_\mu u +\bar{d}\gamma_\mu d
      -2\bar{s}\gamma_\mu s+\ldots),
\ea
where we have used the superscript $\rho$ to denote the iso-vector
contribution. The associated correlator is defined by
\be
  G^{\rho\rho}(x_0)\,\delta_{kl} = -\int d^3x\,\left\langle
  J_k^\rho(x) J_l^\rho(0) 
  \right\rangle.
\ee
The corresponding isospin decomposition of the vector correlator reads
\be\la{eq:isodecomp}
   G(x_0) = G^{\rho\rho}(x_0) + G(x_0)^{(I=0)},
\ee
and it is important to realise that the iso-vector part $G^{\rho\rho}$
is proportional to the quark-connected light quark contribution
$G^{ud}$ defined according to \eq{eq:Gfdef}, i.e.
\be
   G^{\rho\rho}(x_0)= \frac{9}{10}\,G^{ud}(x_0).
\ee
Since the spectral function in the iso-scalar channel vanishes below
the 3-pion threshold, one expects that $G(x_0)$ is dominated by the
lowest-energy state in the iso-vector channel as $x_0\to\infty$. Thus,
the simplest {\it ansatz} for $G(x_0)_{\rm ext}$ is a single
exponential:
\be
   G(x_0)_{\rm ext} = |A_\rho|^2\,\rme^{-m_{\rho}x_0},
\ee
where $m_\rho$ denotes the $\rho$-meson mass and $A_\rho$ is the
matrix element of the vector current and the vacuum. Obviously, this
{\it ansatz} ignores the fact that the iso-vector correlator is
dominated by the two-pion state as $x_0\to\infty$. The starting point
for a rigorous treatment of the long-distance regime of $G^{\rho\rho}$
is the observation that the spectrum in a finite volume of spatial
dimension $L$ is discrete. The iso-vector correlator is then given by
a sum of exponentials
\be\la{eq:GrhorhoL}
   G^{\rho\rho}(x_0,L) = \sum_n\,|A_n|^2\,\rme^{-\omega_n x_0},\quad
   \omega_n=2\sqrt{m_\pi^2+k^2},
\ee
where the argument of $G^{\rho\rho}$ explicitly indicates that we work
in a finite volume. The sum runs over all energy eigenstates, and
$A_n$ is the matrix element of the iso-vector current between the
$n^{\rm th}$ state and the vacuum. The energies $\omega_n$ are related
to the scattering momentum $k$ via the L\"uscher
condition\,\cite{Luscher:1990ux,Luscher:1991cf}
\be\la{eq:Luscher}
   \delta_{1}(k)+\phi(q)=0\;{\rm mod}\;\pi,\quad q=\frac{kL}{2\pi},
\ee
where $\delta_1$ is the infinite-volume scattering phase shift, and
the function $\phi(z)$ is defined by \cite{Luscher:1991cf}
\be
   \phi(z)=-\frac{\pi^{3/2}z}{{\cal{Z}}_{00}(1;z^2)},\quad
       {\cal{Z}}_{00}(s;z^2)=\frac{1}{\sqrt{4\pi}}
       \sum_{\vec{n}\in \dZ^3}\frac{1}{({\vec{n}}^2-z^2)^s}.
\ee
Below the inelastic threshold, i.e. for
$2m_\pi\leq\sqrt{s}\leq4m_\pi$, the amplitudes $A_n$ can be expressed
in terms of the timelike pion form factor \cite{Meyer:2011um} via a
Lellouch-L\"uscher factor~\cite{Lellouch:2000pv}
\be\la{eq:timelikeFF}
   |A_n|^2=\,\frac{2k^5}{3\pi\omega_n^2}\,
   \frac{|F_\pi(\omega_n)|^2}{q\phi^\prime(q)+k\delta_1^\prime(k)}.
\ee
The $p$-wave scattering phase shift can be determined by computing
suitable correlation matrices, followed by the projection onto the
approximate energy eigenstates via the variational
method\,\cite{Michael:1985ne,Luscher:1990ck} and solving for
\eq{eq:Luscher} (see Refs.\,\cite{Aoki:2007rd, Aoki:2011yj,
  Feng:2010es, Lang:2011mn, Pelissier:2012pi, Guo:2016zos,
  Dudek:2012xn, Feng:2014gba, Wilson:2015dqa, Bali:2015gji,
  Bulava:2016mks, Erben:2016zue, Erben:2017hvr}). The matrix elements
$A_n$ can be determined from ratios of correlators involving the
vector current and the linear combination of interpolating operators
that represent the $n^{\rm th}$ energy
eigenstate\,\cite{Feng:2014gba,Bulava:2015qjz,Erben:2017hvr}. As a
side remark we note that the matrix elements $|A_n|$ and the
associated timelike pion form factor allow for a reliable
determination of finite-volume corrections to $\ahvp$ (see
Section\,\ref{sec:FVE}).

\begin{figure}[t]
  \centering
  \includegraphics[width=0.7\textwidth,clip]{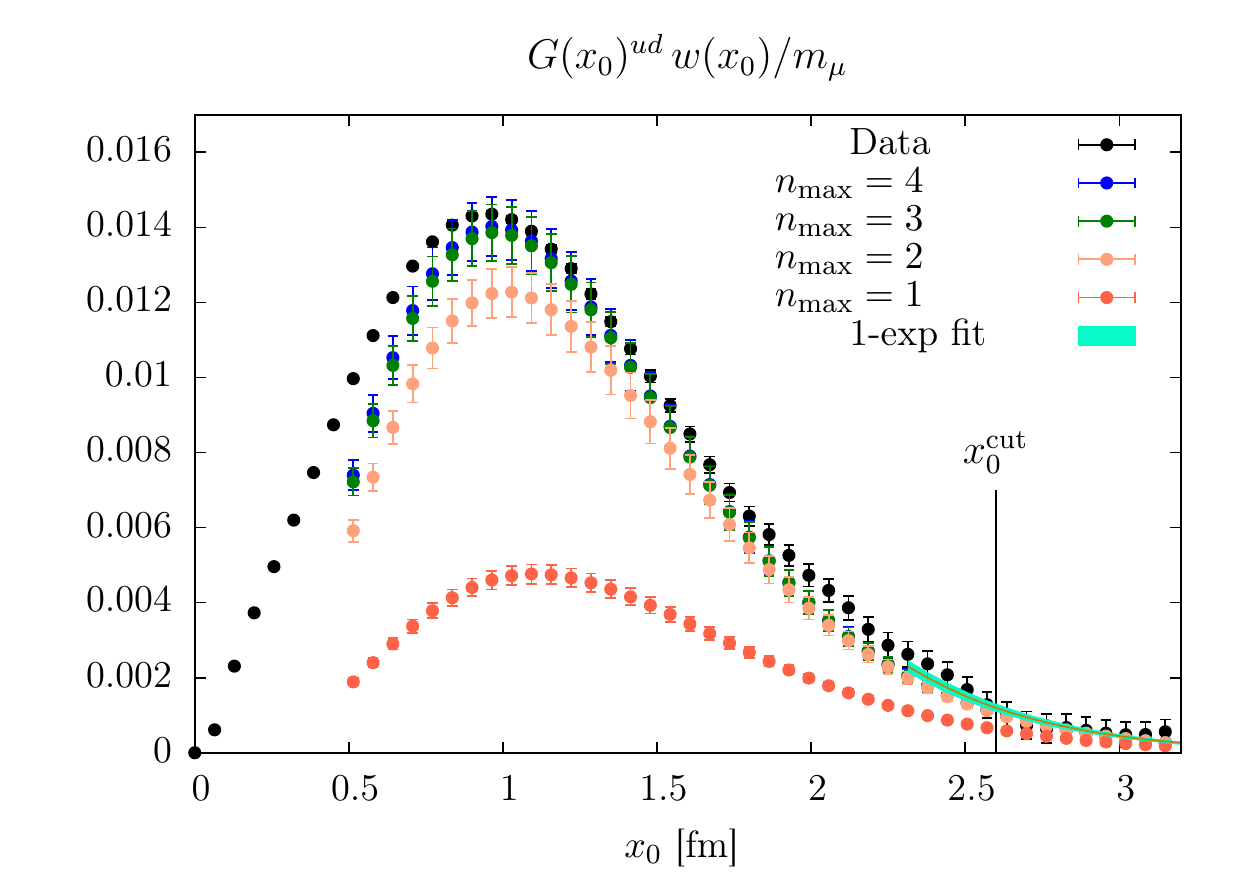}
  \caption{\label{fig:integrand2p1} The light quark contribution to
    the integrand, $w(x_0)G(x_0)^{ud}$, in units of $m_\mu$, computed
    for $\Nf=2+1$ at $m_\pi=200$\,MeV \cite{DellaMorte:2017khn}. Black
    filled squares represent the direct calculation of the spatially
    summed vector correlator. The red circles denote the two-pion
    contribution to the iso-vector correlator $G^{\rho\rho}$, with the
    remaining coloured points showing the accumulated contributions
    from the higher excited states. The green band denotes the naive
    single-exponential ansatz for the extension of $G(x_0)^{ud}$.}
\end{figure}

Replacing the infinite sum in \eq{eq:GrhorhoL} by the sum over a
handful of lowest-lying states is an excellent approximation of the
iso-vector correlator $G^{\rho\rho}(x_0,L)$ for
$x_0\gtrsim1.5$\,fm. Since the iso-scalar contribution to $G(x_0)$ is
sub-dominant, one may replace $G(x_0)^{(I=0)}$ in \eq{eq:isodecomp} by
a single exponential whose fall-off is given by $m_\omega\approx
m_\rho$. In Figure\,\ref{fig:integrand2p1} we show a calculation of
the light-quark contribution $w(x_0)\,G^{ud}(x_0)$ to the integrand in
\eq{eq:TMRamu} by CLS/Mainz\,\cite{DellaMorte:2017khn}. It is obvious
that the statistical accuracy of the direct calculation (represented
by the black points) deteriorates for $x_0\gtrsim2$\,fm. By contrast,
a much more precise determination of the long-distance regime is
obtained through the auxiliary calculation of
$G^{\rho\rho}(x_0,L)$. In particular, one finds that the first four
lowest-lying states saturate the correlator for
$x_0\gtrsim1.2$\,fm. Furthermore, the two-pion contribution, shown in
red, is clearly visible and dominates the correlator for distances
$x_0\gtrsim3.0$\,fm. It is also interesting to note that a naive
single exponential, shown by the green band, provides a fairly good
description of the tail of the correlator.

\begin{figure}[t]
\centering
\includegraphics[width=0.8\textwidth,clip]{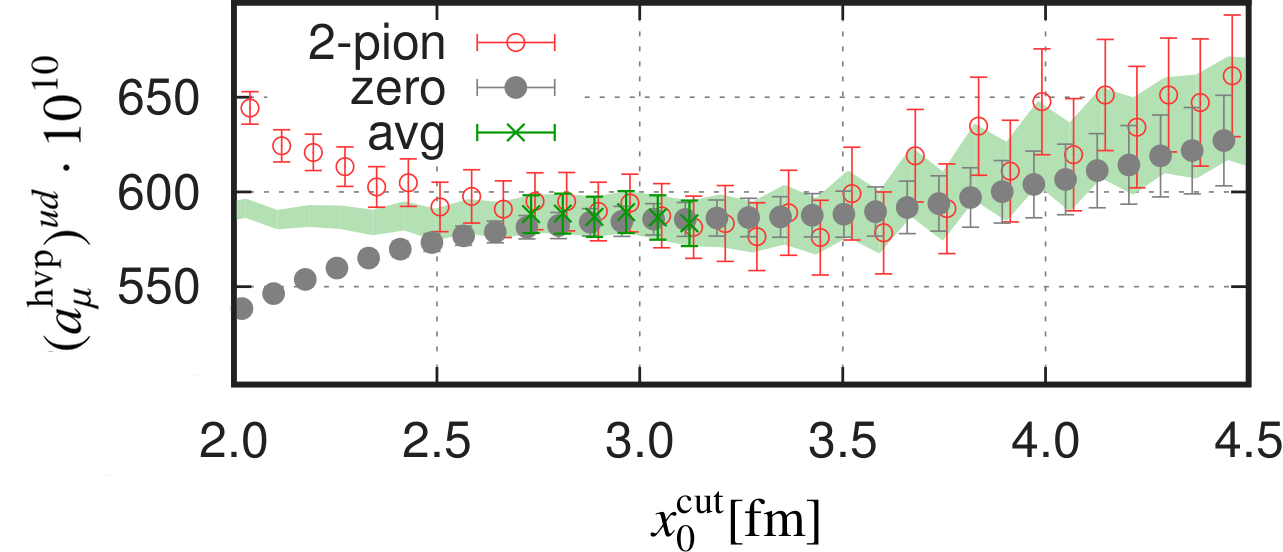}
\caption{\label{fig:bounding} Illustration of the ``bounding method''
  from Ref.\,\cite{Borsanyi:2017zdw}. The red and grey data points
  denote the estimates of the light quark contribution to $\ahvp$
  obtained by inserting the upper and lower bounds in the evaluation
  of the convolution integral for $x_0>\xcut$. Green crosses represent
  the average of the upper and lower bounds in the regime where they
  coincide.}
\end{figure}

A simple method for constraining the large-$x_0$ behaviour of $G(x_0)$
is described in Ref. \cite{Borsanyi:2016lpl}. On a lattice with
temporal and spatial dimensions $T$ and $L$, the correlator $G(x_0)$
is expected to be dominated by a two-pion state as
$x_0\to\infty$. Asymptotically, the corresponding correlator
$G^{2\pi}(x_0)$ has the form
\be
   G^{2\pi}(x_0) \propto
   \left({\rme}^{-E_{2\pi}x_0} + {\rme}^{-E_{2\pi}(T-x_0)} \right).
\ee
For the purpose of constraining the long-distance regime of $G(x_0)$
one may approximate the energy level $E_{2\pi}$ by the energy of two
non-interacting pions whose momenta are each given by the smallest
non-vanishing value $2\pi/L$, i.e.
\be
   E_{2\pi}=\sqrt{m_\pi^2+\left(\frac{2\pi}{L}\right)^2}.
\ee
Since the iso-vector correlator is a sum of exponentials with positive
semi-definite coefficients, it is bounded from below and above
according to
\be\la{eq:bounding}
   0 \leq G(x_0) \leq G(x_0^{\rm cut})
   \frac{G^{2\pi}(x_0)}{G^{2\pi}(x_0^{\rm cut})},
\ee
since $G(x_0)$ must fall off faster than $G^{2\pi}(x_0)$. By
truncating the integration interval in \eq{eq:TMRamu} at $x_0=\xcut$
and inserting the lower and upper bounds in \eq{eq:bounding} to
evaluate the remainder, one can monitor the resulting upper and lower
estimates for $\ahvp$ as a function of
$\xcut$. Figure\,\ref{fig:bounding}, taken from the calculation by the
BMW collaboration\,\cite{Borsanyi:2017zdw}, shows that the upper and
lower bounds agree at $\xcut\approx3.0$\,fm, which coincides with the
observation by CLS/Mainz\,\cite{DellaMorte:2017khn} that the two-pion
states saturates the iso-vector correlator for $x_0\gtrsim3.0$\,fm.

We note that the issue how to constrain the deep infrared regime
concerns all of the methods discussed above: While the direct
calculation of $\Pi(Q^2)$ raises the question how to describe the
low-$Q^2$ regime in an unbiased way, one must address the problem of
describing the long-distance behaviour of $G(x_0)$ when employing the
time-momentum representation or time moments. Moreover, the issue is
intimately linked to the problem of finite-volume effects
\cite{DellaMorte:2017dyu} (see Section\,\ref{sec:FVE}).

\subsubsection{Lorentz-covariant coordinate space representation}

In the time-momentum representation, \eq{eq:TMRamu}, the HVP
contribution is the time integral over the spatially summed vector
correlator $G(x_0)$ multiplied by a weight function $w(x_0)$. As shown
in Ref.\,\cite{Meyer:2017hjv}, $\ahvp$ can also be expressed in terms
of a manifestly Lorentz-covariant integral involving the
point-to-point vector correlator $G_{\mu\nu}(x)\equiv\langle J_\mu(x)
J_\nu(0)\rangle$. A particular benefit of this method may be the
reduction of the noise-to-signal ratio, especially for the
quark-disconnected contribution. The starting point for the derivation
is the representation of $\ahvp$ in terms of the Adler function
$D(Q^2)$:
\be
   D(Q^2) \equiv Q^2\frac{d}{dQ^2}
   \,\Pi(Q^2)=Q^2\int_0^{\infty}ds\,\frac{\rho(s)}{(s+Q^2)^2}\,,
\ee
where the spectral density $\rho(s)$ is related to the $R$-ratio by
\be
  \rho(s)=\frac{R(s)}{12\pi^2},\quad R(s)\equiv
  \frac{\sigma(e^+e^-\to\hbox{hadrons})}{4\pi\alpha^2/(3s)},
\ee
and $\ahvp$ is obtained via the convolution integral as
\cite{Knecht:2003kc}
\be
  \ahvp=2\pi^2\left(\frac{\alpha}{\pi}\right)^2\int_0^1
  \frac{dy}{y}(1-y)(2-y)\,D(Q^2(y)). 
\ee
The integration variable $y$ is related to the Euclidean
four-momentum~$Q$ via
\be
   y=\frac{2|Q|}{|Q|+\sqrt{4m_\mu^2+Q^2}}\quad\Leftrightarrow\quad
   Q^2=\frac{y^2}{1-y}\,m_\mu^2. 
\ee
Equivalently, one can express $\ahvp$ as an integral over $Q^2$ which
can be interpreted as a four-dimensional integral over momentum with
spherical symmetry\,\cite{Meyer:2017hjv}, i.e.
\be\la{eq:Adlerahvp}
  \ahvp=\int_0^{\infty}dQ^2\,D(Q^2)\,g_a(Q^2) = \frac{1}{\pi^2}\int
  \frac{d^4Q}{Q^2}\,D(Q^2)\,g_a(Q^2), 
\ee
where 
\be\la{eq:CCSahvp}
  g_a(Q^2) = 2\alpha^2\frac{m_\mu^4}{|Q|^6}\,(y(|Q|))^4.
\ee
A key observation in Ref.\,\cite{Meyer:2017hjv} is that the Adler
function $D(Q^2)$ is related to the current-current correlator
$G_{\mu\nu}(x)\equiv\langle J_\mu(x)J_\nu(0) \rangle$ via
\be
  D(Q^2)=\frac{1}{3Q^2}\left(\delta_{\mu\nu}-\frac{Q_\mu
    Q_\nu}{Q^2}\right) \int d^4x\,G_{\mu\nu}(x)\,\rme^{iQ\cdot x}\left(
  1-{\textstyle\frac{i}{2}}(Q\cdot x)\right),
\ee
which, when inserted into \eq{eq:Adlerahvp}, yields the HVP
contribution as
\be\la{eq:CCSrep}
  \ahvp = \int d^4x\;G_{\mu\nu}(x)\,H_{\mu\nu}(x).
\ee
The kernel $H_{\mu\nu}(x)$ is given by
\be
  H_{\mu\nu}(x) = \frac{1}{3\pi^2}
      \left(1-\frac{x_\lambda}{2}\frac{\partial}{\partial x_\lambda}
      \right)
      \int\frac{d^4Q}{(Q^2)^2}\,g_a(Q^2)\,\left(\delta_{\mu\nu}-\frac{Q_\mu
      Q_\nu}{Q^2} \right)\,\rme^{iQ\cdot x},
\ee
with $g_a(Q^2)$ specified in \eq{eq:CCSahvp}. In \cite{Meyer:2017hjv}
it was shown that the tensor $H_{\mu\nu}(x)$ can be expressed in terms
of weight functions ${\cal{H}}_1(|x|)$ and ${\cal{H}}_2(|x|)$ that are
analytically computable in terms of Bessel functions. Furthermore, one
finds that, once the space-time indices of $G_{\mu\nu}$ and
$H_{\mu\nu}$ are contracted, the integration over the four-volume
becomes a one-dimensional integral over $|x|$.

Another important result of \cite{Meyer:2017hjv} is the
Lorentz-covariant expression for the slope of the Adler function and,
equivalently, the vacuum polarisation function $\Pi(Q^2)$, i.e.
\be
  D^\prime(0)=\Pi^\prime(0) = \frac{1}{1152}\int
  d^4x\;G_{\mu\nu}(x)\,(x^2)^2\,
  \left(-\frac{7}{4}\delta_{\mu\nu}+\frac{x_\mu x_\nu}{x^2} \right).
\ee
This is the Lorentz-covariant analogue of the relation between the
slope $\Pi^\prime(0)$ and the time-moment $G_4$ (see
\eq{eq:taylorcoeffs}):
\be
  \Pi^\prime(0)=\Pi_1=\frac{1}{4!}\int_{-\infty}^{\infty}dx_0
  \,G(x_0)\,x_0^4.
\ee
The advantage of the covariant integral representation of
\eq{eq:CCSrep} is that only those space-time points are summed over
that contribute to $\ahvp$ up to some particular precision. For
instance, one may define an effective HVP contribution via
\be
   (\ahvp)^{\rm eff}(R) = \int_{|x|<R}
   d^4x\,G_{\mu\nu}(x)\,H_{\mu\nu}(x), 
\ee
in which the integration domain is truncated to a sphere with
radius~$R$. The convergence of $(\ahvp)^{\rm eff}(R)$ towards $\ahvp$
can then be studied systematically by increasing the radius $R$. By
contrast, in the time-momentum representation (and also when computing
$\Pi(Q^2)$ via \eq{eq:PolTens}), the vector correlator is summed over
the entire spatial volume, even though points very far from the origin
barely contribute. This observation also suggests that the estimation
of contributions from quark-disconnected diagrams via the covariant
formulation may be statistically more precise. First results indicate
that this is indeed the case \cite{CCS-inprep}.

\subsubsection{Other methods for determining $\Pi(0)$}

The extensive literature on lattice determinations of $\ahvp$
contains further proposals for computing the additive renormalisation
$\Pi(0)$. 

In Ref.\,\cite{Bali:2015msa} it was noted that the vacuum polarisation
$\Pi(Q^2)$ can be interpreted in terms of magnetic susceptibilities
which, in turn, are defined by taking derivatives of the free energy
with respect to an external magnetic field. For non-zero values of
$Q^2$ the vacuum polarisation is obtained from the susceptibility
derived from a harmonically varying magnetic field. Moreover, the
additive renormalisation $\Pi(0)$ is related to the susceptibility
$\chi_0$ which characterises the response of the system to applying a
homogeneous background field, i.e.
\be
  \chi_0=\Pi(0).
\ee
The main conceptual difficulty arises from the fact that taking
derivatives with respect to a homogeneous magnetic field is not
straightforward, since in a finite volume one has to deal with a
non-vanishing magnetic flux. Several methods have been proposed and
tested \cite{Bonati:2013vba,Bali:2014kia} which give mostly consistent
results. A pilot study using rooted staggered quarks on coarse lattice
spacings shows that this approach yields promising results concerning
the overall accuracy \cite{Bali:2015msa}, yet the method has not been
applied in large-scale calculations of $\ahvp$ aimed at rivalling the
precision of the dispersive method.

A variant of the method that relates $\Pi(0)$ to the time moment $G_2$
via $\Pi(0)\equiv\Pi_0=-G_2/2$ has been proposed in
\cite{deDivitiis:2012vs}. Here the idea is to apply the second
derivative with respect to the momentum directly to the correlation
function of the vector current. The momentum derivatives correspond to
operator insertions in the correlator, so that $\Pi(0)$ can be
computed directly in terms of four-point, three-point and two-point
correlation functions. First results obtained at large pion masses
indicate that $\Pi(0)$ can be obtained with good statistical
precision. The technical challenge of the method consists in isolating
the asymptotic behaviour of three- and four-point correlation
functions.

\subsubsection{Mellin-Barnes representation and time moments
  \la{sec:MB}} 

The difficulty to reach small values of the squared Euclidean momentum
in lattice simulations has been the motivation for several recent
analyses, aimed at providing an alternative representation of $\ahvp$
in terms of quantities that can easily be computed in lattice
calculations
\cite{deRafael:2014gxa,deRafael:2017gay,Benayoun:2016krn}. The
starting point is the Mellin-Barnes representation of the hadronic
vacuum polarisation 
\be\la{eq:MBrep}
   \ahvp = \left(\frac{\alpha}{\pi}\right)^2
   \frac{1}{2\pi i}\int_{c-i\infty}^{c+i\infty}
   \,{\cal{F}}(s){\cal{M}}(s), 
\ee
where the exact kernel function ${\cal{F}}(s)$ is given in terms of
Euler $\Gamma$-functions
\be
   {\cal{F}}(s) = -\Gamma(3-2s)\Gamma(-3+s)\Gamma(1+s),
\ee
and ${\cal{M}}(s)$ denotes the Mellin transform of the hadronic
spectral function\footnote{Here and in \eq{eq:MBrep} we use our
  definition of the electromagnetic current and the vacuum
  polarisation $\hat\Pi(Q^2)$ according to eqs. (\ref{eq:emcurrent})
  and (\ref{eq:Pihat}). This accounts for an extra factor of
  $(\alpha/\pi)$ in the Mellin-Barnes representation compared with
  \cite{deRafael:2014gxa,deRafael:2017gay,Benayoun:2016krn}.}
\be
   {\cal{M}}(s) = \int_{t_0=4m_\pi^2}^\infty
   \frac{dt}{t}\,\left(\frac{t}{t_0}\right)^{s-1}\frac{1}{\pi}\,
   {\rm Im}\hat\Pi(t).
\ee
As proposed in \cite{deRafael:2014gxa} one can perform a low-momentum
expansion of the kernel function ${\cal{F}}(s)$ by calculating its
residues and poles. This yields the expansion of $\ahvp$ in terms of
the Mellin moments \cite{deRafael:2017gay,Charles:2017snx}
\ba
  \ahvp&=& \left(\frac{\alpha}{\pi}\right)^2 \frac{m_\mu^2}{t_0}
  \left\{ \frac{1}{3}{\cal{M}}(0)
  +\frac{m_\mu^2}{t_0} \left[\left(
     \frac{25}{12}-\ln\frac{t_0}{m_\mu^2}\right){\cal{M}}(-1) 
    +\widetilde{\cal{M}}(-1)\right] \right. \nonumber \\
  & &
  +\left(\frac{m_\mu^2}{t_0}\right)^2 \left[\left(
     \frac{97}{10}-6\ln\frac{t_0}{m_\mu^2}\right){\cal{M}}(-2) 
    +6\widetilde{\cal{M}}(-2)\right]  \nonumber  \\
  & &
  +\left(\frac{m_\mu^2}{t_0}\right)^3 \left[\left(
     \frac{208}{5}-28\ln\frac{t_0}{m_\mu^2}\right){\cal{M}}(-3) 
    +28\widetilde{\cal{M}}(-3)\right]
  +\,\rmO\left(\left(m_\mu^2/t_0\right)^4\right) 
  \Bigg\}. \la{eq:MBexp} 
\ea
The key observation is that the moments ${\cal{M}}(-n)$ are related to
the derivatives of $\hat\Pi(Q^2)$ which can be computed on the lattice
from time moments, i.e.
\ba
  {\cal{M}}(-n) \equiv \int_0^\infty\frac{dt}{t}
  \,\left(\frac{t_0}{t}\right)^{n+1} \frac{1}{\pi}\,
  {\rm Im}\,\hat\Pi(t) = \frac{(-1)^{n+1}}{(n+1)!}\,t_0^{n+1}
  \left. \frac{\partial^{n+1}}{(\partial Q^2)^{n+1}}
  \,\hat\Pi(Q^2)\right|_{Q^2=0}. 
\ea
In other words, the determination of the first few terms in the Taylor
expansion of $\hat\Pi(Q^2)$ yields the Mellin transform of the
spectral function at negative integer argument
\cite{deRafael:2017gay}. Computing the slope $\Pi_1$ and the curvature
$\Pi_2$ via \eq{eq:taylorcoeffs} should already provide a precise
estimate of $\ahvp$ due to the good convergence property of the
expansion in terms of the Mellin moments. When applied to
phenomenological models for $\hat\Pi(Q^2)$ such as the one described
in \cite{Bernecker:2011gh}, one finds that the expansion up to
$\rmO((m_\mu^2/t_0)^2)$ already provides an excellent approximation
\cite{deRafael:2014gxa,deRafael:2017gay}.

The expression in \eq{eq:MBexp} also contains the first derivatives of
${\cal{M}}(s)$, defined by
\be
   \widetilde{\cal{M}}(s)\equiv -\frac{d}{ds}{\cal{M}}(s) =
   \int_0^\infty\frac{dt}{t}
   \,\left(\frac{t_0}{t}\right)^{1-s}\ln\frac{t_0}{t} \frac{1}{\pi}\,
  {\rm Im}\hat\Pi(t).
\ee
Their determination is, however, more involved and requires the
evaluation of an integral over the subtracted vacuum polarisation
$\Pi(Q^2)$ weighted by inverse powers of $Q^2$. Lattice calculations
of $\widetilde{\cal{M}}(s)$ will thus be confronted with similar
problems as those encountered for the integral representation of
\eq{eq:amublum2}, but for a convolution function which is not as
strongly peaked at low momenta as $f(Q^2)$. Concrete proposals for the
determination of the log-weighted moments $\widetilde{\cal{M}}(s)$
from lattice data are described in \cite{Benayoun:2016krn}.

In Ref.~\cite{Benayoun:2016krn} the Mellin moments were determined
using experimental data for $e^+ e^-\to\hbox{hadrons}$, and the
resulting values can be used to infer the Taylor coefficients $\Pi_1$
and $\Pi_2$ which can be directly confronted with lattice
calculations. We will present a more detailed discussion in
Section\,\ref{sec:results}. Moreover, in Ref.~\cite{Charles:2017snx}
the Mellin-Barnes technique was advocated as a viable method to derive
a highly precise estimate for $\ahvp$, using the Taylor coefficients
of $\Pihat(Q^2)$ determined either in lattice QCD or from the
experimental spectral function.

\subsubsection{QCD sum rules and the slope of
  $\Pi(Q^2)$. \la{sec:QCDSR}}

Lattice QCD also plays a central role in an approach that combines QCD
sum rules with lattice calculation of the slope of the vacuum
polarisation function $\Pi(Q^2)$ at $Q^2=0$ (i.e. the Taylor
coefficient $\Pi_1$) as well as experimental data for the hadronic
cross section data \cite{Bodenstein:2011qy,Dominguez:2017yga}. The
resulting expression for $\ahvp$ ensures that the latter contribute
only a small part to the overall result, making experimental
uncertainties quite irrelevant. It starts with the observation that
the QED kernel function $\hat{K}(s)$ in \eq{eq:dispersion} varies only
slowly with $s$ \cite{Jegerlehner:2009ry}. One may therefore
approximate it with a meromorphic function $\hat{K}_1(s)$ in the
low-energy region \cite{Bodenstein:2011qy}, e.g.
\be
   \frac{\hat{K}(s)}{s^2} \;\longrightarrow\;
   \frac{\hat{K}_1(s)}{s^2} = \frac{\hat{c}_{-2}}{s^2}
  +\hat{c}_0+\hat{c}_1s,\quad m_{\pi^0}^2\leq s < s_0,
\ee
where $s_0 \approx 4\,{\gev}^2$ delineates the low-energy from the
perturbative region. The coefficients $\hat{c}_{-2}, \hat{c}_0$ and
$\hat{c}_1$ may be determined by requiring
\be
  \int_{m_{\pi^0}^2}^{s_0}\,\hat{K}(s)\,s^{n-2}\,ds =
  \int_{m_{\pi^0}^2}^{s_0}\,\hat{K}_1(s)\,s^{n-2}\,ds, 
\ee
for suitably chosen integers $n$. As shown in \cite{Dominguez:2017yga}
the sum of the contributions from up, down and strange quarks to
$\ahvp$ can be separated into four terms:
\be
   (\ahvp)^{uds} = a_\mu^{\rm SR} +a_\mu^{\rm Lat} +a_\mu^{\rm Exp}
+a_\mu^{\rm Pert},
\ee
where
\ba
  & & a_\mu^{\rm SR} = \left(\frac{\alpha m_\mu}{3\pi}\right)^2
6{\pi}i \oint_{|s|=s_0}ds\,\frac{\hat{K}_1(s)}{s^2}\,\Pi(s), \quad
  a_\mu^{\rm Lat} =\left(\frac{\alpha m_\mu}{3\pi}\right)^2
12\pi^2\,\hat{c}_{-2}\,\Pi_1, \\[0.5ex]
  & & a_\mu^{\rm Exp} =\left(\frac{\alpha m_\mu}{3\pi}\right)^2
\int_{4m_\pi^2}^{s_0} ds
\frac{\hat{K}(s)-\hat{K}_1(s)}{s^2}\,R(s)^{\rm data}, \quad
  a_\mu^{\rm Pert} =\left(\frac{\alpha m_\mu}{3\pi}\right)^2
\int_{s_0}^{\infty} ds \frac{\hat{K}(s)}{s^2}\,R(s)^{\rm
  pQCD}. \nonumber  
\ea
The integral that appears in the expression for the low-energy
contribution $a_\mu^{\rm SR}$ can be evaluated using QCD sum rules
\cite{Dominguez:2017yga}. While $a_\mu^{\rm Exp}$ must be determined
using experimental data for the hadronic cross section ratio $R(s)$,
the influence of experimental uncertainties is greatly diminished
relative to the standard dispersive approach, since $R(s)$ is
multiplied by the difference of kernel functions,
$\hat{K}(s)-\hat{K}_2(s)$, in the integrand. By far the largest
contribution to $\ahvp$ comes from the term $a_\mu^{\rm Lat}$, which
contains the slope of $\Pi(s)$ at $s=0$, a quantity that can be
obtained in lattice QCD, either from the Pad\'e approximation of the
vacuum polarisation function or from time moments. Without going into
further detail concerning the evaluation of $a_\mu^{\rm SR},
a_\mu^{\rm Exp}$ and $a_\mu^{\rm Pert}$, we refer to
Ref.\,\cite{Dominguez:2017yga} and simply quote the final result as
\be
  (\ahvp)^{uds} = \left\{ (183.2\pm2.1)+
  \left(\frac{\alpha
  m_\mu}{3\pi}\right)^2\,12\pi^2\,\hat{c}_{-2}\, \Pi_1 \right\}\cdot
  10^{-10}. 
\ee
After inserting the numerical value for $\hat{c}_{-2}$ determined in
\cite{Dominguez:2017yga},
i.e. $m_\mu^2\,\hat{c}_{-2}/3=2.36\cdot10^{-3}\,{\gev}^2$, one obtains
\be
  (\ahvp)^{uds} = \left\{ (183.2\pm2.1)+
  5027\,\left(\Pi_1\,[{\gev}^{-2}]\right)\,\right\}\cdot
  10^{-10},
\ee
which is easily converted into a estimate for the hadronic vacuum
polarisation, by providing a lattice result for the Taylor coefficient
$\Pi_1$ in units of ${\gev}^{-2}$. The contribution from the charm
quark must also be added before confronting this method with results
from the standard dispersive approach, direct determinations of
$\ahvp$ in lattice QCD and from the approach based on Mellin-Barnes
moments.

\subsection{Quark-disconnected diagrams \la{sec:disc}}

The correlator of the electromagnetic current contains both
quark-connected and quark-disconnected contributions, as depicted in
\fig{fig:conndisc}. Despite the fact that the latter occur frequently
in lattice calculations of a variety of hadronic observables involving
flavour-singlet contributions, they have often been ignored for
technical reasons related to the large level of statistical noise
encountered when the standard techniques for computing quark
propagators are employed. Obviously, neglecting this class of diagrams
amounts to an uncontrolled approximation, and their inclusion is
indispensable if one strives for sub-percent accuracy. For
concreteness, we consider the electromagnetic current of
\eq{eq:emcurrent}, which we write as\footnote{For simplicity, we omit
  the multiplicative renormalisation factor $\zv$ of the local vector
  current on the lattice. See~\ref{app:vector} for details.}
\be
   J_\mu(x)=\sum_{f=u,d,s,\ldots}{\cal{Q}}_f\,
   \psibar_f(x)\gamma_\mu\psi_f(x),
\ee
where ${\cal{Q}}_f$ denotes the electric charge of quark flavour $f$. After
inserting the current into the correlation function and performing the
Wick contractions, one obtains
\ba
  \left\< J_\mu(x)J_\nu(y)\right\> &=& \phantom{+} 
  \sum_f {\cal{Q}}_f^2 \left\< {\rm Tr\,}
  \left\{ \gamma_\nu\gamma_5 S^f(x,y)^\dagger
          \gamma_\mu\gamma_5 S^f(x,y) \right\} \right\>
  \nonumber\\
  & & +\sum_{f,f^\prime} {\cal{Q}}_f {\cal{Q}}_{f^\prime} \left\< 
      {\rm Tr\,}\left\{\gamma_\mu S^f(x,x)\right\}\,
      {\rm Tr\,}\left\{\gamma_\nu S^{f^\prime}(y,y)\right\} \right\>,
\ea
where $S^f$ denotes the quark propagator of flavour $f$, and the
second line corresponds to the diagram depicted on the right in
\fig{fig:conndisc}.

The standard technique for computing the quark propagator $S(x,y)$
amounts to fixing the coordinate $y$ (i.e. the source point) and
inverting the lattice Dirac operator $D$, by solving the linear system
\be\la{eq:Dphieta}
  \sum_z D(x,z)\,\phi(z)=\delta_{xy} \quad\Rightarrow\quad
  \phi(x)=\sum_z D^{-1}(x,z)\,\delta_{zy}=D^{-1}(x,y)\equiv S(x,y).
\ee
The solution $S(x,y)$ is interpreted as the ``point-to-all''
propagator, starting from the (fixed) point $y$ to any space-time
point $x$ on the lattice. Let us now consider the spatially summed
vector correlator $G(x_0)$, which plays a central role for determining
the vacuum polarisation using the time-momentum representation or time
moments. Its connected part is easily obtained from the point-to-all
propagator via
\be
   G_{\rm con}(x_0)=-\frac{a^3}{3}\sum_{k=1}^3  \sum_f {\cal{Q}}_f^2
   \sum_{\vec{x}}\left\< {\rm Tr\,}  
   \left\{ \gamma_k\gamma_5 S^f(x,0)^\dagger
           \gamma_k\gamma_5 S^f(x,0) \right\} \right\>,
\ee
where we have explicitly chosen $y=0$. The disconnected part of
$G(x_0)$ involves the quantity
\be\la{eq:Deltaf}
  \Delta^f(x_0)\equiv a^3\sum_{\vec{x}} {\rm Tr\,}
  \left\{\gamma_k S^f(x,x) \right\},\quad f=ud, s.
\ee
In order to sum over $\vec{x}$ one has to solve the linear system in
\eq{eq:Dphieta} for every spatial coordinate $\vec{x}$ and repeat this
for every timeslice $x_0$ to obtain $\Delta^f(x_0)$. Thereby the
numerical effort is increased by a factor proportional to the 4-volume
of the lattice, which is of order $10^7$. This is prohibitively
costly, and one usually resorts to stochastic techniques in order to
compute the ``all-to-all'' propagator $S(x,y)$ in which the source
point $y$ runs over all points of the lattice. To this end one
generalises \eq{eq:Dphieta} according to
\be\la{eq:alltoall}
  \sum_z\,D(y,z)\phi^{(r)}(z)=\eta^{(r)}(y),
  \quad r=1,\ldots,N_{\rm r}, 
\ee
where $\eta^{(r)}(y)$ is a random noise vector which satisfies
\be\la{eq:stochav}
   \<\<\eta(x)\,\eta^\dagger(y)\>\>\equiv\lim_{N_{\rm r}\to\infty}
   \frac{1}{N_{\rm r}} \sum_{r=1}^{N_{\rm r}}\eta^{(r)}(x)
   \,\eta^{(r)}(y)^\dagger=\delta_{xy}. 
\ee
By $\<\<\cdots\>\>$ one denotes the stochastic average over a sample
of $N_{\rm r}$ random noise vectors. A few lines of straightforward
algebra show that the solution of \eq{eq:stochav}, i.e.
\be
   \phi^{(r)}(x)=\sum_y S^{f}(x,y)\,\eta^{(r)}(y),
\ee
yields $\Delta^f(x_0)$ via the stochastic average involving the
original noise vector $\eta^{(r)}$
\be
   \Delta^f(x_0)=a^3\sum_{\vec{x}} \lim_{N_{\rm r}\to\infty}
   \frac{1}{N_{\rm r}} \sum_{r=1}^{N_{\rm r}}\,{\rm Tr}\,\left\{
   \eta^{(r)}(x)^\dagger \gamma_k \phi^{(r)}(x) \right\}.
\ee

We now return to the spatially summed vector correlator $G(x_0)$ which
is the main quantity for the determination of $\ahvp$ based either on
the time-momentum representation or time moments. We restrict the
discussion to the case of the $u,d,s$ quarks, and hence the current
components with $\mu=k=1, 2, 3$ are given by
\be\la{eq:emcurrentk}
  J_k= {\textstyle\frac{2}{3}}\overline{u}\gamma_k u 
      -{\textstyle\frac{1}{3}}\overline{d}\gamma_k d
      -{\textstyle\frac{1}{3}}\overline{s}\gamma_k s.
\ee
Furthermore, we ignore isospin breaking and set $m_u=m_d$. The
correlator then assumes the form
\ba
  G(x_0)&=& G_{\rm con}^{ud}(x_0)
           +G_{\rm con}^{s}(x_0)
           -G_{\rm disc}(x_0), \la{eq:uds_disc} \\
  G_{\rm disc}(x_0)&=& G_{\rm disc}^{ud}(x_0)
  +G_{\rm disc}^{s}(x_0)-2G_{\rm disc}^{ud, s}(x_0).
\ea
Here we have made the distinction between quark-connected and
-disconnected contributions explicit by using the the subscripts
``con'' and ``disc'', while the superscripts indicate whether the
contribution involves only light $(ud)$, strange $(s)$ or both $(ud,
s)$ quark flavours (for the definition of the connected single-flavour
contribution $G^f(x_0)$, see \eq{eq:Gfdef}). In
Ref.\,\cite{Francis:2014hoa} it was shown that $G_{\rm disc}(x_0)$
factorises according to
\be
  G_{\rm disc}(x_0)=-\frac{1}{9}\left\< 
  \left(\Delta^{ud}(x_0)-\Delta^{s}(x_0)\right)
  \left(\Delta^{ud}(0)-\Delta^{s}(0)\right) \right\>.
\ee
It is now important to realise that the stochastic noise in the
evaluation of the disconnected part largely cancels in the difference
$(\Delta^{ud}-\Delta^{s})$, provided that the same noise vectors
$\eta^{(r)}$ are used to compute the individual estimates for
$\Delta^{ud}$ and $\Delta^{s}$. In
refs. \cite{Francis:2014hoa,Gulpers:PhD2015} it was demonstrated that
this is indeed the case and that the gain in statistical precision
amounts to almost two orders of magnitude.

There are several refinements of the method, designed to suppress the
intrinsic stochastic noise. One is based on the hopping parameter
expansion (HPE) of the quark propagator: The Wilson-Dirac operator can
be expressed as a sum of two terms, one of which is diagonal in
coordinate space, while the other one, the hopping term $H$, encodes
the nearest-neighbour interactions
\be
   D_{\rm w} = \frac{1}{2\kappa}\mathbb{1}-\frac{1}{2}H.
\ee
The hopping parameter $\kappa$ is related to the bare quark mass of
flavour $f$, $m_f$, via
\be
   \kappa=\frac{1}{2am_f+8}.
\ee
With these definitions one can express the quark propagator as
\cite{Thron:1997iy,Bali:2009hu} 
\be
   S(x,y)=2\kappa\sum_{k=0}^{N-1}\,(\kappa H)^k+(\kappa H)^N
   D_{\rm w}^{-1}(x,y). 
\ee
When $D_{\rm w}^{-1}$ is computed using the noise sources as described
above, the stochastic noise is further suppressed by a factor
$\kappa^N$, where $N$ denotes the order of the HPE. The factors
$(\kappa H)^k$, on the other hand, contain only products of the
hopping matrix $H$ and are cheap to evaluate. The HPE can also be
adapted to the case of $\rmO(a)$ improved Wilson fermions
\cite{Gulpers:2013uca,Gulpers:2015bba}.

Another method for achieving stochastic noise cancellation makes use
of the spectral decomposition of the quark propagator in terms of
eigenvectors of the lattice Dirac operator
\cite{Foley:2005ac,Giusti:2004yp}
\be\la{eq:specdecomp}
   S(x,y)=\sum_{k=1}^{N_{\rm ev}} 
   \frac{v_k(x)\otimes v_k(y)^\dagger}{\lambda_k} +S_\perp(x,y)
\ee
where the sum runs over the $N_{\rm ev}$ lowest eigenmodes $v_k(x)$ of
the Dirac operator with eigenvalue $\lambda_k$. Stochastic sources are
only applied in the calculation of $S_\perp(x,y)$, which is the
propagator restricted to the orthogonal complement of the subspace
spanned by the low modes. If one chooses $N_{\rm ev}$ large enough, so
that $\lambda_{N_{\rm{ev}}}\simeq m_s$ then the stochastic noise in
the calculation of $\Delta^{ud}$ will be suppressed by a factor of
$\rmO(m_s^{-1})$, and the signal for $G_{\rm disc}(x_0)$ will be
dominated by the low-mode contribution \cite{Blum:2015you}.

Several methods have been developed and tested in order to minimise
the stochastic noise in the calculation of the individual flavour
contribution $\Delta^f$, such as the application of suitable
``dilution schemes'' \cite{Wilcox:1999ab,Foley:2005ac} which improve
the convergence towards the right-hand-side of
\eq{eq:stochav}. Furthermore it was found that the use of
four-dimensional random noise vectors, either at fixed momentum
\cite{Bali:2015msa} or in combination with hierarchical probing
\cite{Stathopoulos:2013aci,Green:2015wqa} is particularly efficient in
suppressing stochastic noise.

\begin{figure}[t]
\begin{center}
\includegraphics[width=0.7\textwidth]{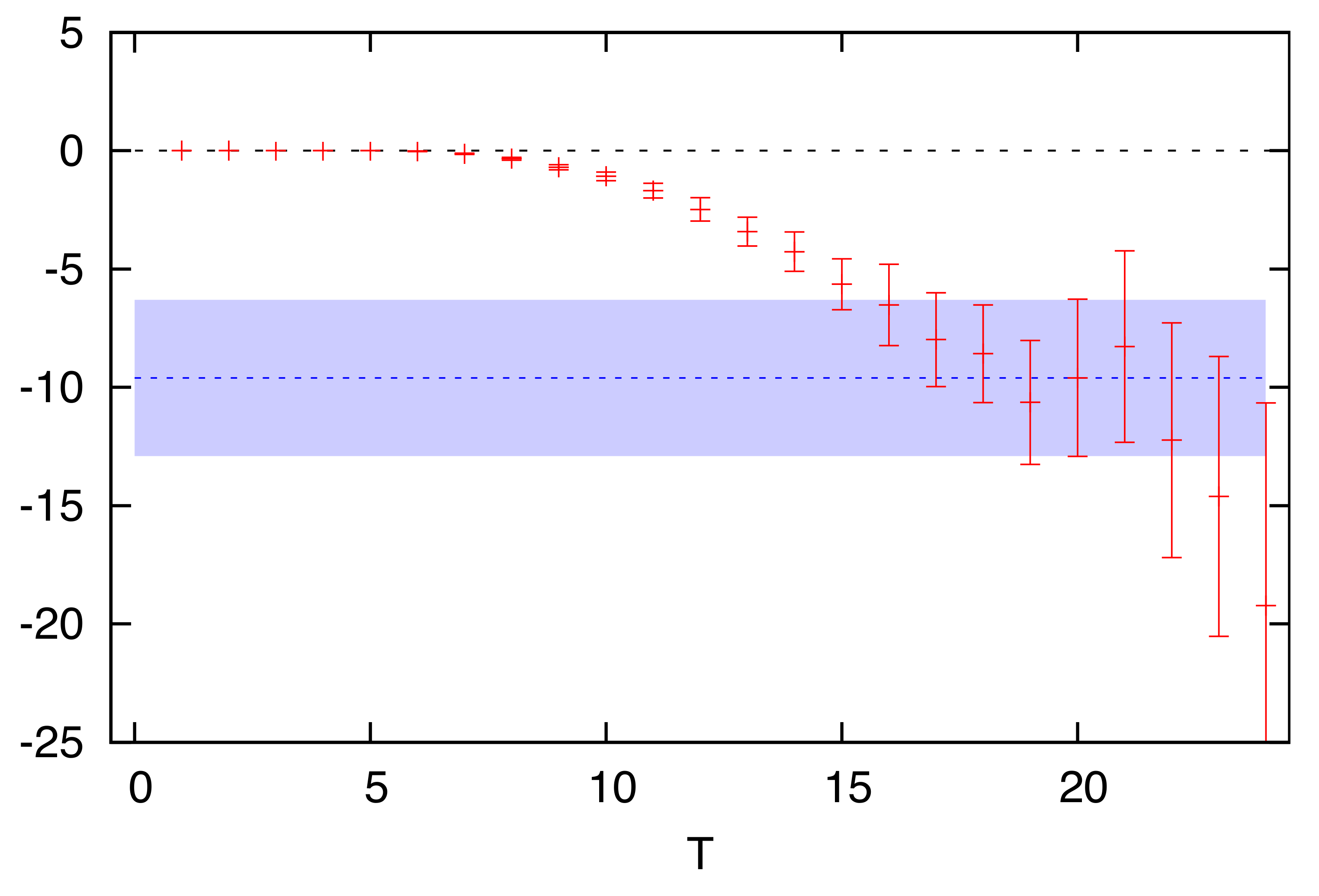}
\caption{The effective disconnected contribution $(\ahvp)_{\rm
    disc}^{\rm eff}(T)$ to the hadronic vacuum polarisation (see
  \eq{eq:ahvpdisc}), as computed in \cite{Blum:2015you} plotted versus
  $T$ in units of $10^{-10}$. The blue band denotes the value at
  $T=20$. \la{fig:RBCdisc}}
\end{center}
\end{figure}

We will now discuss three specific calculations
\cite{Blum:2015you,DellaMorte:2017dyu,Borsanyi:2016lpl} of the
quark-disconnected contribution which are all based on the
time-momentum representation. The RBC/UKQCD
Collaboration\,\cite{Blum:2015you} have employed the spectral
decomposition of \eq{eq:specdecomp} to compute the disconnected part
$G_{\rm disc}(x_0)$ on gauge ensembles generated using domain wall
fermions at the physical pion mass and a lattice spacing of
$0.114\,\fm$. In order to quantify the contribution from
quark-disconnected diagrams, defined by
\be\la{eq:ahvpdisc}
  (\ahvp)_{\rm disc}=\left(\frac{\alpha}{\pi}\right)^2 \int_0^\infty
  dx_0\,(-G_{\rm disc}(x_0))\,w(x_0),
\ee
with $w(x_0)$ given in \eq{eq:TMRkernel}, one may consider the
effective disconnected contribution
\be
   (\ahvp)_{\rm disc}^{\rm eff}(T) \equiv \left(\frac{\alpha}{\pi}\right)^2
   \int_0^T\,(-G_{\rm disc}(x_0))\,w(x_0).
\ee
As $T\to\infty$ this quantity converges towards $(\ahvp)_{\rm
  disc}$. A plateau in a plot of $(\ahvp)_{\rm disc}^{\rm eff}(T)$
versus $T$ would signal that the sum in \eq{eq:ahvpdisc} is
saturated. As indicated in \fig{fig:RBCdisc}, RBC/UKQCD find that the
asymptotic value is reached for $T/a\gtrsim 17$, and the resulting
estimate for the quark-disconnected contribution is
\cite{Blum:2015you}
\be
   (\ahvp)_{\rm disc}=-(9.6\pm3.3\pm2.3)\cdot 10^{-10}.
\ee
Here the first error is statistical, and the second is an estimate of
systematic effects such as discretisation and finite-volume effects.

\begin{figure}[t]
\begin{center}
\leavevmode
\includegraphics[width=6.5cm]{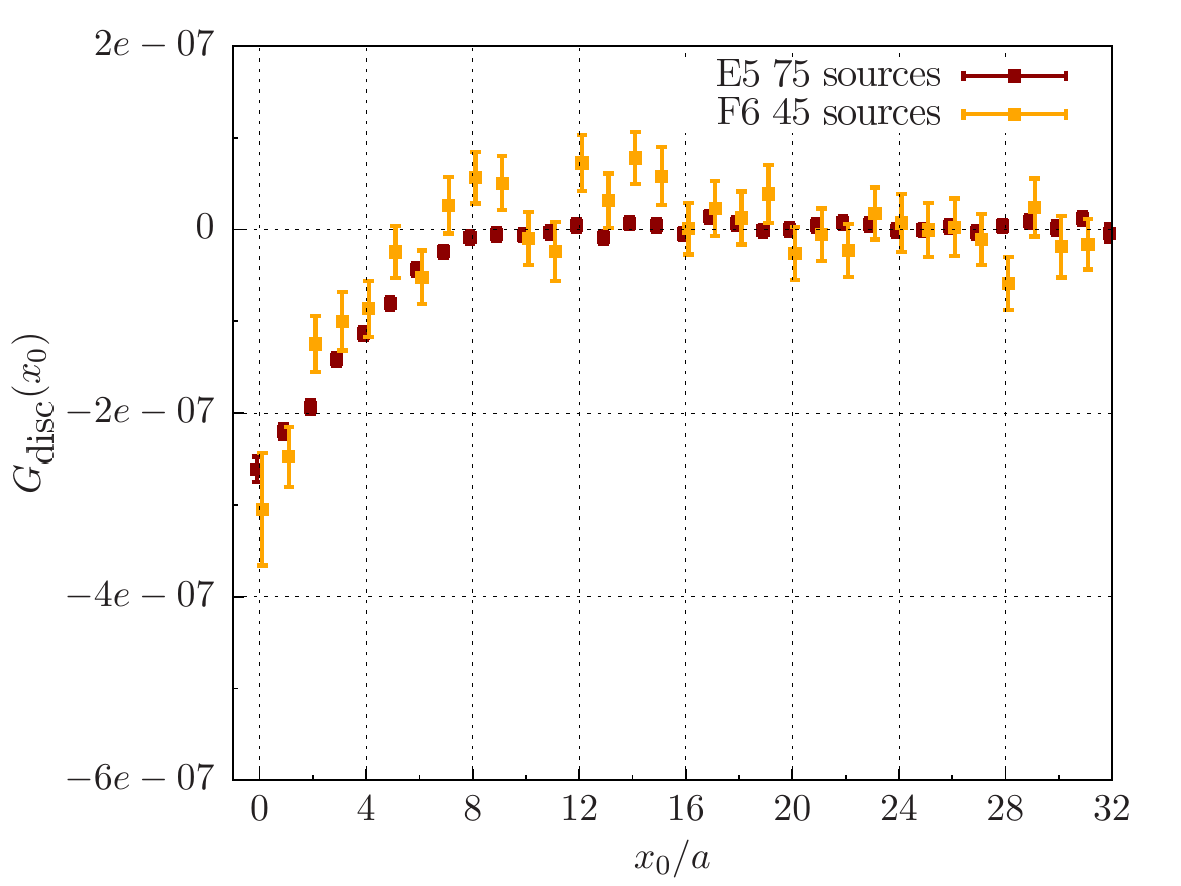}
\includegraphics[width=6.5cm]{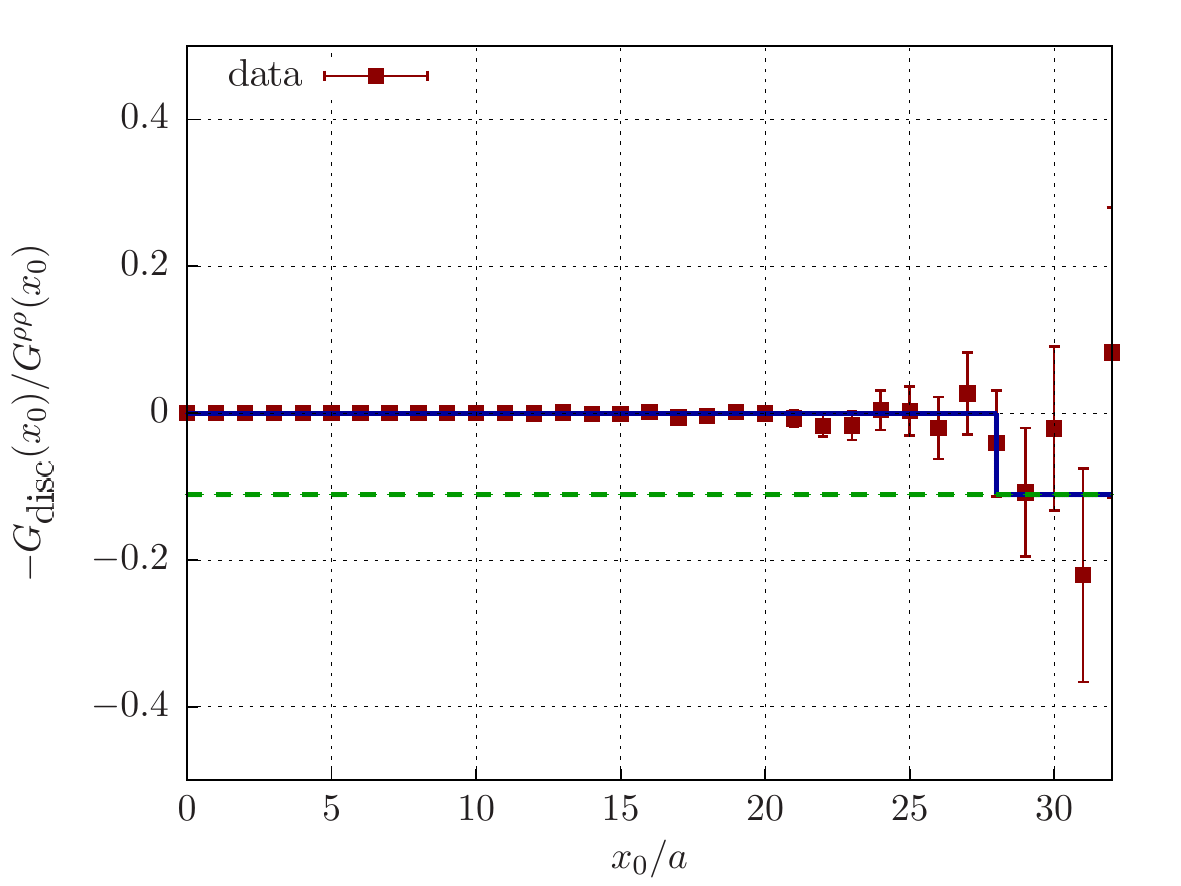}
\caption{Left: the disconnected contribution $G_{\rm{disc}}$ in
  two-flavour QCD at pion masses of $440\,\mev$ (red squares) and
  $310\,\mev$ (yellow squares) \cite{DellaMorte:2017dyu}. Right: the
  ratio defined in \eq{eq:discasympdef} at $m_\pi=440\,\mev$. The data
  are compatible with \eq{eq:discasympdef} which is represented by the
  blue line.\la{fig:MZdisc}}
\end{center}
\end{figure}

Another determination of the disconnected contribution was performed
by CLS/Mainz \cite{DellaMorte:2017dyu}, using two flavours of
$\rmO(a)$ improved Wilson fermions at pion masses of 440 and
$310\,\mev$ and a lattice spacing of $a=0.066\,\fm$. Here the
disconnected part was computed employing stochastic noise cancellation
as described above. In
Refs.\,\cite{Francis:2014hoa,DellaMorte:2017dyu} the Mainz group
describes how an upper bound on the magnitude of the disconnected
contribution can be derived.\footnote{See also contribution 2.16 in
  \cite{Benayoun:2014tra}.} Recalling the isospin decomposition of the
vector correlator in \eq{eq:isodecomp}, i.e. $G(x_0) =
G^{\rho\rho}(x_0) + G(x_0)^{(I=0)}$, one can identify the iso-vector
and iso-scalar contributions as
\ba
  G^{\rho\rho}(x_0) &=& \frac{9}{10}G^{ud}_{\rm con}(x_0),
      \nonumber \\
  G^{(I=0)}(x_0) &=& \frac{1}{10}G^{ud}_{\rm con}(x_0)
     +G^{s}_{\rm con}(x_0) -G_{\rm disc}(x_0). \la{eq:Gx0isoscalar}
\ea
Using \eq{eq:uds_disc} one can then derive the relation
\be\la{eq:discratio}
   -\frac{G_{\rm disc}(x_0)}{G^{\rho\rho}(x_0)}=
   \frac{G(x_0)-G^{\rho\rho}(x_0)}{G^{\rho\rho}(x_0)} -\frac{1}{9}
   \left(1+9\frac{G^s_{\rm con}(x_0)}{G^{\rho\rho}(x_0)} \right).
\ee
Since the iso-scalar spectral function vanishes below the three-pion
threshold, the long-distance behaviour of the iso-scalar correlator is
given by $G^{(I=0)}(x_0)\sim \rme^{-3m_\pi x_0}$. According to
\eq{eq:Gx0isoscalar} this implies
\ba
  G_{\rm disc}(x_0) &=& \left(
    {\frac{1}{10}}G^{ud}_{\rm con}(x_0)+G^{s}_{\rm con}(x_0) \right)
    \cdot(1+\rmO(\rme^{-m_\pi x_0})), \\
  G(x_0) &=& G^{\rho\rho}(x_0)\cdot (1+\rmO(\rme^{-m_\pi x_0}))
\ea
in the deep infrared. In this way one can derive the asymptotic
behaviour of the ratio in \eq{eq:discratio} in the long-distance
regime as
\be
   -\frac{G_{\rm disc}(x_0)}{G^{\rho\rho}(x_0)}
   \stackrel{x_0\to\infty}{\longrightarrow} -\frac{1}{9},
\ee
where it is taken into account that $G^s_{\rm con}(x_0)$ drops off
faster than $G^{\rho\rho}(x_0)$ due to the heavier mass of the strange
quark. 

Data for the correlator ratio $-{G_{\rm disc}}/{G^{\rho\rho}}$ are
shown in the right-hand panel of \fig{fig:MZdisc}
\cite{DellaMorte:2017dyu}. While there is no visible trend that the
ratio approaches the asymptotic value of $-1/9$ for $x_0/a\lesssim 26$
or $x_0\lesssim1.7$\,fm, one may still derive an upper bound on the
magnitude of the disconnected contribution: To this end one assumes
that the ratio drops to $-1/9$ at $x_0=x_0^\ast$, which marks the
timeslice where the precision is insufficient to distinguish between
zero and the expected asymptotic value. In other words, $x_0^\ast$ is
chosen such that the data are statistically compatible with
\be\label{eq:discasympdef}
   -\frac{G_{\rm disc}(x_0)}{G^{\rho\rho}(x_0)}=
   \left\{ \begin{array}{cl} 0, & x_0\leq x_0^\ast, \\
     -1/9, & x_0 > x_0^\ast
     \end{array} \right.
\ee
One can now define the relative size of the connected and disconnected
contributions to $\ahvp$ via
\be\la{eq:discbound}
  \Delta\ahvp\equiv -\frac{(\ahvp)_{\rm disc}}{(\ahvp)_{\rm con}} > 0,
\ee
with $(\ahvp)_{\rm disc}$ defined in \eq{eq:ahvpdisc}. After inserting
\eq{eq:discasympdef} one obtains the upper bound on the magnitude of
$(\ahvp)_{\rm disc}$ as
\be
   \left|(\ahvp)_{\rm disc}\right|^{\rm (max)} =
   \frac{1}{10}\left(\frac{\alpha}{\pi}\right)^2
   \int_{x_0^\ast}^\infty dx_0\,G^{ud}_{\rm con}(x_0))\,w(x_0).
\ee
In their two-flavour calculation\,\cite{DellaMorte:2017dyu} CLS/Mainz
find that $\Delta\ahvp$ is less than 1\% for a pion mass of 440\,MeV
but that the magnitude increases to 2\% for $m_\pi=310$\,MeV, which is
the estimate quoted in \tab{tab:hvpdisc} and represented by the red
band in \fig{fig:hvpdisc}.


Not only the fraction of the disconnected contribution to $\ahvp$ has
been the subject of recent investigations, but also the determination
of the ratio $\Pi^{\rm disc}/\Pi^{\rm con}$ of the subtracted vacuum
polarisation function $\Pihat(Q^2)$ itself. An analytic study based on
chiral perturbation theory (ChPT) at NLO \cite{DellaMorte:2010aq} has
found the result $\Pi^{\rm disc}/\Pi^{\rm con}=-1/10$, implying that
disconnected contributions reduce the vacuum polarisation function by
10\%. Moreover, general arguments based on properties of spectral
functions that are related to $\Pi(Q^2)$ via the optical theorem also
produce the value $-1/10$ for the long-distance part in $\Pi^{\rm
  disc}/\Pi^{\rm con}$ \cite{Francis:2013fzp}. Recently, the ChPT
calculation has been extended by including some of the two-loop
contributions\,\cite{Bijnens:2016ndo,Bijnens:2017esv}. In this way the
Taylor expansion of the ratio $\Pi^{\rm disc}/\Pi^{\rm con}$ is
obtained as
\be
   \frac{\Pi^{\rm disc}}{\Pi^{\rm con}}(Q^2) = -0.0353+0.031\left(
   Q^2\,[\gev]^2 \right),
\ee
which implies that higher-order corrections reduce the magnitude of
the disconnected contributions by roughly a factor three relative to
the NLO estimate.

\begin{figure}[t]
\begin{center}
\includegraphics[width=11cm,trim=-0.25cm 0 0 0]{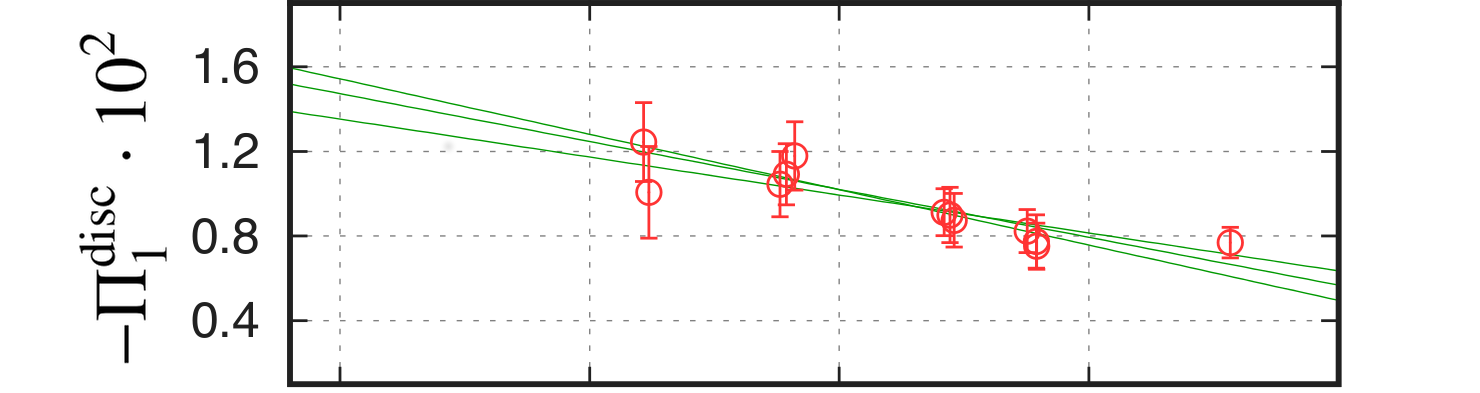}\\[0.3cm]
\hspace*{-0.16cm}\includegraphics[width=10.17cm,trim=0 0 0 0]{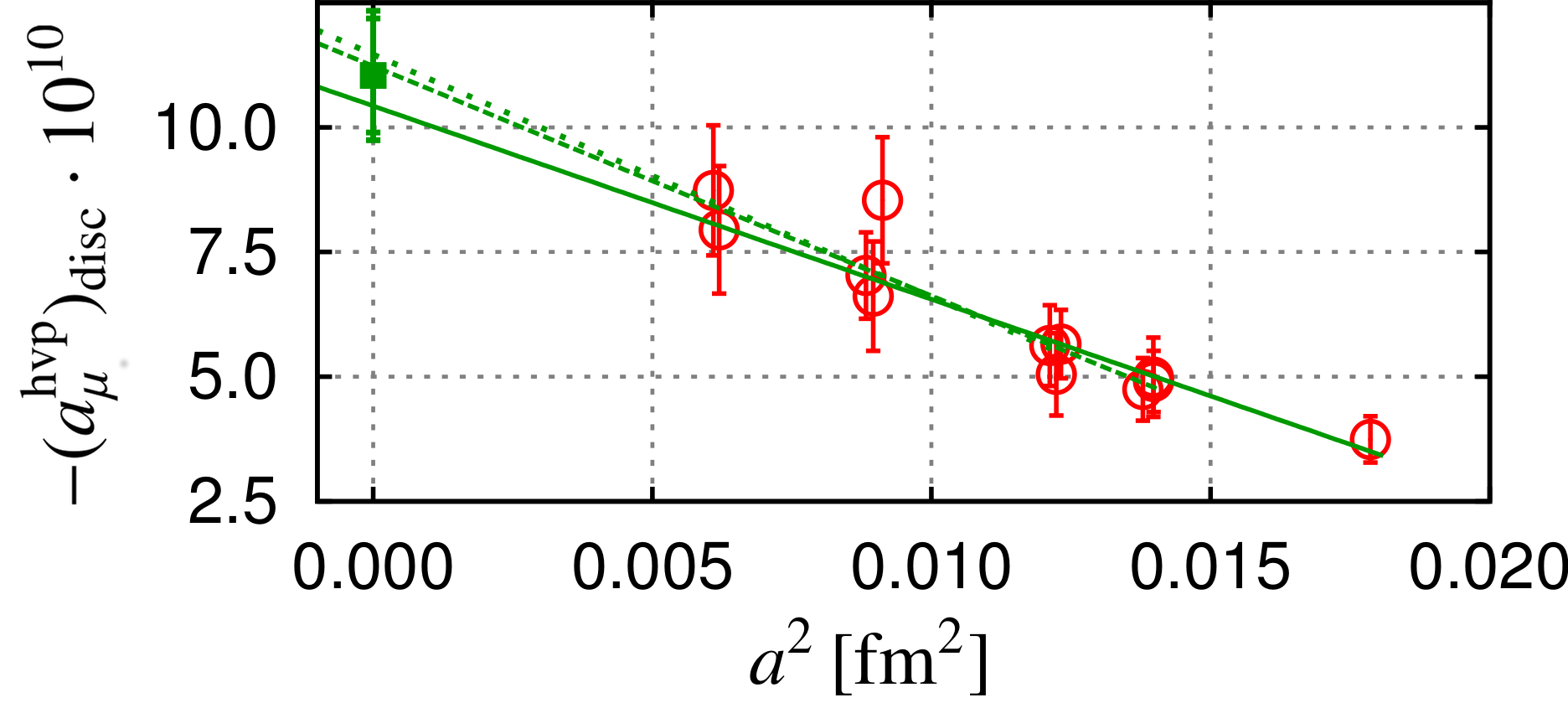}
\caption{The continuum extrapolation of the quark-disconnected
  contribution to the leading time moment $\Pi_1$ of the vacuum
  polarisation function (top) and to $\ahvp$ (bottom) computed by the
  BMW Collaboration
  \cite{Borsanyi:2016lpl,Borsanyi:2017zdw}. \la{fig:BMWdisc}}
\end{center}
\end{figure}

These results can be contrasted with a direct determination of the
connected and disconnected contributions to the lowest two time
moments, $\Pi_1$ and $\Pi_2$, calculated by the BMW
Collaboration\,\cite{Borsanyi:2016lpl} at the physical pion mass. By
employing a massive 6000 stochastic sources and exploiting stochastic
noise cancellation between light and strange quark contributions
\cite{Francis:2014hoa}, the disconnected contributions $\Pi_1^{\rm
  disc}$ and $\Pi_2^{\rm disc}$ could be well resolved. The upper
panel of \fig{fig:BMWdisc} shows the continuum extrapolation of
$\Pi_1^{\rm disc}$ computed at five different lattice spacings. One
clearly sees that the leading disconnected contributions is
negative. From the results of the two leading moments listed in
Table\,II of Ref.\,\cite{Borsanyi:2016lpl} one can infer the Taylor
expansion of the ratio $\Pi^{\rm disc}/\Pi^{\rm con}$ in the continuum
limit as
\be
   \frac{\Pi^{\rm disc}}{\Pi^{\rm con}}(Q^2) = -0.0166(25)+0.021(13)
   \left(Q^2\,[\gev^2]\right) +\rmO(Q^4).
\ee
Hence, for $Q^2=0$, the ratio is given by $\Pi^{\rm disc}/\Pi^{\rm
  con}= -0.0166(25)$ which is $\approx -1/60$ and thus another factor
two smaller in magnitude than the ChPT estimate of
Ref.\,\cite{Bijnens:2016ndo}.


In two recent calculations the quark-disconnected contribution was
computed at the physical value of the pion mass: The absolute
magnitude of the disconnected part was found to be
$(\ahvp)^{\rm{disc}}=-12.8(1.0)\cdot10^{-10}$ by the BMW
Collaboration\,\cite{Borsanyi:2017zdw} (see the lower panel of
\fig{fig:BMWdisc} for a plot of the continuum extrapolation), while
RBC/UKQCD\,\cite{Blum:2018mom} reported
$(\ahvp)^{\rm{disc}}=-11.2(4.0)\cdot10^{-10}$.

\begin{sidewaystable}
\begin{center}
\begin{tabular}{lllccl}
\hline\hline
Collab. && $(\ahvp)_{\rm disc}\cdot10^{10}$
 & $(\ahvp)_{\rm disc}/(\ahvp)_{\rm con}^{\ell\ell}$
 & $\Pi^{\rm disc}/\Pi^{\rm con}$ & Comments \\ 
\hline
RBC/UKQCD &
 & $-11.2(3.3)(2.3)$ & $-1.6(6)\%$ & & $m_\pi=139\,\mev$, \\
 \cite{Blum:2015you,Blum:2018mom} & & & & & domain wall fermions \\[2.0ex]
BMW \cite{Borsanyi:2016lpl,Borsanyi:2017zdw} &
 & $-12.8(1.0)(1.6)$ & $-1.8(3)\%$ & $-0.0166(25)$ &
$m_\pi=139\,\mev$, cont. limit,\\
 & & & &          & staggered fermions \\[2.0ex]
CLS/Mainz \cite{DellaMorte:2017dyu} &
 & & $>-2\%$ & & $m_\pi=437$ and $311\,\mev$, \\
 & & & & & Clover fermions \\[2.0ex]
Bali \& Endr\H{o}di &
 & & & $-0.00036(45)$ & $m_\pi=139\,\mev$, $a=\,0.29\,\fm$,\\
 \cite{Bali:2015msa} & & & &          & staggered fermions \\[2.0ex]
HSC\,15 \cite{Chakraborty:2015ugp} &
 & $-0.8(3)$ & $-0.14(5)\%$ & & $m_\pi=391\,\mev$, Clover fermions,\\
 & & & & & smeared vector current \\
\hline\hline
\end{tabular}
\caption{Compilation of recent results for the quark-disconnected
  contribution to the hadronic vacuum polarisation. The result marked
  by an asterisk has been obtained from $(\ahvp)_{\rm disc}$ assuming
  $(\ahvp)_{\rm{con}}^{\ell\ell}=600\cdot10^{-10}$. \la{tab:hvpdisc}}
\end{center}
\end{sidewaystable}

\begin{figure}[t]
\begin{center}
\leavevmode
\includegraphics[width=10cm]{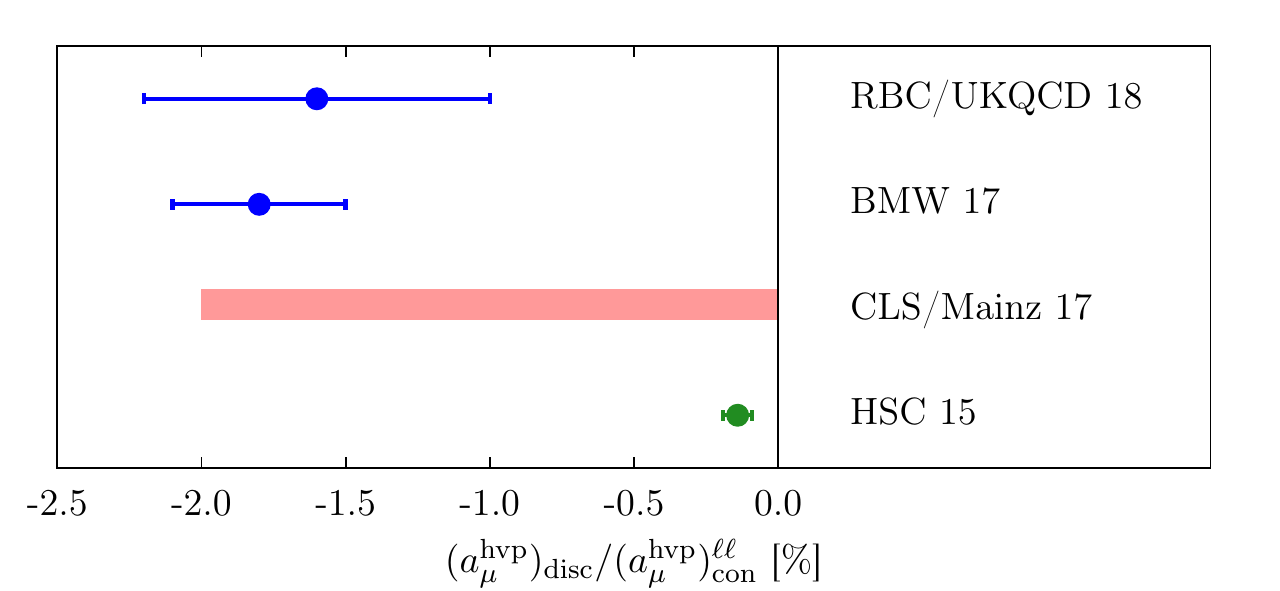}
\caption{Ratio of the quark-disconnected and light quark-connected
  contribution to $\ahvp$ as determined by RBC/UKQCD
  \cite{Blum:2018mom}, BMW \cite{Borsanyi:2017zdw}, CLS/Mainz
  \cite{DellaMorte:2017dyu} and HSC \cite{Chakraborty:2015ugp}. The
  red band denotes the range allowed by the lower bound determined via
  \eq{eq:discbound}. The green circle indicates that the result has
  been obtained using smeared sources.\la{fig:hvpdisc}}
\end{center}
\end{figure}

A compilation of recent results for the quark-disconnected
contribution $(\ahvp)_{\rm disc}$, as well as the ratio $\Pi^{\rm
  disc}/\Pi^{\rm con}$ is shown in Table\,\ref{tab:hvpdisc}. With the
recent determinations of Refs. \cite{Borsanyi:2017zdw} and
\cite{Blum:2018mom} a consistent picture emerges: It is obvious that
disconnected diagrams have only a minor influence on the total value
of $\ahvp$: their contribution is negative and amounts to $1.5-2$\% in
magnitude. This is also demonstrated by the plot shown in
Figure\,\ref{fig:hvpdisc}. 

In summary one finds that quark-disconnected contributions to $\ahvp$
can nowadays be quantified reliably thanks to a number of technical
improvements. Their overall magnitude is estimated to be at the level
of a percent, which implies that they are important regarding the
overall target precision. The accuracy achieved in the most recent
determinations shows that they do not represent a serious obstacle for
reaching the goal of making lattice calculations of $\ahvp$ at least
as precise as the dispersive analysis.

\subsection{Finite-volume effects \la{sec:FVE}}

As we shall see below, the effects induced by performing calculations
in a finite volume lead to sizeable corrections when the minimum pion
mass in units of the spatial box length satisfies $m_\pi L\approx4$
. Obviously, a very accurate determination of finite-volume
corrections is necessary, in order to estimate $\ahvp$ with the
desired level of overall precision.

The empirical evidence from calculations of hadron masses and decay
constants suggests that finite-volume effects are negligibly small if
the spatial box size $L$ satisfies $m_\pi L \gtrsim 4$, where $m_\pi$
is the actual value of the pion mass in the simulation. By contrast,
the determination of $\ahvp$ in lattice QCD appears to be much more
sensitive to finite-size effects such that volumes in excess of
$L=6\,\fm$ may be required.\footnote{At the physical pion mass of
  $m_\pi=139\,\mev$ the condition $m_\pi L=4$ implies $L=5.7\,\fm$.}

Most estimates of finite-volume corrections that enter the current
lattice QCD estimates of $\ahvp$ are based on chiral effective field
theory (EFT). There are also efforts to confront EFT estimates of
finite-volume corrections with lattice data
\cite{Aubin:2015rzx,Chakraborty:2016mwy}, as well as scaling studies
employing several different volumes \cite{Malak:2015sla}. There are
some arguments that suggest that $m_\pi L \geq 6$ is necessary to
suppress finite-volume sufficiently
\cite{DellaMorte:2017dyu,Izubuchi:2018tdd}, although more detailed
studies are required to corroborate this.

One particular method to quantify finite-volume corrections is to
consider anisotropy effects in the vacuum polarisation function
$\Pi(Q^2)$. The paper by Aubin et al. \cite{Aubin:2015rzx} starts from
the observation that the vacuum polarisation tensor,
$\Pi_{\mu\nu}(Q)$, does not vanish for $Q=0$ in a finite
volume\,\cite{Bernecker:2011gh}, contrary to what is expected from the
tensor structure in \eq{eq:PimunuQ}, which is valid in infinite
volume. It is then possible to construct the tensor
$\overline\Pi_{\mu\nu}(Q)$ which has the zero mode subtracted and
which satisfies the Ward-Takahashi
identities. $\overline\Pi_{\mu\nu}(Q)$ contains five irreducible
substructures that do not transform into each other under cubic
rotations and which differ by finite-volume effects. In
Ref. \cite{Aubin:2015rzx} the different irreducible substructures were
computed in a lattice calculation employing rooted staggered quarks
with $m_\pi=220\,\mev$ and $m_\pi L=4.0$, as well as in chiral
perturbation theory. While the effective chiral theory fails to
reproduce the absolute value of the vacuum polarisation function, it
describes the difference between different irreducible substructures
quite well within the quoted statistical errors. The difference in the
vacuum polarisation function due to finite volume effects can then be
inserted into the convolution integral (see \eq{eq:amublum2}), in
order to determine the corresponding shift in $\ahvp$. One finds that
the correction amounts to $10-15$\%. Thus, one concludes that the
condition $m_\pi L=4$ at $m_\pi=220\,\mev$ is not sufficient to
guarantee that finite-volume effects are suppressed below the
percent-level.

The observation that the assumption $\Pi_{\mu\nu}(0)=0$ does not hold
in a finite volume has inspired the common practice of subtracting the
zero mode via a simple modification of the phase factor in the Fourier
transform of the vector correlator
\cite{Bernecker:2011gh,Lehner:2015bga}, i.e.
\be
  \Pi_{\mu\nu}(Q)-\Pi_{\mu\nu}(0) = \int d^4x 
  \,\left( \rme^{iQ\cdot x}-1 \right) \left\langle J_\mu(x)J_\nu(0)
  \right\rangle \stackrel{!}{=}(Q_\mu Q_\nu-\delta_{\mu\nu}) \Pi(Q^2).
\ee
As discussed in \cite{Malak:2015sla}, the subtraction of the zero mode
leads to much smaller finite-volume effects in the determination of
$\Pi(Q^2)$ and, in turn, the estimate of $\ahvp$.

Another approach to quantify finite-volume corrections and effects
arising from the mass splitting between different ``tastes'' in the
rooted staggered fermion formulation was presented in
\cite{Chakraborty:2016mwy}. The starting point is an effective theory
of photons, pions and $\rho$-mesons, similar to that used in
\cite{Jegerlehner:2011ti}. This set-up can be used to compute the
subtracted vacuum polarisation function in terms of an integral over
the four-momentum. The coefficients in the Taylor expansion are
related to the time moments. Their shift due to the finite volume can
then be worked out by replacing the integral with a discrete sum over
the Fourier modes and averaging over the multiplets related by the
taste symmetry. The overall finite-volume correction to the estimate
of $\ahvp$ is estimated to be 7\%.

We are now going to present a more detailed discussion of a dynamical
theory of finite-volume effects, which uses as input the mass ratio
$m_\pi/m_\rho$, as well as the box size in units of the pion mass,
$m_\pi L$. This method is based on the time-momentum representation,
and a detailed account can be found
in\,\cite{Francis:2013fzp,DellaMorte:2017dyu}. The goal is to compute
the difference of the spatially summed vector correlator in infinite
and finite volume, $G(x_0,\infty)-G(x_0,L)$. When inserted in
\eq{eq:TMRamu}, the finite-volume shift is determined. At short
distances, i.e. for $x_0\lesssim 1\,\fm$ the Poisson resummation
formula based on non-interacting pions should provide a good
approximation for $G(x_0,\infty)-G(x_0,L)$. The long-distance
contribution ($x_0\gtrsim1\,\fm$) to the finite-size effect can be
determined by invoking the L\"uscher formalism using the low-lying
energy eigenstates on a torus.

The integral representation for the short-distance part reads
\cite{Francis:2013fzp,DellaMorte:2017dyu}
\ba
& & G(x_0,L)-G(x_0,\infty) = \frac{1}{3}\left\{
\frac{1}{L^3}\sum_{{\vec{n}}}-\frac{1}{(2\pi)^3}\int d^3k \right\}
\frac{{\vec{k}^2}}{{\vec{k}^2+m_\pi^2}}
\rme^{-2x_0\sqrt{{{\vec{k}}}^2+m_\pi^2}} \\
& & \phantom{=} =\frac{m_\pi^4 x_0}{3\pi^2}\sum_{\vec{n} \neq \vec{0}}\left\{
\frac{K_2\left(m_\pi\sqrt{L^2{\vec{n}}^2+4x_0^2}\,\right)}{m_\pi^2(L^2{\vec{n}}^2+4x_0^2)}
\right. \\
& & \left. \phantom{\sum_{\vec{n} \neq \vec{0}}\frac{K_2\left(m_\pi\sqrt{L^2}\right)}{m_\pi^2(L^2)}}
-\frac{1}{m_\pi L|\vec{n}|} \int\limits_1^\infty dy\,
 K_0\left(m_\pi y\sqrt{L^2{\vec{n}}^2+4x_0^2}\,\right)\,\sinh\left(m_\pi L|\vec{n}|(y-1)\right)
 \right\}, \nonumber
\ea
where $K_0, K_2$ denote modified Bessel functions of the second kind.
Numerical estimates for the finite-volume shift in $\ahvp$ have been
worked out for the two-flavour CLS ensembles used in
\cite{DellaMorte:2011aa,DellaMorte:2017dyu}. It turns out that in the
region where $x_0\lesssim1\,\fm$ finite-volume corrections are
negligibly small for $m_\pi L \geq 4$ and $m_\pi\lesssim 300\,\mev$.

In order to determine finite-volume effects at large distances,
i.e. in the case of interacting pions, we can rely on our earlier
discussion in Section~\ref{sec:TMR} on constraining the long-distance
part of the iso-vector correlator. The iso-vector vector correlator in
infinite volume is expressed in terms of the spectral function
$\rho(\omega)$ as
\be\la{eq:Gx0infty}
   G^{\rho\rho}(x_0,\infty) = \int_0^\infty
   d\omega\,\omega^2\rho(\omega^2)\,\rme^{-\omega|x_0|},
\ee
with the $\pi\pi$ contribution given by
\be
   \rho(\omega^2)=\frac{1}{48\pi^2}
   \left(1-\frac{4m_\pi^2}{\omega^2}\right)^{3/2} 
   \left|F_\pi(\omega)\right|^2.
\ee
Above the threshold $\omega=2m_\pi$ the phase of the timelike pion
form factor $F_\pi(\omega)$ is equal to the $p$-wave pion scattering
phase shift $\delta_{11}(k)$, according to Watson's theorem:
\be
   F_\pi(\omega)=\left|F_\pi(\omega)\right|\rme^{i\delta_{11}(k)}.
\ee
In finite volume the correlator is given by \eq{eq:GrhorhoL},
i.e. $G^{\rho\rho}(x_0,L)=\sum_n |A_n|^2\,\rme^{-\omega_n x_0}$. The
discrete energy levels $\omega_n$ are related to the infinite-volume
phase shifts by the L\"uscher condition (see \eq{eq:Luscher}), while
the amplitudes $|A_n|^2$ are related to the timelike pion form factor
according to \eq{eq:timelikeFF}. The determination of both
$G^{\rho\rho}(x_0,\infty)$ and $G^{\rho\rho}(x_0,L)$ rely on input
data for $F_\pi(\omega)$. The Gounaris-Sakurai parameterisation
\cite{Gounaris:1968mw} of $F_\pi(\omega)$ proves helpful in this
case. It describes the $\rho$-resonance in terms of two free
parameters, $m_\rho$ and $\Gamma_\rho$. The expression for
$F_\pi(\omega)$ reads
\be
   F_\pi(\omega)={f_0}\left/{\frac{k^3}{\omega}[\cot\delta_{11}(k)-i]}
   \right.,
\ee
and if one defines $k_\rho$ via $m_\rho=2\sqrt{k_\rho^2+m_\pi^2}$ one
can express the phase shift $\delta_{11}(k)$ and the quantity $f_0$ in
terms of $k_\rho$ and $\Gamma_\rho$.

In order to determine the iso-vector correlator
$G^{\rho\rho}(x_0,\infty)$ in infinite volume, one can use the values
of $m_\pi$ and $m_\rho$ computed on a given ensemble and evaluate
$k_\rho$ and $\Gamma_\rho$, assuming that $\Gamma_\rho\propto
k_\rho^3/m_\rho^2$. Since $k$ is related to $\omega$ by
$\omega=2\sqrt{k^2+m_\pi^2}$ in infinite volume, one can then evaluate
$k^3\cot\delta_{11}(k)/\omega$ and $f_0$, and insert the result for
$|F_\pi(\omega)|$ into \eq{eq:Gx0infty}.

In finite volume one uses $m_\pi, m_\rho$ as before to determine
$\delta_{11}(k)$ and $|F_\pi(\omega)|$. Both quantities then serve as
input to solve the L\"uscher condition, \eq{eq:Luscher}, as well as
\eq{eq:timelikeFF} for $n=1,2,\ldots.$ In this way one obtains both the
finite-volume energy levels $\omega_n$ and the corresponding matrix
elements $A_n$, which are used to evaluate $G^{\rho\rho}(x_0,L)$ of
\eq{eq:GrhorhoL}. The resulting difference
$G^{\rho\rho}(x_0,\infty)-G^{\rho\rho}(x_0,L)$ can then be used to
determine the finite-volume shift $\ahvp(\infty)-\ahvp(L)$.

\begin{figure}[t]
\begin{center}
\includegraphics[width=10cm]{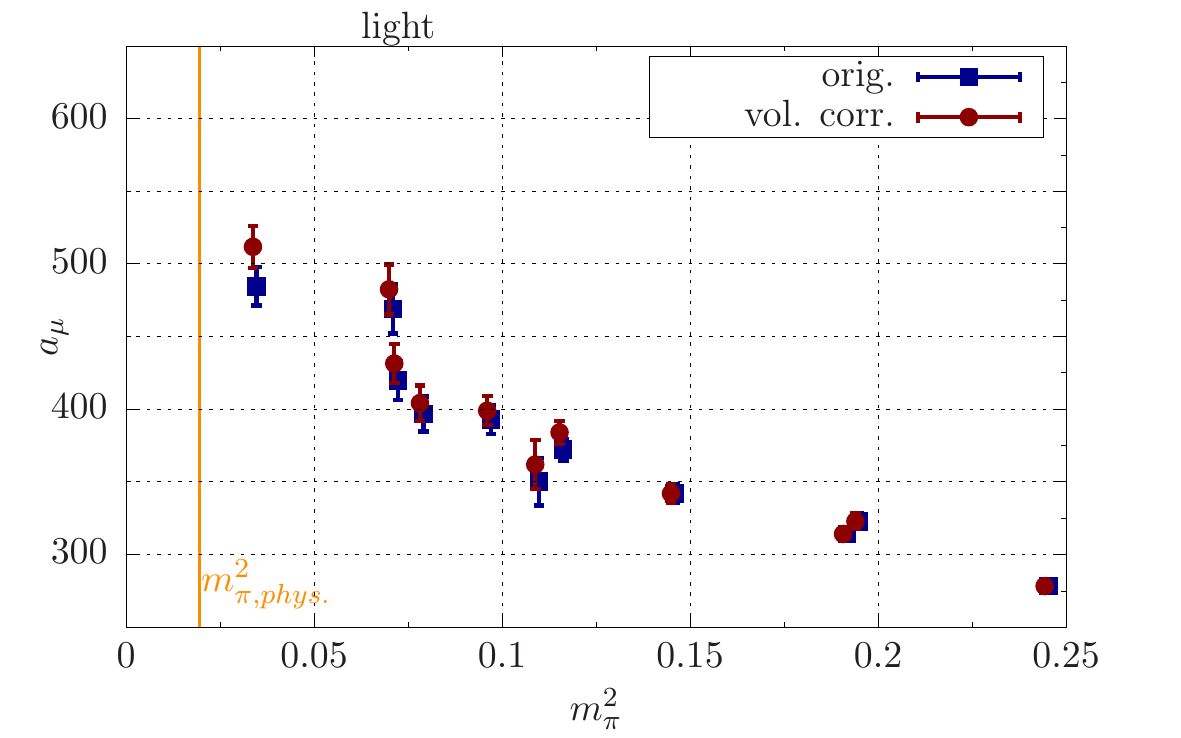}
\caption{The ight quark contribution to the hadronic vacuum
  polarisation, $(\ahvp)^{ud}$, computed by CLS/Mainz
  \cite{DellaMorte:2017dyu} in two-flavour QCD. Filled red circles and
  blue squares denote the results with and without the finite-volume
  shift, respectively. The largest correction of $5.2\%$ is observed
  at the lightest pion mass of $\approx190\,\mev$. \la{fig:CLSFVE}}
\end{center}
\end{figure}

This procedure has been applied by the Mainz group to determine the
finite-volume shifts for $\ahvp$ determined using the time-momentum
representation \cite{DellaMorte:2017dyu}. Figure\,\ref{fig:CLSFVE}
shows the estimates for the hadronic vacuum polarisation contribution
of the light quarks as a function of the pion mass, with and without
the finite-volume shift. The largest correction of
$(\ahvp(\infty)-\ahvp(L))/\ahvp(\infty)=5.2\%$ is encountered for
$m_\pi\approx190\,\mev$ at $m_\pi L=4.0$. In order to check the
stability of the finite-volume shifts one can consider variations of
the Gounaris-Sakurai parameters $m_\rho, \Gamma_\rho$, as well as the
Euclidean time at which one switches from considering non-interacting
pions and the Poisson formula to the interacting case. In this way one
can assign a 20\% systematic uncertainty to the estimate of
$\ahvp(\infty)-\ahvp(L)$. After applying the correction to each
ensemble and performing a simultaneous chiral and continuum
extrapolation one finds a finite-volume shift of 6.7\% at the physical
point.

Preliminary results by CLS/Mainz obtained in QCD with $2+1$ flavours
of O($a$) improved Wilson fermions suggest that the above procedure is
able to quantify the finite-volume correction quite accurately. By
computing the correlator $G^{ud}(x_0)$ on two different volumes,
corresponding to $L=2.7$\,fm and 4.1\,fm, respectively one observes a
significant finite-volume shift for the integrand
$w(x_0)\,G^{ud}(x_0)$. Applying the finite-volume correction
determined via the timelike pion form factor and the L\"uscher
procedure makes the results obtained on the two volumes compatible
within statistical errors.

Another direct comparison of results for $\ahvp$ obtained on two
different volumes has been reported by the PACS
collaboration\,\cite{Izubuchi:2018tdd}. Employing $2+1$ flavours of
O($a$) Wilson quarks on lattice sizes of $96^4$ and $64^4$ at
$a=0.085$\,fm, which correspond to box lengths of $L=8.1$\,fm and
5.4\,fm, respectively, they study the volume dependence of the
integrand $w(x_0)\,G(x_0)$ of the time-momentum representation, as
well as the estimate for $\ahvp$ resulting from the integration up to
$\xcut\approx3$\,fm. At their reference pion mass of 146\,MeV they
find an absolute finite-volume correction of
\be
   \ahvp(L=8.1\,\fm)-\ahvp(L=5.4\,\fm) = (10\pm26)\cdot10^{-10}
\ee
for the light quark contribution to $\ahvp$. The central value of this
estimate agrees well with the result obtained using chiral EFT
\cite{Aubin:2015rzx}. One concludes that, at near-physical pion mass,
the finite-volume correction between ensembles with $m_\pi L\approx4$ and
$m_\pi L\approx6$ amounts to 1.5\%, although the shift is not
statistically significant.

\subsection{Chiral extrapolation of $\ahvp$}

Simulations of lattice QCD at parameter values that correspond to the
physical pion mass have become routine. However, in some cases the
result at the physical point is still obtained by chirally
extrapolating the data obtained for pion masses in the range of
$200-400\,\mev$. Furthermore, results computed directly at the
physical value of $m_\pi$ are often combined with data at larger
masses in order to increase the overall accuracy, and in many cases
the final estimate is obtained through a simultaneous chiral and
continuum extrapolation. Results from the first comprehensive
calculations of $\ahvp$ on the lattice
\cite{Feng:2011zk,Boyle:2011hu,DellaMorte:2011aa} showed a strong
dependence of $\ahvp$ on $m_\pi^2$. In Ref.\,\cite{Feng:2011zk} it was
observed that this behaviour was correlated with a strong variation of
the $\rho$-meson mass with $m_\pi$. Hence, in order to produce a
milder dependence of $\ahvp$ on the pion mass, the authors of
Ref. \cite{Feng:2011zk} proposed a rescaling of the momentum $Q^2$ of
the subtracted vacuum polarisation according to
\be
  \la{eq:rescale} \hat\Pi(Q^2) \to \hat\Pi(hQ^2),\quad
  h=\frac{m_{\rm V}^2}{m_\rho^2}, 
\ee
where $m_\rho$ is the physical $\rho$-meson mass, while $m_{\rm V}$
denotes its value computed for the actual pion mass of a given gauge
ensemble. The fact that the rescaling factor $h$ approaches unity as
$m_\pi$ is tuned towards its physical value implies that the limits of
$\ahvp$ computed with or without rescaling are the same.

The motivation for the rescaling can be derived from a simple
consideration based on vector meson dominance
\cite{DellaMorte:2011aa,Golterman:2017njs}. In the VMD model the
$Q^2$-dependence of the hadronic vacuum polarisation in the iso-vector
channel is given by
\be
   \hat\Pi(Q^2)_{\rm VMD}(Q^2) \sim g_{\rm V}
   \frac{Q^2}{Q^2+m_{\rm V}^2},
\ee
where $g_{\rm V}$ is related to the $\rho$-meson decay
constant. Assuming that $m_{\rm V}$ depends strongly on $m_\pi$ while
$g_{\rm V}$ does not, one easily sees that the rescaling of $Q^2$
according to \eq{eq:rescale} makes $\hat\Pi(Q^2)_{\rm VMD}$
broadly independent of the pion mass, and the same will be true for
the resulting value of $\ahvp$ after evaluating the convolution
integral of \eq{eq:amublum2}. In \cite{Chakraborty:2016mwy}, a variant
of the method was considered, which combines the rescaling with the
subtraction of the relative pion loop correction computed in chiral
effective theory at NLO between the physical and actual values of
$m_\pi$. Indeed, one finds that the chiral behaviour of $\ahvp$ is
much flatter as a result of multiplying $Q^2$ by
${m_{\rm{V}}^2}/{m_\rho^2}$ \cite{Feng:2011zk, Chakraborty:2016mwy}.

The stability of the chiral extrapolation of $\ahvp$ was analysed
extensively in Ref. \cite{Golterman:2017njs} using a model for the
iso-vector vacuum polarisation derived from ChPT at two loops and the
experimentally determined spectral function. By comparing several {\it
  ans\"atze} for the chiral extrapolation of the hadronic vacuum
polarisation determined for pion masses in the range $200-400\,\mev$
it was found that the typical spread of $\ahvp$ at the physical pion
mass is of the order of 5\%. Importantly, while the rescaling helps to
produce a flatter pion mass dependence, there remains an ambiguity
arising from different model functions for the chiral fit, none of
which is clearly preferred on theoretical grounds. The authors of
Ref. \cite{Golterman:2017njs} conclude that extrapolations from pion
masses larger than 200\,\mev\ are not reliable enough to achieve
sub-percent level precision.

\subsection{Scale setting}

A source of systematic uncertainty that received little attention in
early calculations of $\ahvp$ is the error on the lattice scale
\cite{DellaMorte:2017dyu,DellaMorte:2017khn}. Although $\ahvp$ is a
dimensionless quantity, there are two ways in which the lattice scale
enters the calculation. This is most easily explained in the framework
of the time-momentum representation of $\ahvp$ defined in
\eq{eq:TMRamu}: Firstly, the muon mass $m_\mu$ enters the kernel
function $w(x_0)$ via the dimensionless combination $x_0
m_\mu$. Secondly, the masses of the dynamical quarks enter implicitly
via the lattice evaluation of the vector correlator. Therefore,
$\ahvp$ can be thought of as a function in the dimensionless variables
$M_\mu\equiv m_\mu/\Lambda, M_u\equiv m_u/\Lambda, M_d\equiv
m_d/\Lambda,\ldots$, where $\Lambda$ is the quantity that sets the
lattice scale. The scale setting error $\Delta\Lambda$ then induces a
corresponding uncertainty in $\ahvp$, i.e.
\be
   \Delta\ahvp = \bigg|\Lambda\frac{d\ahvp}{d\Lambda}\bigg|\,
   \frac{\Delta\Lambda}{\Lambda} = \bigg|
   M_\mu\frac{\partial\ahvp}{\partial M_\mu} + \sum_{f=1}^{N_{\rm
       f}}M_{\rm f}\frac{\partial \ahvp}{\partial M_{\rm f}} \bigg|\,
   \frac{\Delta\Lambda}{\Lambda}.
\ee
Often one employs a hadronic renormalisation scheme in which the quark
masses are expressed in terms of suitable meson masses. For instance,
in the isospin limit one can fix the average light quark mass
$m_{ud}\equiv\frac{1}{2}(m_u+m_d)$ by the pion mass $m_\pi$, which can
be easily generalised to apply to heavier quark flavours, too. The
uncertainty $\Delta\ahvp$ can then be written as
\be
   {\Delta \ahvp} = \bigg|M_\mu\frac{\partial \ahvp}{\partial M_\mu} +
   M_\pi\frac{\partial \ahvp}{\partial M_\pi}+ M_{\rm K}\frac{\partial
     \ahvp}{\partial M_{\rm K}}+\ldots \bigg|\,
   \frac{\Delta\Lambda}{\Lambda},
\ee
where $M_\pi, M_{\rm K},\ldots$ denote the meson masses in units of
$\Lambda$. In the time-momentum representation one can determine the
derivative term involving the muon mass via \cite{DellaMorte:2017dyu}
\be
   M_\mu\frac{\partial a_\mu^{\rm hvp}}{\partial M_\mu} = -\ahvp
   +\left(\frac{\alpha}{\pi}\right)^2\int_0^\infty 
   dx_0\,G(x_0)\,J(x_0),\quad
   J(x_0)=x_0 w^\prime(x_0)-w(x_0),
\ee
where $w^\prime(x_0)$ denotes the derivative of the kernel function
$w(x_0)$ in \eq{eq:TMRkernel}. Both $w(x_0)$ and $w^\prime(x_0)$ can
be easily computed using the series expansion from appendix~B
in\,\cite{DellaMorte:2017dyu}. Moreover, in the same paper the
derivative with respect to the pion mass $M_\pi$ has been determined
from the slope of the chiral extrapolation at
$m_\pi=m_\pi^{\rm{phys}}$. In this way one finds
\be
   \frac{\Delta \ahvp}{\ahvp} = \Bigg|
   \underbrace{\frac{M_\mu}{\ahvp}\frac{\partial \ahvp}{\partial M_\mu}}_{\displaystyle{1.8}}+
   \underbrace{\frac{M_\pi}{\ahvp}\frac{\partial \ahvp}{\partial M_\pi}}_{\displaystyle{-0.18(6)}}
   \Bigg|\,\frac{\Delta\Lambda}{\Lambda}.
\ee
Thus, the factor multiplying the scale setting uncertainty is
dominated by the contribution from the muon mass, with only a 10\%
reduction coming from the light quarks. Heavier quark flavours are
likely to have an even smaller effect. One concludes from this
analysis that the proportionality between the relative uncertainties
of $\ahvp$ and the lattice scale $\Lambda$ is a number of
order~one. Therefore, the lattice scale must be known to within a
fraction of a percent, if one is to reach the precision goal in the
determination of $\ahvp$.

\subsection{Isospin-breaking effects\la{sec:IB}}

Controlling and quantifying the effects from isospin breaking is of
major importance for the determination of $\ahvp$. In the
phenomenological approach based on the hadronic cross section ratio
$R(s)$, it is necessary to include the contributions from final state
radiation \cite{Jegerlehner:2009ry}, $\pi^0\gamma$ and $\eta\gamma$
channels \cite{Davier:2010nc}, as well as $\rho\omega$ mixing
\cite{Davier:2009ag,Jegerlehner:2011ti}. The latter, in particular,
played a vital r\^ole in arriving at a consistent estimate of the
iso-vector $\pi\pi$-contribution to $\ahvp$ using data from either
$e^+e^-\to\pi^+\pi^-$ or, alternatively, from hadronic $\tau$-decays
\cite{Jegerlehner:2011ti}. In total, isospin breaking effects account
for 1.3\% of the dispersive estimate for $\ahvp$ and represent a
crucial ingredient for reaching the current level of precision.

The treatment of isospin breaking in lattice QCD has been a major
focus of recent activity. The inclusion of isospin breaking effects in
lattice calculations of $\ahvp$ is indispensable for the goal of
reaching sub-percent precision. There are two sources of isospin
breaking: (1) the strong interaction contribution that arises from the
mass splitting between the up and down quarks, $m_u\neq m_d$, which is
of order $\alpha^2(m_d-m_u)$, and (2) electromagnetic corrections of
O($\alpha^3$) due to the different electric charges of the quarks.

The inclusion of electromagnetism in a manner that is consistent with
the lattice formulation of QCD is technically challenging. As
discussed in a recent review article \cite{Patella:2017fgk}, Gauss's
law forbids the existence of states with non-zero electric charge in a
finite volume with periodic boundary conditions. Another way of
expressing this obstacle is the statement that charged states do not
propagate in a finite periodic box: Large gauge transformations that
are admitted by the boundary conditions cannot be eliminated by any
local gauge-fixing procedure.

Since the photon is a massless unconfined particle, the finite-volume
effects of lattice QCD in the presence of electromagnetism must be
reassessed. In particular, one expects finite-volume effects to be
more severe, since the leading corrections fall off as powers of the
inverse volume instead of exponentially \cite{Duncan:1996xy}. In order
to circumvent these problems several prescriptions have been
pursued. The starting point is the Euclidean path integral of QCD and
QED
\be\la{eq:QCDQED}
   Z_{\rm QCD+QED}=\int D[U]D[A]D[\psibar,\psi]\,
   {\rm e}^{-S_\gamma[A]-S_{\rm G}[U]-S_{\rm F}[U,A,\psibar,\psi]},
\ee
where $S_\gamma[A]$ denotes the photon action, and $A_\mu(x)$
represents the photon field. Usually one employs the non-compact
formulation of QED in which the action -- including a gauge-fixing
term -- is given by
\be
  S_\gamma[A] = a^4\sum_x \left\{ \frac{1}{4}\sum_{\mu,\nu} \left(\nabla_\mu
  A_\nu - \nabla_\nu A_\mu\right)^2 +\frac{1}{2\xi} \left(\sum_\mu
  \nabla_\mu A_\mu(x) \right)^2 \right\},
\ee
where $\nabla_\mu$ denotes the forward lattice derivative. Below we
outline several prescriptions that have been pursued to circumvent the
conceptual problems of QED in a finite volume.

\begin{itemize}

\item 
In the QED$_{\rm TL}$ prescription, originally proposed in
Ref.\,\cite{Duncan:1996xy}, the zero modes of the photon field are
explicitly set to zero by imposing
\be
   \left.\widetilde{A}_\mu(p)\right|_{p=0} \equiv
   \int d^4x\,A_\mu(x) = 0, 
\ee
where $\widetilde{A}_\mu(p)$ denotes the photon field in Fourier
space. As this is a non-local constraint, the path integral in
\eq{eq:QCDQED} does not admit a representation in terms of the
transfer matrix, and hence the Hamiltonian cannot be defined.

\item
The QED$_{\rm L}$ prescription \cite{Hayakawa:2008an} imposes a
different constraint, i.e.
\be
   \int d^3x\,A_\mu(x_0,\vec{x}) = 0,
\ee
which corresponds to setting all spatial zero modes of the photon
field to zero. Since this condition is local in time, the theory does
admit a Hamiltonian. However, the non-locality in space spoils the
renormalisation of local composite operators with dimensions larger
than~four\,\cite{Patella:2017fgk}.

\item
In the formulation of Ref.\,\cite{Gockeler:1989wj}, proposed
originally to study the infrared properties of QED, one imposes a
cutoff on the value of the zero mode. The existence of a transfer
matrix is not guaranteed.

\item
Alternative treatments include the QED$_{\rm m}$ prescription which
introduces a massive photon\,\cite{Endres:2015gda}, and the
formulation based on the introduction of C-parity boundary conditions,
called QED$_{\rm C}$\,\cite{Polley:1990tf,Lucini:2015hfa}.  While both
are consistent quantum field theories, one finds that QED$_{\rm m}$
requires a careful treatment of the limits $m_\gamma\to0$ and
$L\to\infty$. The QED$_{\rm C}$ formulation breaks flavour symmetry,
but as the breaking is local, the effects of flavour symmetry breaking
are exponentially suppressed.
\end{itemize}

Two distinct approaches are widely applied in calculations of
observables in the presence of strong and electromagnetic isospin
breaking. The first is the ``stochastic method'': It is based on the
direct Monte Carlo evaluation of the path integral
in\,\eq{eq:QCDQED}. The coupling of the photon $A_\mu(x)$ to the quark
fields is accomplished by augmenting the link variables describing the
gluons by a U(1) phase factor according to
\be
   U_\mu(x)\to\rme^{ieA_\mu(x)}U_\mu(x),
\ee
where $e$ is the electric charge. In order to facilitate the
stochastic calculation of observables defined with respect to $Z_{\rm
  QCD+QED}$ the sea quarks are assumed to be electrically neutral, so
that the U(1) gauge field is generated independently from the SU(3)
gauge fields. This defines the so-called ``electro-quenched''
approximation which has been applied very successfully to determine
electromagnetic mass splittings among hadrons, as well as the up-down
quark mass difference (see Refs.\,\cite{Duncan:1996xy, Blum:2007cy,
  Blum:2010ym, Borsanyi:2013lga, Borsanyi:2014jba, Fodor:2016bgu,
  Horsley:2015eaa, Horsley:2015vla, Basak:2016jnn}). Strong isospin
breaking is either incorporated by choosing different up and down
quark masses, as was done, for instance, in\,\cite{Borsanyi:2014jba}
or via reweighting techniques (an example is discussed in
Ref.\,\cite{Aoki:2012st}).

The second method for determining isospin breaking effects arising from
electromagnetic corrections is based on the perturbative expansion of
the path integral of \eq{eq:QCDQED} in powers of the fine structure
constant $\alpha$ \cite{deDivitiis:2013xla}. In a similar manner,
strong isospin breaking effects can be treated in this framework by
expanding the path integral in powers of the light quark mass
difference $(m_d-m_u)$ \cite{deDivitiis:2011eh}.

We will now discuss several recent lattice calculations of
isospin-breaking effects in $\ahvp$. The RBC/UKQCD Collaboration has
studied electromagnetic corrections using the QED$_{\rm L}$
prescription at a pion mass of 340\,MeV and focussing on connected
diagrams only\,\cite{Boyle:2017gzv}. By performing a detailed
comparison of the stochastic and perturbative methods they conclude
that, while both yield consistent results of similar accuracy, the
stochastic approach fares slightly better in terms of numerical
accuracy. In a follow-up paper\,\cite{Blum:2018mom} they present a
calculation of both strong and electromagnetic isospin-breaking
contributions based on the perturbative method and at the physical
pion mass. The finite-volume corrections of order $1/L$ and $1/L^2$
are removed, a subset of QED-disconnected graphs is included, and
strong isospin-breaking effects have been included by computing the
leading-order graphs arising in the expansion in $(m_u-m_d)$. The
isospin-breaking corrections to the renormalisation factor of the
electromagnetic current have also been included. In this way they
obtain a total isospin-breaking correction to $\ahvp$ of
$(9.5\pm10.4)\cdot10^{-10}$, which amounts to $+(1.5\pm1.6)$\% of the
iso-symmetric contribution from up and down quarks. It is interesting
to note that the contribution from strong isospin breaking alone is 
$(10.6\pm8.0)\cdot10^{-10}$, indicating that the dominant effect is
due to $m_u\neq m_d$.

The ETM Collaboration\,\cite{Giusti:2017jof} has used the QED$_{\rm
  L}$ prescription together with the perturbative approach to
determine the electromagnetic corrections to the strange and charm
quark contributions, $(\ahvp)^s$ and $(\ahvp)^c$. Neglecting
disconnected diagrams and extrapolating results to the physical pion
mass and vanishing lattice spacing, they find that electromagnetic
corrections to $(\ahvp)^s$ and $(\ahvp)^c$ amount to $-0.03$\,\% and
$-0.21$\,\%, respectively, which is negligible relative to the
statistical error of the iso-symmetric contribution.

The calculation by the Fermilab-HPQCD-MILC Collaboration
\cite{Chakraborty:2017tqp} is focussed on the determination of the
strong isospin-breaking correction alone. Using two ensembles at the
physical pion mass -- one in the iso-symmetric limit, i.e. with
$m_u=m_d$, and another one that realises a splitting between up and
down quark masses consistent with an earlier calculation
\cite{Chakraborty:2017tqp} -- they are able to test strong
isospin-breaking effects arising from the quark sea as well. Their
main result for the strong isospin correction in $\ahvp$ is
$(9.0\pm2.3\pm3.1)\cdot 10^{-10}$, where the second error is the total
systematic error which is dominated by the neglected
quark-disconnected contribution. Combining this result with the
iso-symmetric light quark contribution to $\ahvp$ of
Ref.\,\cite{Chakraborty:2016mwy}, the relative shift is
$+(1.5\pm0.7)$\%.

\begin{figure}[t]
\begin{center}
\includegraphics[width=10cm]{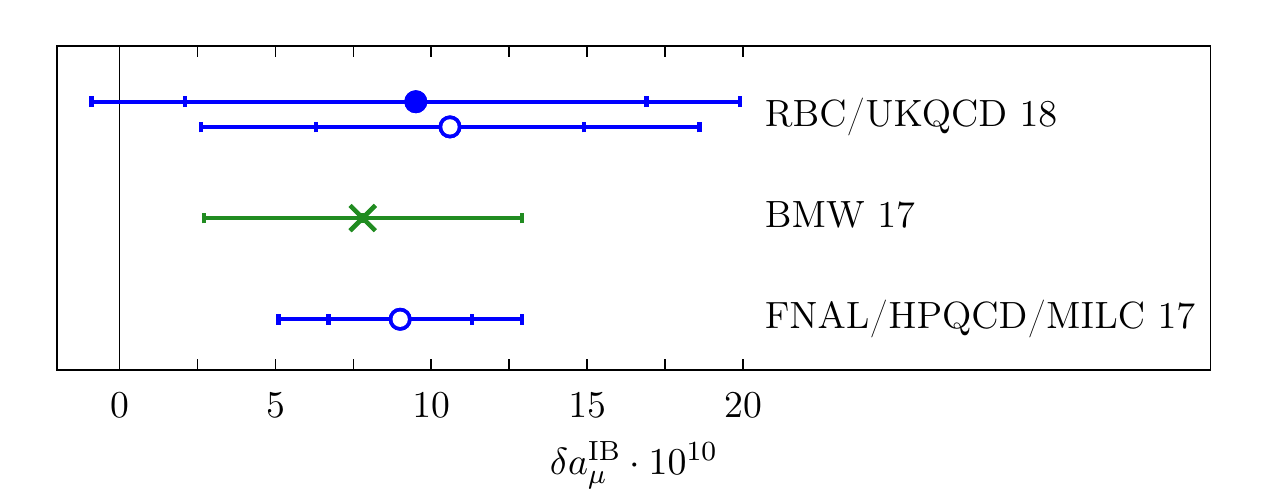}
\caption{Compilation of determinations of the isospin-breaking
  correction $\delta a_\mu^{\rm IB}$ to the hadronic vacuum
  polarisation, computed by RBC/UKQCD\,\cite{Blum:2018mom},
  BMW\,\cite{Borsanyi:2017zdw} and Fermilab/HPQCD/MILC
  \cite{Chakraborty:2017tqp}. Circles denote results computed in
  lattice QCD, while the green cross represents the phenomenological
  estimate used in the calculation by BMW. Strong isospin-breaking
  contributions are shown as open
  circles.\la{fig:IBcomp}}
\end{center}
\end{figure}

The estimate of isospin-breaking effects in the result of the BMW
Collaboration \cite{Borsanyi:2017zdw} has not been determined by a
lattice calculation but instead by phenomenology. The contributions
from the $\pi^0\gamma$ and $\eta\gamma$ channel have been taken over
from the dispersive approach, final-state radiation has been estimated
using a combination of data and point-particle QED corrections, and
hadronic models have been used to estimate the contribution from
$\rho$-$\omega$ mixing. The slight detuning of the pion mass has been
corrected for using leading-order chiral EFT. The total correction due
to isospin breaking is found to be $(7.8\pm5.1)\cdot10^{-10}$ or
$+(1.2\pm0.8)$\,\% of the light-quark iso-symmetric contribution.

Thanks to the considerable effort invested, a coherent picture emerges
regarding the isospin-breaking contribution $\delta a_\mu^{\rm IB}$ to
$\ahvp$: as is evident from the compilation in
Figure\,\ref{fig:IBcomp}, the correction is positive and of order
$10\cdot 10^{-10}$, which amounts to about 1.5\% of the total
leading-order hadronic vacuum polarisation. In view of the target
precision, this is a significant correction. Moreover, the
calculations of RBC/UKQCD \cite{Blum:2018mom} and FNAL/HPQCD/MILC
\cite{Chakraborty:2017tqp} show that the size of $\delta a_\mu^{\rm
  IB}$ is dominated by strong isospin-breaking effects. The fact that
electromagnetic corrections are small and negligible relative to the
statistical errors in current calculations has also been confirmed by
ETMC\,\cite{Giusti:2017jof}. It is also interesting to note that
isospin-breaking corrections to $\ahvp$ are similar in size compared
to the quark-disconnected contribution, but come in with the opposite
sign.

Clearly, the precision of these calculations must be further
increased, since the errors quoted for $\delta a_\mu^{\rm IB}$ amount
to $50-100$\,\%. It is also important to note that the mass difference
between the charged and neutral pions must be treated in a consistent
manner, in order to guarantee a reliable determination of
isospin-breaking effects.

\subsection{Results for $\ahvp$ \la{sec:results}}

We will now discuss the available results from lattice QCD
calculations of $\ahvp$, assess their level of accuracy and compare
them to estimates based on dispersion relations. We begin by
presenting short accounts of the individual calculations whose results
are listed in Table\,\ref{tab:hvp}, with additional information on
simulation details given in Table\,\ref{tab:hvpdetails}.

{\sc{Aubin}} and {\sc{Blum}}\,\cite{Aubin:2006xv} performed the first
determination of $\ahvp$ using $\Nf=2+1$ flavours of staggered quarks,
following the strategy outlined in\,\cite{Blum:2002ii}. By fitting the
$Q^2$-dependence of $\Pi(Q^2)$ to the functional form predicted by
staggered ChPT, they determined $\ahvp$ at a single value of the
lattice spacing and three pion masses in the range
$240-470\,\mev$. The results were extrapolated to the physical pion
mass using either a linear or quadratic ansatz in
$m_\pi^2$. Contributions from the charm quark were not included, and
the quoted errors are purely statistical.

The calculations by the {\sc{ETM Collaboration}}
\cite{Feng:2011zk,Burger:2013jya} were performed using twisted mass
QCD as the discretisation of the quark action. One of the main
features of their analysis is the rescaling of the squared momentum
$Q^2$ in the argument of $\Pi(Q^2)$ according to \eq{eq:rescale}. In
Ref.\,\cite{Feng:2011zk} it was argued (see
also\,\cite{DellaMorte:2011aa}) that this rescaling results in a much
milder pion mass dependence, leading to a more stable chiral
extrapolation to the physical point. The first calculation
\cite{Feng:2011zk} was performed in two-flavour QCD, considering only
the $u,d$ contributions to the hadronic vacuum polarisation. No
significant dependence on the lattice spacing, the volume and the
details of the chiral extrapolation were observed at the level of
statistical precision. Quark-disconnected contributions were computed
but found to be zero within errors. The second calculation was
performed for tmQCD with $\Nf=2+1+1$ dynamical flavours at three
different lattice spacings. Systematic uncertainties were estimated
from variations in the fit ansatz used to describe the low-$Q^2$
regime and by considering different fit intervals in the determination
of the vector meson mass $m_{\rm V}$ which enters the rescaling factor
in \eq{eq:rescale}. No appreciable change in the result was observed
when ensembles with $m_\pi L<4$ were excluded from the analysis, and
hence they concluded that volume effects were small.  In their 2017
paper \cite{Giusti:2017jof}, ETM presented results for the strange and
charm quark contributions (which had not been quoted as separate
quantities in \cite{Burger:2013jya}), including the corrections from
electromagnetism. As was already discussed in Section\,\ref{sec:IB},
the latter were found to be much smaller than the overall statistical
uncertainty.

{\sc{CLS/Mainz}} \cite{DellaMorte:2011aa,DellaMorte:2017dyu} have
determined $\ahvp$ using two flavours of $\rmO(a)$ improved Wilson
fermions at three lattice spacings and pion masses from
$190-500\,\mev$. Twisted boundary conditions were employed to realise
smaller values of $Q^2$ and a high density of points to describe the
$Q^2$-dependence of the vacuum polarisation function. The overall
precision of the earlier result\,\cite{DellaMorte:2011aa} amounts to
10\% with the main systematic uncertainty arising from the uncertainty
in the lattice scale. The more recent publication
\cite{DellaMorte:2017dyu} contains a detailed comparison of the
different methods to determine $\ahvp$, i.e. the hybrid method, the
time-momentum representation and time moments. Finite-volume
corrections were determined using the Gounaris-Sakurai model for the
timelike pion form factor. The uncertainty in the final result is
dominated by statistics, while the systematic error estimate includes
the uncertainties due to variations in the analysis procedure,
finite-volume corrections, scale setting and quark-disconnected
diagrams. The calculations have now been extended to QCD with
$\Nf=2+1$ flavours of dynamical quarks\,\cite{DellaMorte:2017khn},
including ensembles at the physical value of the pion mass. Special
attention is given to constraining the long-distance regime of the
spatially summed vector correlator via a dedicated calculation of the
energies $\omega_n$ and associated amplitudes $A_n$ for the first four
lowest-lying states in the iso-vector channel (see \eq{eq:GrhorhoL}),
which also allows for the determination of the finite-volume shift via
the timelike pion form factor.

The {\sc{RBC/UKQCD Collaboration}}
\cite{Boyle:2011hu,Blum:2016xpd,Blum:2018mom} employs domain wall
fermions for the discretisation of the quark action. In their initial
study \cite{Boyle:2011hu} they studied eight different ensembles at
three different lattice spacings to obtain a result for $\ahvp$, yet
without contributions from the charm quark and disconnected
diagrams. By comparing the results from a chiral extrapolation with
and without the rescaling factor of \eq{eq:rescale} no statistically
significant difference between the two procedures was detected at the
physical pion mass. Results were found to be stable against variations
in the lattice spacing and volume in the accessible range. The
calculation of the strange quark contribution $(\ahvp)^s$ reported in
\cite{Blum:2016xpd} was based on ensembles at the physical pion mass
and two lattice spacings. The $Q^2$-dependence was studied extensively
by employing the hybrid method of
Ref.\,\cite{Golterman:2014ksa}. Systematic errors were estimated by
considering a large number of procedural variations.
In their latest paper\,\cite{Blum:2018mom} RBC/UKQCD presented a
determination of $\ahvp$ at two values of the lattice spacing and at
the physical pion mass. Employing the time-momentum representation,
the long-distance regime was constrained by the ``bounding method''
used also by the BMW Collaboration (see \fig{fig:bounding}). The total
precision of the final result, which includes finite-volume
corrections, quark-disconnected diagrams and isospin-breaking
(QCD+QED) effects, is 2.6\%. They also quote a far more precise
estimate which is obtained through a combination of their lattice
calculation of the correlator $G(x_0)$ and experimentally determined
cross section ratio $R(s)$. This is discussed further below.

\begin{sidewaystable}
\begin{center}
\begin{tabular}{lllllll}
\hline\hline
Collaboration & $\ahvp$ & $(\ahvp)^{uds}$
 & $(\ahvp)^{ud}$ & $(\ahvp)^{s}$ & $(\ahvp)^{c}$ & Method \\
\hline
$N_{\rm f}=2+1+1:$ & & & & & & \\
BMW 17 \cite{Borsanyi:2017zdw} &
711.0(7.5)(17.3)${}^{\ast\dag}$ & 696.3(7.5)(17.3)${}^{\ast\dag}$ & & 
53.7(0)(4) & 14.7(0)(1) & TMR\\
ETMC 17 \cite{Giusti:2017jof} &  & & & 
53.1(2.5)${}^{\dag}$ & 14.72(56)${}^{\dag}$ & TMR\\
HPQCD 16 \cite{Chakraborty:2016mwy} &
667(6)(12)${}^{\ast\dag}$ & & 599(11)${}^{\ast\dag}$ & & & Moments\\
HPQCD 14 \cite{Chakraborty:2014mwa} & & & &
53.4(6) & 14.4(4) & Moments\\
ETMC 13 \cite{Burger:2013jya,Burger:priv} &
674(21)(18) & 655(21) & & 53(3) & 14.1(6) & Fits in $Q^2$\\
\hline
$N_{\rm f}=2+1:$ & & & & & & \\
 & 715.4(16.3)(9.2)${}^{\ast\dag}$ & 701.2(16.3)(9.2)${}^{\ast\dag}$ & &
53.2(4)(3) & 14.3(0)(7) & TMR\\
\rb{RBC/UKQCD 18 \cite{Blum:2018mom}} &
692.5(1.4)(0.5)(0.7)(2.1)${}^{\ast\dag}$ & & & & & $R$-ratio, TMR\\
RBC/UKQCD 16 \cite{Blum:2016xpd} & & & & 53.1(9)(${}^{1}_{3}$) &
& Hybrid \\
RBC/UKQCD 11 \cite{Boyle:2011hu} & & 641(33)(32) & & & & Fits in
$Q^2$\\ 
Aubin \& Blum 07 \cite{Aubin:2006xv} & & 713(15) / 748(25) & & &
& Fits in $Q^2$ \\
\hline
$N_{\rm f}=2:$ & & & & & & \\
CLS/Mainz 17 \cite{DellaMorte:2017dyu} &
654(32)(${}^{21}_{23}$)${}^\ast$ &
639(32)(${}^{21}_{23}$)${}^\ast$ & 588(32)(${}^{21}_{23}$) & 51.1(1.7)(0.4) &
14.3(2)(1) & TMR \\
CLS/Mainz 11 \cite{DellaMorte:2011aa} & & 618(64) & & & & Fits in $Q^2$ \\
ETMC 11 \cite{Feng:2011zk} & & & 572(16)${}^\ast$ & & & Fits in $Q^2$ \\
\hline\hline
\end{tabular}
\caption{Compilation of recent results for the hadronic vacuum
  polarisation contribution in units of $10^{-10}$. The method used to
  determine $\ahvp$ is specified in the last column. Whenever
  contributions from quark-disconnected diagrams enter the estimate or
  the quoted error this is marked by an asterisk. Entries marked by a
  dagger have either been corrected for isospin-breaking effects or
  the corresponding uncertainty has been included in the
  error. Additional information on simulation details are given in
  Table\,\ref{tab:hvpdetails}.\la{tab:hvp}}
\end{center}
\end{sidewaystable}

\begin{sidewaystable}
\begin{center}
\begin{tabular}{lccccc}
\hline\hline
Collaboration & Action & $a$\,[fm] & $m_\pi^{\rm min}$\,[MeV] &
$m_\pi^{\rm min}L$ & FV corr. \\
\hline
$N_{\rm f}=2+1+1:$ & & & & & \\
BMW 17 \cite{Borsanyi:2017zdw} & Staggered & $0.064, 0.095,
0.111, 0.118, 0.134$ & 139 & 4.3
& ChEFT \\
ETMC 17 \cite{Giusti:2017jof} & tmQCD & $0.062, 0.082, 0.089$ &
223 & 3.4 & {\sffamily X} \\
HPQCD 16 \cite{Chakraborty:2016mwy} & HISQ & $0.09, 0.12, 0.15$ & 133
& 3.9 & ChEFT \\
HPQCD 14 \cite{Chakraborty:2014mwa} & HISQ & $0.09, 0.12, 0.15$ & 133
& 3.9 & ChEFT \\
ETMC 13 \cite{Burger:2013jya,Burger:priv} & tmQCD & $0.062, 0.078,
0.086$ & 227 & 3.3 & {\sffamily X} \\
\hline
$N_{\rm f}=2+1:$ & & & & & \\
RBC/UKQCD 18 \cite{Blum:2018mom} & DWF & $0.084, 0.114$ & 139 & 3.8 &
ChEFT \\
RBC/UKQCD 16 \cite{Blum:2016xpd} & DWF & $0.084, 0.114$ & 139 & 3.8 &
{\sffamily X} \\
RBC/UKQCD 11 \cite{Boyle:2011hu} & DWF &  $0.084, 0.114, 0.143$ & 180
& 4.0 & {\sffamily X} \\
Aubin \& Blum 07 \cite{Aubin:2006xv} & Staggered & $0.086$ & 241
& 4.2 & {\sffamily X} \\
\hline
$N_{\rm f}=2:$ & & & & & \\
CLS/Mainz 17 \cite{DellaMorte:2017dyu} & Clover & $0.049, 0.066, 0.076$
& 185 & 4.0 & GS \\
CLS/Mainz 11 \cite{DellaMorte:2011aa} & Clover &  $0.05, 0.06, 0.08$
& 277 & 4.0 & {\sffamily X} \\
ETMC 11 \cite{Feng:2011zk} & tmQCD & $0.063, 0.079$ & 290 & 3.7 &
{\sffamily X} \\
\hline\hline
\end{tabular}
\caption{Simulation details for the calculations of the hadronic
  vacuum polarisation contribution $\ahvp$ listed in
  Table~\ref{tab:hvp}. The discretisation of the quark action is
  indicated in the second column (HISQ: highly improved staggered
  quarks; tmQCD: twisted mass QCD; DWF: domain wall fermions; Clover:
  O$(a)$ improved Wilson quarks). The column labelled ``FV corr.'',
  indicate whether finite-volume corrections have been included,
  either via chiral effective field theory (ChEFT) or the formalism
  based on the Gounaris-Sakurai (GS)
  parameterisation.\la{tab:hvpdetails}}
\end{center}
\end{sidewaystable}

The {\sc{HPQCD Collaboration}}
\cite{Chakraborty:2014mwa,Chakraborty:2016mwy} use ensembles with
$\Nf=2+1+1$ flavours of ``highly improved staggered quarks'' (HISQ) to
determine $\ahvp$ by computing time moments. The large-$x_0$ regime of
the vector correlator $G(x_0)$ was constrained by performing
multi-exponential fits using Bayesian priors. Ensembles at three
different lattice spacings including physical pion masses enable a
simultaneous chiral and continuum extrapolation of the moments from
which the vacuum polarisation function $\Pi(Q^2)$ was determined. The
light quark contribution was corrected for finite-volume effects as
well as taste-symmetry violations that manifest themselves as mass
splittings among pions within a taste multiplet. Contributions from
quark-disconnected diagrams were included in the final result and
error. The effects of strong isospin-breaking were reported in a
companion paper together with the Fermilab-HPQCD-MILC
Collaboration\,\cite{Chakraborty:2017tqp}. The findings were already
discussed in Section\,\ref{sec:IB} of this review. The overall
precision of the result quoted in \cite{Chakraborty:2016mwy} is 2.0\%.

The {\sc{BMW Collaboration}} \cite{Borsanyi:2016lpl,Borsanyi:2017zdw}
has performed a comprehensive study using $\Nf=2+1+1$ flavours of
staggered quarks at six values of the lattice spacing including
physical pion masses and volumes that satisfy $m_\pi L\approx
4.3$. Thanks to the massive accumulated statistics (i.e. more than
$10^6$ individual measurements for the contributions from up and down
quarks and about $10^5$ measurements for strange and charm), they
obtained very precise results and were also able to compute
quark-disconnected diagrams for all four flavours. The infrared regime
of the correlator was constrained via the ``bounding method'' (see
\eq{eq:bounding}). The first paper \cite{Borsanyi:2016lpl} was
focussed on the leading two time moments, $\Pi_1$ and $\Pi_2$, which
can be used to constrain $\ahvp$ via the Mellin moments discussed in
Section\,\ref{sec:MB} and the QCD sum rule approach of
Section\,\ref{sec:QCDSR}. The more recent publication
\cite{Borsanyi:2017zdw} contains the result for $\ahvp$ extrapolated
to the continuum limit, including finite-volume corrections,
quark-disconnected diagrams and a correction for isospin-breaking
effects estimated from phenomenology (see Section\,\ref{sec:IB}). The
overall precision is 2.7\%.

\begin{figure}[t]
\begin{center}
\includegraphics[width=0.7\textwidth]{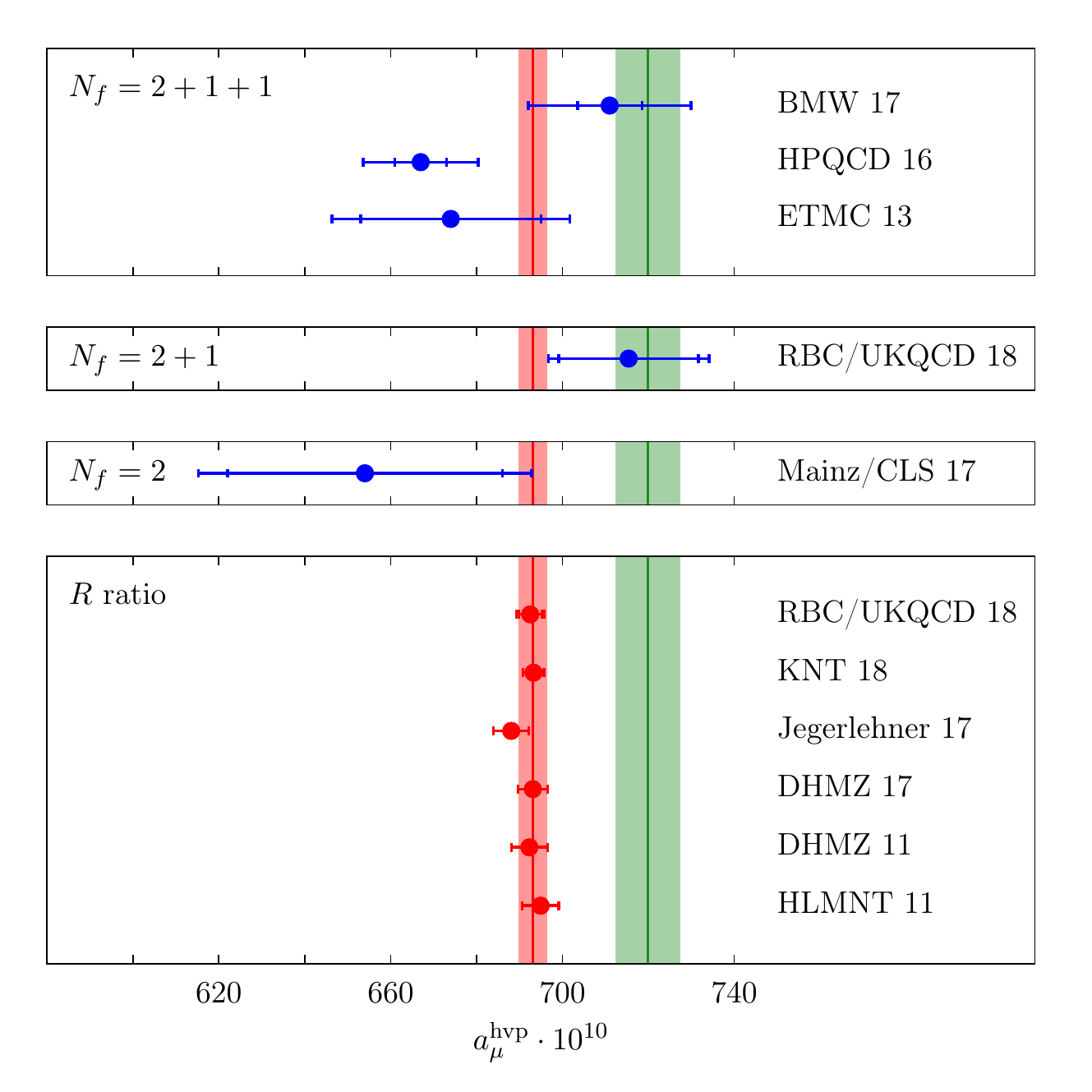}
\vspace{-0.5cm}
\caption{Compilation of recent results for the hadronic vacuum
  polarisation contribution in units of $10^{-10}$. The three panels
  represent calculations with different numbers of sea
  quarks. References for individual lattice calculations are listed in
  Table\,\ref{tab:hvp}. The red vertical band denotes the estimate
  from dispersion theory quoted in DHMZ\,17 \cite{Davier:2017zfy},
  while the green band represents the ``no new physics'' scenario (see
  text). The other phenomenological determinations based on the
  $R$-ratio are labelled as HLMNT\,11 \cite{Hagiwara:2011af}, DHMZ\,11
  \cite{Davier:2010nc}, Jegerlehner\,17 \cite{Jegerlehner:2017lbd} and
  KNT\,18 \cite{Keshavarzi:2018mgv}. \la{fig:hvp}}
\end{center}
\end{figure}

In\,\fig{fig:hvp} we show a compilation of recent results for the
total leading-order hadronic vacuum polarisation contribution, while
Figures\,\ref{fig:hvpstr} and \ref{fig:hvpch} show recent estimates of
the individual contributions from the strange and charm quarks to
$\ahvp$. The overall precision of current lattice calculations of
$\ahvp$ is at the level of 2.5\%. Estimates are mostly dominated by
systematics, with the largest uncertainties associated with
finite-volume effects, the continuum extrapolation, isospin-breaking
and scale setting. Constraining the infrared regime of the vector
correlator also makes a sizeable contribution to the total error
budget of the calculations shown and listed in Figure\,\ref{fig:hvp}
and Table\,\ref{tab:hvp}.

It is interesting to note that the results of ETMC\,13
\cite{Burger:2013jya}, HPQCD\,16 \cite{Chakraborty:2016mwy} and
CLS/Mainz\,17 \cite{DellaMorte:2017dyu} are 3--5\% lower than the
phenomenological estimate, while the most recent estimates by BMW\,17
\cite{Borsanyi:2017zdw} and RBC/UKQCD\,18 \cite{Blum:2018mom} are
larger by 3\%. By contrast, lattice estimates for the individual
strange and charm quark contributions are quite compatible among
different collaborations, as indicated in Figures\,\ref{fig:hvpstr}
and\,\ref{fig:hvpch}. This confirms that the contribution from the
light quark flavours not only accounts for the bulk of the HVP
contribution but is also mainly responsible for the accuracy of
current lattice calculations. In Figure\,\ref{fig:hvp} the lattice
results for $\ahvp$ are also compared to the estimates obtained by
evaluating the dispersion integral using the experimentally measured
cross section ratio for $e^+e^-\to\rm hadrons$ in the low-energy
regime. Clearly, the uncertainties of the phenomenological approach
are much smaller, implying that the overall precision of lattice
calculations must be improved by a factor $\sim5$.

Figure\,\ref{fig:hvp} also shows that the estimates from BMW\,17 and
RBC/UKQCD\,18 are compatible with the ``no new physics'' (NNP)
scenario. The quantity $(\ahvp)_{\rm NNP}$ is defined as the value of
$\ahvp$ which would make the currently observed discrepancy of 3.5
standard deviations between SM prediction and experimental measurement
of $a_\mu$ disappear, i.e.
\be
   (\ahvp)_{\rm NNP} \equiv (\ahvp)_{\rm disp}+\Delta a_\mu,\quad
   \Delta a_\mu\equiv a_\mu^{\rm exp}-a_\mu^{\rm SM}
   = 26.6(6.3)_{\rm exp}(4.3)_{\rm theo}\cdot 10^{-10},
\ee
where $(\ahvp)_{\rm disp}$ is the result from dispersion theory.
After adding $\Delta a_\mu$ to the result for $(\ahvp)_{\rm disp}$
from Ref.\,\cite{Davier:2017zfy} one finds $(\ahvp)_{\rm NNP} =
719.7(7.6)\cdot 10^{-10}$, which yields the green vertical band in
Figure\,\ref{fig:hvp}. One concludes that current lattice estimates of
$\ahvp$ are not yet precise enough to distinguish between the ``no new
physics'' scenario and a clear deviation between Standard Model and
experiment.

\begin{figure}[t]
\begin{center}
\includegraphics[width=0.7\textwidth]{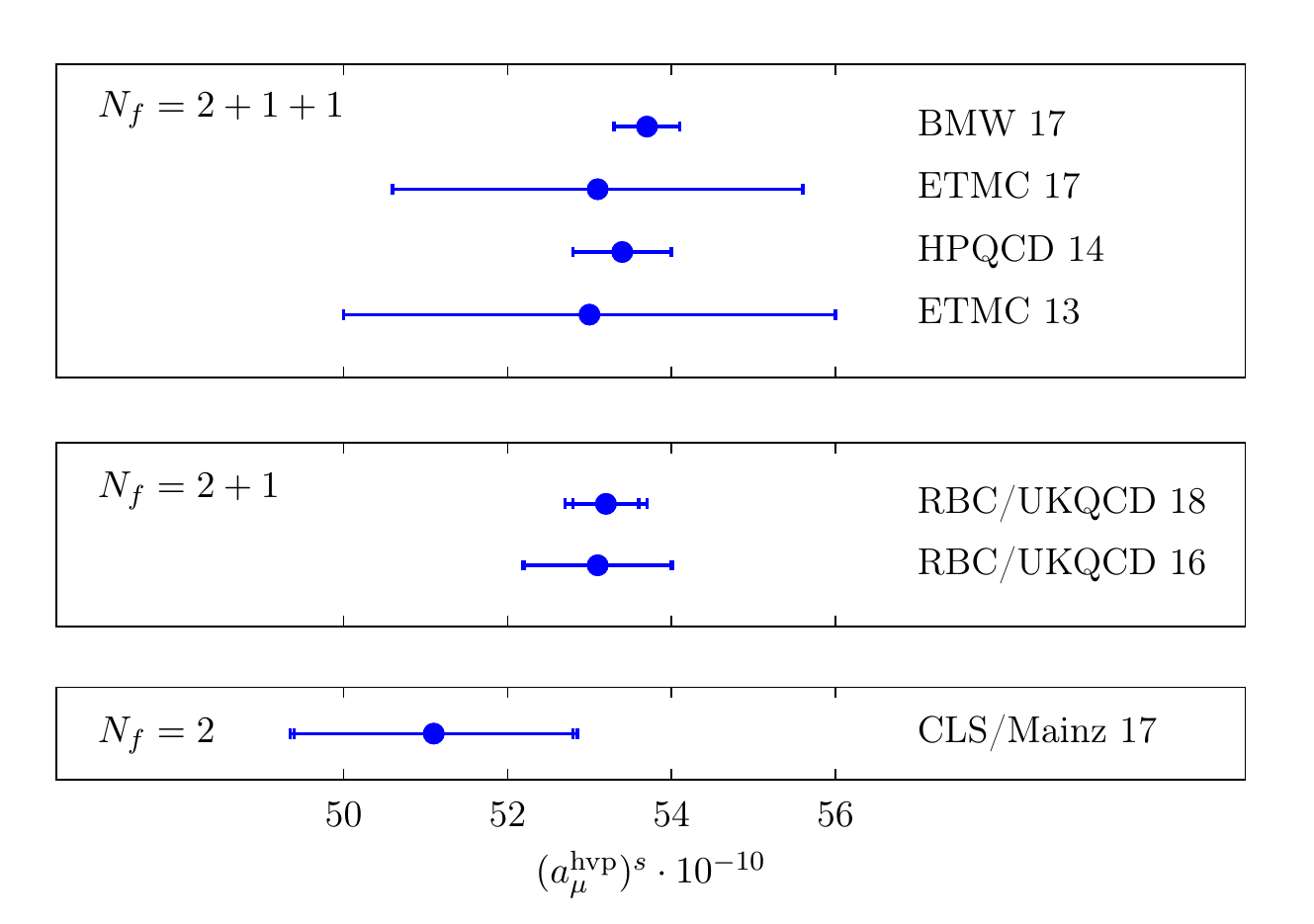}
\vspace{-0.5cm}
\caption{The hadronic vacuum polarisation contribution from the
  strange quark in units of $10^{-10}$.  References for individual
  calculations are listed in Table\,\ref{tab:hvp}.  \la{fig:hvpstr}}
\end{center}
\end{figure}

\begin{figure}
\begin{center}
\includegraphics[width=0.7\textwidth]{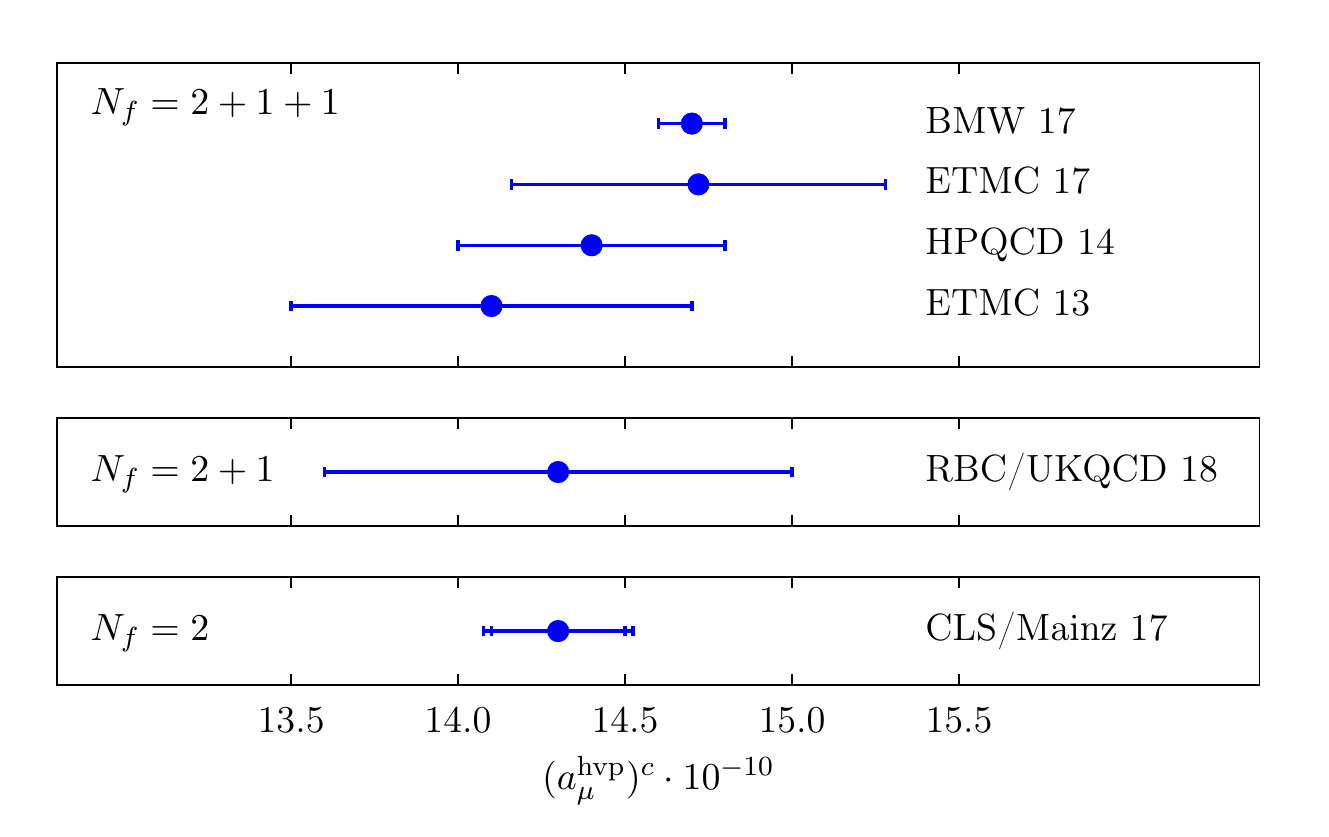}
\vspace{-0.5cm}
\caption{The hadronic vacuum polarisation contribution from the charm
  quark in units of $10^{-10}$.  References for individual
  calculations are listed in Table\,\ref{tab:hvp}.  \la{fig:hvpch}}
\end{center}
\end{figure}

An alternative method for determining $\ahvp$ from a combination of
the experimentally measured $R$-ratio and lattice data was used by
RBC/UKQCD\,\cite{Blum:2018mom}. It is based on the relation between
the vector correlator $G(x_0)$ and the hadronic cross section ratio
$R(s)$ derived in Ref.\,\cite{Bernecker:2011gh}, i.e.
\be\la{eq:Gx0Rratio}
  G(x_0) = \frac{1}{12\pi^2}\int_0^{\infty} d(\sqrt{s})\,R(s)s\,
  e^{-\sqrt{s}x_0},
\ee
with $R(s)$ defined in \eq{eq:Rratio}. When multiplied by the kernel
function $w(x_0)$ of the time-momentum representation one obtains
$\ahvp$ via \eq{eq:TMRamu}. As explained in detail in
\cite{Lehner:2017kuc,Blum:2018mom}, replacing the lattice
determination of the vector correlator $G(x_0)$ by the expression in
\eq{eq:Gx0Rratio} results in a statistically more precise integrand
$w(x_0)G(x_0)$ for $x_0\lesssim0.4$\,fm and
$x_0\gtrsim1.0$\,fm. Replacing $G(x_0)$ by its representation in terms
of the $R$-ratio for $x_0\leq0.4$\,fm reduces the influence from
discretisation errors, while for $x_0\geq1.0$\,fm the uncertainties
associated with the modelling of the long-distance behaviour can be
eliminated. By splitting the integration over $x_0$ into three
intervals and using the lattice data for $G(x_0)$ only for
$0.4\,\fm\leq x_0 \leq 1.0$\,fm, RBC/UKQCD\,18 obtain
\be
   \ahvp = (692.5\pm2.7)\cdot10^{-10},
\ee
where statistical and systematic errors from the lattice calculation
and the $R$-ratio have been combined in quadrature. This estimate,
shown in the lower panel of Figure\,\ref{fig:hvp}, not only agrees
very well with the results from the most recent phenomenological
analyses \cite{Davier:2017zfy,Jegerlehner:2017lbd,Keshavarzi:2018mgv},
but is also slightly more precise. Owing to the use of the
experimentally determined $R$-ratio and keeping in mind that the value
of $\ahvp$ is dominated by the low-energy regime, it is not too
surprising that this method produces a central value that agrees so
well with dispersion theory.

\begin{sidewaystable}
\begin{center}
\begin{tabular}{llll}
\hline\hline
Collaboration & $\phantom{0.0}\Pi_1 [\gev^{-2}]$
&  $\phantom{-0.}\Pi_2 [\gev^{-4}]$ & Comments \\
\hline
$N_{\rm f}=2+1+1:$ & & & \\
BMW 16 \cite{Borsanyi:2016lpl}
& $0.0999(10)(9)(23)_{\rm FV}\,(13)_{\rm IB}$
& $-0.181(6)(4)(10)_{\rm FV}\,(2)_{\rm IB}$
& Continuum limit \\
& $0.0889(16)$ & $-0.206(10)$ & $a=0.15\,\fm$ \\
\rb{HPQCD 16 \cite{Chakraborty:2016mwy}}
& $0.0892(14)$ & $-0.204(9)$ & $a=0.12\,\fm$ \\
\hline
$N_{\rm f}=2:$ & & & \\
CLS/Mainz 17 \cite{DellaMorte:2017dyu} & $0.0883(59)$ & & Continuum limit
\\
\hline
Benayoun  & & & \\
et al. $16\qquad$ \rb{\cite{Benayoun:2016krn,Borsanyi:2016lpl}}
 & \rb{$0.0990(7)$} & \rb{$-0.2057(16)$}
 & \rb{$R(e^+ e^-\to\hbox{hadrons})$} \\
\hline\hline
\end{tabular}
\caption{Results for the first two time moments of the hadronic vacuum
  polarisation function $\hat\Pi(Q^2)$. Lattice estimates have been
  computed at or extrapolated to the physical pion
  mass. The subscript ``FV'' on the results from
  \cite{Borsanyi:2016lpl} indicates the uncertainty in the applied
  volume correction, while the label ``IB'' denotes a phenomenological
  estimate of isospin breaking corrections. \la{tab:moments}}
\end{center}
\end{sidewaystable}

Instead of focussing directly on $\ahvp$ it is also instructive to
discuss the individual time moments which also provide useful
information, since they can be linked to the rapidly converging
expansion of $\ahvp$ in terms of the Mellin moments discussed in
Section\,\ref{sec:MB}. Results for the leading two time moments
published by BMW \cite{Borsanyi:2016lpl} are listed in
Table\,\ref{tab:moments} together with the estimates by
HPQCD\,\cite{Chakraborty:2016mwy} and CLS/Mainz
\cite{DellaMorte:2017dyu}.

BMW correct their results for finite-volume effects, computed in
chiral effective theory at one loop. They quote corrections of 2\% and
10\% for $\Pi_1$ and $\Pi_2$, respectively. They also add an
uncertainty reflecting the fact that their results are not corrected
for isospin breaking effects. HPQCD apply a correction for
finite-volume effects and effects relating to the breaking of taste
symmetry, which amounts to 7\% for $\Pi_1$. The results from CLS/Mainz
have been extrapolated to the continuum limit without applying any
finite-volume corrections. While the estimates for $\Pi_1$ determined
by HPQCD and CLS/Mainz agree within errors, the result from BMW is
larger by 10\%. A similar observation applies to $\Pi_2$. It is
instructive to compare these findings to a recent study
\cite{Benayoun:2016krn} in which the moments are determined from the
experimentally determined $R$-ratio. While the phenomenological
estimate for $\Pi_1$ agrees well with the determination from BMW,
$\Pi_2$ from \cite{Benayoun:2016krn} compares more favourably to the
estimate by HPQCD. Obviously, the current situation calls for further
investigations into the systematics of lattice calculations, even more
so since both BMW and HPQCD employ staggered fermions as their
discretisation of the quark action.  It should also be noted that the
overall precision of the phenomenological determination of the moments
is higher than that of current lattice calculations.

\subsection{Concluding remarks on the hadronic vacuum polarisation}

Recent years have seen enormous progress concerning a first-principles
determination of the O($\alpha^2$) hadronic vacuum polarisation
contribution. Lattice QCD is not only capable of providing direct
determinations of $\ahvp$, but also provides estimates that combine
experimental information and/or other theoretical methods with lattice
results. In addition to controlling the standard systematic effects
arising in any lattice calculation, isospin breaking corrections,
contributions from quark-disconnected diagrams, finite-volume
corrections and the contributions from the IR regime must all be
quantified at the desired level of precision if lattice QCD is to have
a decisive impact on understanding the observed discrepancy between
experiment and SM prediction. The currently available direct
determinations of $\ahvp$ carry an overall uncertainty of $2-3$\%
which is not yet sufficient to discriminate between the dispersive
estimate and the ``no new physics'' scenario which assumes that the
observed discrepancy is due to some unknown or uncontrolled hadronic
effect. Ongoing efforts to provide more precise lattice QCD estimates
focus on increasing the overall statistical precision, including the
accuracy of the lattice scale which is a limiting factor, as well as
constraining finite-volume effects and the deep IR region of the
vector correlator.

\section{Hadronic light-by-light scattering in $(g-2)_\mu$\la{sec:HLbL}}

At O($\alpha^3$), the theoretically most challenging hadronic
contribution to $(g-2)_\mu$ is the scattering of light by light via
QCD degrees of freedom. In this section we review the progress made in
formulating the calculation of $\amuhlbl$ for a lattice QCD treatment,
the first numerical results and discuss some of the sources of
systematic error. The first proposal to compute $\amuhlbl$ in lattice QCD
was made in 2005~\cite{Hayakawa:2005eq}, but this activity picked up 
momentum only about five years ago. 
Prior to that, the only approach pursued to
estimate $\amuhlbl$ was the use of a model mainly based on the
exchange of hadronic resonances with various quantum numbers
(mainly $J^{PC}=0^{-+}$, $0^{++}$, $1^{++}$, $2^{++}$). In addition
to the ``Glasgow consensus'' value $\amuhlbl = (105\pm26)\cdot
10^{-11}$ quoted in Table \ref{tab:amustatus}, an alternative estimate
is $\amuhlbl = (102\pm39)\cdot 10^{-11}$
\cite{Jegerlehner:2015stw}. The largest contribution by far comes
from the exchange of the lightest pseudoscalar mesons, $\pi^0$, $\eta$
and $\eta'$. In recent years, major progress has also been made in
developing a dispersive approach to $\amuhlbl$~\cite{Colangelo:2014dfa,
  Colangelo:2014pva, Colangelo:2015ama, Colangelo:2017qdm,
  Colangelo:2017fiz}.

\subsection{General considerations}
\la{sec:seqprop}
We begin with a review of the initial steps in setting up the
calculation; most of this material is standard and has been known for
a long time. We follow the treatment given in~\cite{Knecht:2001qf}, with the difference
that we work directly in the Euclidean theory. This approach is appropriate because we want to 
compute an electromagnetic form factor at an (infinitesimal) \emph{spacelike} momentum transfer.
It is well known in the lattice community that the calculation of spacelike form factors can be formulated
directly in Euclidean space~\cite{Martinelli:1988rr}, thus making them accessible to lattice QCD simulations. Let 
\be
J_\rho = {\textstyle\frac{2}{3}} \bar u\gamma_\rho u -
{\textstyle\frac{1}{3}} \bar d\gamma_\rho d - {\textstyle\frac{1}{3}} \bar s\gamma_\rho s 
\ee
be the contribution of the light quarks to the electromagnetic current (in units of $-e$,
$e$ being the electric charge of the electron).
Here we are interested in the matrix element of $J_\rho$ between single-muon states,
$\<\mu^-(p',s')|J_\rho(0)|\mu^-(p,s)\>$. 
In Euclidean space, we therefore set 
\be
p_\mu = (iE_{\vec p},\vec p),\qquad \quad p'_\mu = (iE_{\vec p'},\vec p'),\qquad \quad  k_\mu= p_\mu'-p_\mu.
\ee
Lorentz symmetry implies generically
\be\la{eq:F1F2}
\Big\<\mu^-(p',s')\big|J_\rho(0)\big|\,\mu^-(p,s)\Big> 
= - \bar u^{s'}(p') \Big[\gamma_\rho \hat F_1(k^2) + \frac{\sigma_{\rho\tau}k_\tau }{2m}\hat F_2(k^2)\Big] u^{s}(p),
\ee
with $\sigma_{\rho\tau} \equiv \frac{i}{2} \left[\gamma_\rho,\gamma_\tau\right]$ in terms of the Euclidean Dirac matrices,
$\left\{\gamma_\rho,\gamma_\sigma\right\} = 2\delta_{\rho\sigma}$. The matrix element is thus parameterised in terms
of contributions (indicated by the hat) to the Dirac form factor $F_1(k^2)$ and the Pauli form factor $ F_2(k^2)$ of the muon.
The spinors $u^s(p)$ are the usual plane-wave solutions to the Dirac equation.

\begin{figure}
\centerline{\includegraphics[width=0.82\textwidth]{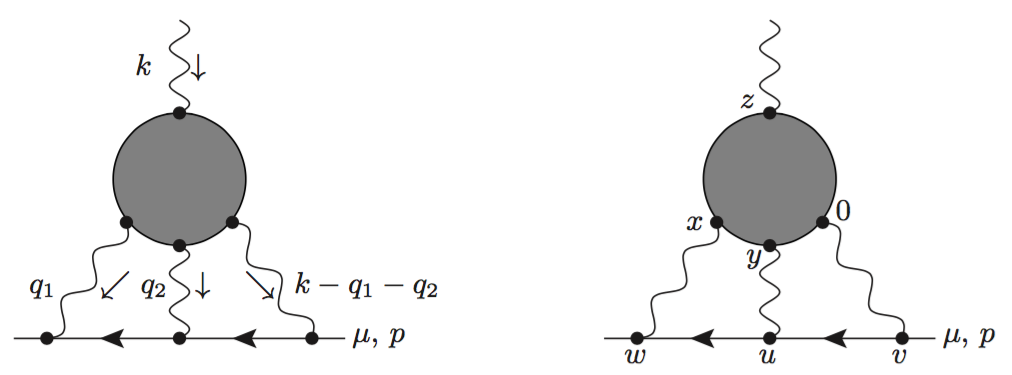}}
\caption{The hadronic light-by-light contribution to $(g-2)_\mu$. }
\label{fig:hlbl_p_and_x}
\end{figure}

Using the Feynman rules for QED in Euclidean space (see the left panel of Fig.\ \ref{fig:hlbl_p_and_x}), 
we get for the leading contribution in powers of $e$ to the matrix element
\ba
 (ie)\Big\<\mu^-(p')\big|J_\rho(0)\big|\,\mu^-(p)\Big> &=&  (-ie)^3  \,(ie)^4
\int \frac{d^4q_1}{(2\pi)^4} \int \frac{d^4q_2}{(2\pi)^4}  
\frac{1}{q_1^2 q_2^2 (q_1+q_2-k)^2}\cdot \\
&& 
\cdot\; \frac{-1}{(p'-q_1)^2+m^2}\,  \frac{-1}{(p'-q_1-q_2)^2+m^2}\cdot
\nonumber 
\phantom{\frac{1}{1}} \\ &&  
\cdot\; \bar u(p') \gamma^\mu (ip\!\!\!/'-iq\!\!\!/{}_1 - m) \gamma^\nu
 (ip\!\!\!/'-iq\!\!\!/{}_1 -iq\!\!\!/{}_2 - m) \gamma^\lambda u(p)\cdot 
\nonumber
\\
&& \cdot\; \Pi_{\mu\nu\lambda\rho}(q_1,q_2,k-q_1-q_2),
\nonumber
\ea
with 
\be\la{eq:PiMomSp}
\Pi_{\mu\nu\lambda\rho}(q_1,q_2,q_3) = \int d^4x_1\int d^4x_2 \int d^4x_3 \; \rme^{-i(q_1x_1+q_2x_2+q_3x_3)}
\Big\< J_\mu(x_1) J_\nu(x_2) J_\lambda(x_3) J_\rho(0)\Big\>_{\rm QCD}.
\ee
The tensor $\Pi_{\mu\nu\lambda\rho}$ satisfies 
\ba
(q_1)_\mu\, \Pi_{\mu\nu\lambda\rho}(q_1,q_2,q_3)=0,
&\qquad& 
(q_2)_\nu\, \Pi_{\mu\nu\lambda\rho}(q_1,q_2,q_3)=0,
\\
(q_3)_\lambda\, \Pi_{\mu\nu\lambda\rho}(q_1,q_2,q_3)=0,
&\qquad &
(q_1+q_2+q_3)_\rho\, \Pi_{\mu\nu\lambda\rho}(q_1,q_2,q_3)=0.
\ea
The last equation implies in particular 
\be
k_\sigma \; \Pi_{\mu\nu\lambda\sigma}(q_1,q_2,k-q_1-q_2)  = 0 \qquad \forall k,
\ee
and therefore 
\be
\frac{\partial}{\partial k_\rho}\Big(k_\sigma \; \Pi_{\mu\nu\lambda\sigma}(q_1,q_2,k-q_1-q_2)\Big)
= 0, 
\ee
implying 
\be
 \Pi_{\mu\nu\lambda\rho}(q_1,q_2,k-q_1-q_2) = 
- k_\sigma \frac{\partial}{\partial k_\rho}\Pi_{\mu\nu\lambda\sigma}(q_1,q_2,k-q_1-q_2).
\ee
We can therefore write
\be
\Big\<\mu^-(p',s')\big|J_\rho(0)\big|\,\mu^-(p,s)\Big> = k_\sigma \hat\gamma_{\rho\sigma}^{s's}(p',p),
\qquad 
\hat\gamma_{\rho\sigma}^{s's}(p',p) = \bar u^{s'}(p') \Gamma_{\rho\sigma}(p',p) u^s(p)
\ee
with 
\ba
\la{eq:Grhosig}
\Gamma_{\rho\sigma}(p',p)
  &=&   - e^6 \, 
\int \frac{d^4q_1}{(2\pi)^4} \int \frac{d^4q_2}{(2\pi)^4}  
\frac{1}{q_1^2 q_2^2 (q_1+q_2-k)^2}\cdot \\
&& 
\cdot\; \frac{1}{(p'-q_1)^2+m^2}\,  \frac{1}{(p'-q_1-q_2)^2+m^2}\cdot
\nonumber 
\phantom{\frac{1}{1}} \\ &&  
\cdot\;  \gamma^\mu (ip\!\!\!/'-iq\!\!\!/{}_1 - m) \gamma^\nu
 (ip\!\!\!/'-iq\!\!\!/{}_1 -iq\!\!\!/{}_2 - m) \gamma^\lambda \cdot 
\nonumber
\\
&& \cdot\; \frac{\partial}{\partial k_\rho}\Pi_{\mu\nu\lambda\sigma}(q_1,q_2,k-q_1-q_2).
\nonumber
\ea
This expression is well known and has been the starting point for many calculations.

Note that due to the property $k_\rho k_\sigma \Gamma_{\rho\sigma}(p',p) = 0$, one finds that $\hat F_1(0) = 0$.
Identifying terms to linear order in $k$, we have 
\be\la{eq:gF2}
 \hat\gamma_{\rho\tau}^{s's}(p,p) = -\bar u^{s'}(p) \frac{\hat F_2(0)}{2m}\sigma_{\rho\tau}u^s(p).
\ee
By writing $ -ip\!\!\!/+m = \sum_{s} u^s(p) \bar u^s(p) $ and using Eq.\ (\ref{eq:gF2}), one obtains the expression~\cite{Aldins:1970id}
\be\la{eq:amuhlbl}
\amuhlbl=\hat F_2(0) = -\frac{i}{48m} {\rm Tr}\left\{ [\gamma_{\rho},\gamma_{\tau}]\,(-ip\!\!\!/+m)\Gamma_{\rho\tau}(p,p) (-ip\!\!\!/+m)\right\}.
\ee
In summary, it suffices to know $\Gamma_{\rho\sigma}(p,p)$ to
determine $\hat F_2(0)$. This observation is interesting, because in
the loop integral (\ref{eq:Grhosig}), it means that the integrand is
now function of three ($p,q_1,q_2$) rather than four ($p,p',q_1,q_2$)
momenta. 

The number of relevant four-momenta can be further reduced, down to
two, by realizing that $\hat F_2(0)$ is a Lorentz scalar, and therefore
does not depend on the direction of the muon's momentum. In the rest
frame of the muon, its four-momentum has the form $p_\mu = (im,\vec
0)$. Therefore, in Euclidean space, the momentum may be parameterised as
\be
p = i\,m\,\hat\epsilon,
\ee
with $\epsilon\in\mathbb{R}^4$ a unit vector. The integrand in
Eq.\ (\ref{eq:Grhosig}) projected onto the anomalous magnetic moment via Eq.\ (\ref{eq:amuhlbl}), 
can therefore be averaged over the direction of $\hat\epsilon$,
\be
\<f(\hat\epsilon)\>_{\hat\epsilon} \equiv \frac{1}{2\pi^2} \int d\Omega_\epsilon\; f(\hat\epsilon),
\ee
$2\pi^2$ being the surface of the unit-sphere embedded in four dimensions.
The integrand for $\amuhlbl$ can thus be brought into the form
\be
\hat F_2(0) = e^6 \int \frac{d^4q_1}{(2\pi)^4}\int \frac{d^4q_2}{(2\pi)^4}\; 
\underbrace{{\cal K}_{\mu\nu\lambda[\rho\sigma]}(q_1,q_2)}_{{\displaystyle{\rm QED}}}
\underbrace{\Big[\frac{\partial}{\partial k_\rho}\Pi_{\mu\nu\lambda\sigma}(q_1,q_2,k-q_1-q_2)\Big]_{k=0}}_{{\displaystyle{\rm QCD}}}.
\ee
After the contraction of the five Lorentz indices of the QED kernel
${\cal K}_{\mu\nu\lambda\rho\sigma}$ with those of the QCD four-point
function, the integrand is a Lorentz scalar. It is therefore a
function of the three invariants $q_1^2$, $q_2^2$ and $q_1\cdot q_2$.
Performing the integral in hyperspherical coordinates, only the
integrals over these variables are non-trivial. The reduction to a
three-dimensional integral can be made explicit if the QCD four-point
function is decomposed into a number of tensor structures with
associated form factors, which are functions of $q_1^2,q_2^2,q_1\cdot
q_2$. This program has been carried out in
\cite{Colangelo:2015ama}. Taking into account crossing symmetry, a
total of twelve form factors characterizing $\frac{\partial}{\partial
  k_\rho}\Pi_{\mu\nu\lambda\sigma}(q_1,q_2,k-q_1-q_2)_{k=0}$
contribute to $\amuhlbl$. From a lattice point of view, a possible
strategy is thus to provide a parameterisation of each of these twelve
form factors, which can then be fed into the integral over
($q_1^2,q_2^2,q_1\cdot q_2$) to obtain $\hat F_2(0)$. No attempt has
been made yet to implement this strategy.

As will be described in Section \ref{sec:semia}, it is also possible to 
write the desired quantity in terms of a position-space integral,
\ba\la{eq:MasterPosSpace}
\hat F_2(0) &=&  \frac{m e^6}{3}\int d^4y \int d^4x\;
\underbrace{\bar {\cal L}_{[\rho,\sigma];\mu\nu\lambda}(x,y)}_{{\displaystyle{\rm QED}}}\; 
\underbrace{i\widehat\Pi_{\rho;\mu\nu\lambda\sigma}(x,y)}_{{\displaystyle{\rm QCD}}}\;,
\\
i\widehat \Pi_{\rho;\mu\nu\lambda\sigma}( x, y)  &=& 
-\int d^4z\; z_\rho\, \Big\<\,J_\mu(x)\,J_\nu(y)\,J_\sigma(z)\, J_\lambda(0)\Big\>.
\la{eq:PihatDef}
\ea
From the point of view of lattice calculations, one advantage of this
representation is that for fixed $y$, the $x$-integral over the fully
connected contribution to the four-point function can be evaluated with
a computational effort of order volume by using the technique of
sequential propagators -- see the next paragraph. An explicit
decomposition of the QCD four-point function into form factors is thus
not necessary. After the $x$-integral is performed, by Lorentz
invariance of $\amuhlbl$, the $y$ integral reduces to a
one-dimensional integral over $|y|$. A second advantage is that on
the torus, the asymptotic finite-size effects are determined by the
longest QCD correlation length; power-law corrections in the box size
$L$ are thus avoided altogether.

\begin{figure}[tb]
 \begin{center}
\includegraphics[width=0.8\textwidth]{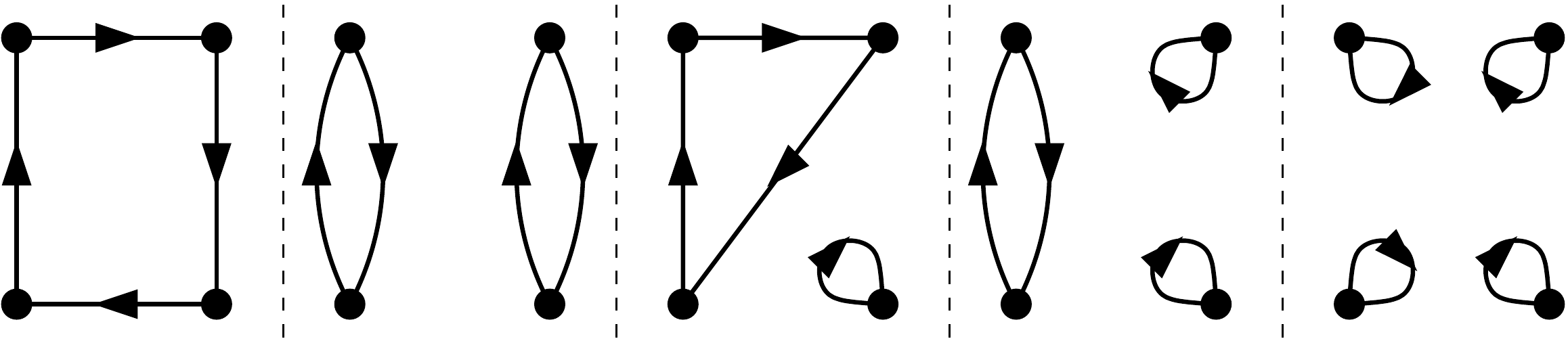}
\caption{The five classes of quark-field Wick contractions contributing to HLbL scattering.
From left to right we refer to them as (4), (2,2), (3,1), (2,1,1) and (1,1,1,1). Each class contains a number of 
actual contractions. Figure from~\cite{Green:2015sra}.
\label{fig:Wtopo}}
 \end{center}
\end{figure}

The HLbL scattering amplitude, which is determined by the four-point
function (\ref{eq:PiMomSp}) of the vector current, can be computed in
lattice QCD by constructing all possible Wick contractions of the
quark fields -- their interaction with the SU(3) gauge fields being
taken into account non-perturbatively by the importance-sampling of
the gauge fields. A complete list of the Wick contraction topology
classes is given in Fig.\ \ref{fig:Wtopo}. Each class consists of a number 
of Wick contractions; for the fully connected class, labelled ``(4)''
 in Fig.\ \ref{fig:Wtopo}, there are six of them.
An important computational aspect in lattice QCD is then the following:
consider an $n$-point correlation function of the type
\be\la{eq:exampcorr}
\sum_{x_2} f(x_2)\; \left\<{\cal O}_0(0) \,{\cal O}_1(x_1) \,{\cal O}_2(x_2)\right\>,
\ee
with ${\cal O}_i(x) = \bar\psi(x)\Gamma_i \psi(x)$. One of the fully connected contributions to 
the Wick contractions reads
\be
\sum_{x_2} f(x_2)\; \left\<\Gamma_0 S^f(0,x_1) \Gamma_1 S^f(x_1,x_2)\Gamma_2 S^f(x_2,0)\right\>,
\ee
where $S^f$ is the quark propagator of flavour $f$ and the average is taken over the SU(3) gauge field.
The key to computing the quantity
\be
v(x_1)\equiv\sum_{x_2} f(x_2)S^f(x_1,x_2)\Gamma_2 S^f(x_2,0)
\ee
is to note that it satisfies 
\be
\sum_z D^f(x,z) v(z) = f(x) \Gamma_2 S^f(x,0).
\ee
 Thus the spinor field
$v(z)$, called a sequential propagator, is obtained by inverting the Dirac operator on a single 
given ``source'' field, $D^f v = \eta$, with $\eta(x) =f(x) \Gamma_2 S^f(x,0)$. This technique 
allows one to calculate the correlation function (\ref{eq:exampcorr}) with O($V$) operations.
More generally, correlation functions of the type 
\be
C[f_1,\dots,f_{n-1}](x_n) = \sum_{x_1,\dots,x_{n-1}} \Big(\prod_{i=1}^{n-1} f_i(x_i)\Big)\; \< {\cal O}_1(x_1)\dots {\cal O}_n(x_n)\>
\ee
can be computed simultaneously for all $x_n$ with O($V$) operations using this technique. The important point is that 
the weighting function of the coordinates $(x_1,\dots,x_{n-1})$ must be factorised, $\prod_{i=1}^{n-1} f_i(x_i)$.
This observation will be used repeatedly in the forthcoming sections, particularly for computing
the fully connected class of Wick contractions of the vector four-point function.

\subsection{Stochastic U(1) field}

In this subsection we describe a class of lattice methods to compute
$\amuhlbl$ whose common point is the (at least partially) stochastic
treatment of the electromagnetic field. We follow the treatment of Ref.\ \cite{Blum:2015gfa}.
Consider then the Euclidean correlation function
\be\la{eq:calMdef}
-{\cal M}^{\rm l+h}_\nu(x_{\rm src},x_{\rm op},x_{\rm snk}) 
= \left\<\mu(x_{\rm snk})\;J^{\rm l+h}_\nu(x_{\rm op})\; \bar\mu(x_{\rm src})\right\>
\ee
involving the muon field $\mu(x)$ and the full electromagnetic current $J^{\rm l+h}_\nu(x)$ in units of $(-e)$.
For $t_{\rm src}\to-\infty$ and $t_{\rm snk}\to+\infty$, 
the Fourier-transform of ${\cal M}_\nu(x_{\rm src},x_{\rm op},x_{\rm snk}) $,
projecting the initial-state muon on momentum $\vec p$ and the final-state muon on momentum $\vec p'$,
is proportional to the matrix element (\ref{eq:F1F2}), a linear combination of the Dirac and Pauli form factors
at momentum transfer $q=p'-p$. 

However, treating QED perturbatively, there are many Feynman diagrams
contributing to ${\cal M}_\nu(x_{\rm src},x_{\rm op},x_{\rm snk})$,
and we are only interested in the HLbL contribution. Therefore, we
keep only the hadronic contribution $J_\nu$ of $J^{\rm l+h}_\nu$ and
select the relevant Feynman diagrams. Call this correlation function ${\cal M}_\nu$
and consider the fully connected HLbL contribution ${\cal M}^{(4)}_\nu$, 
\ba\la{eq:M4}
{\cal M}^{(4)}_\nu(x_{\rm src},x_{\rm op},x_{\rm snk}) &=& \sum_{x,y,z} {\cal F}_\nu(x,y,z,x_{\rm src},x_{\rm op},x_{\rm snk}),
\\
-ie {\cal F}_\nu(x,y,z,x_{\rm src},x_{\rm op},x_{\rm snk})
&=&
- (-ie)^3 (ie)^4\sum_{f=u,d,s} {\cal Q}_f^4\;
\nonumber \\ && \Big\<{\rm Tr}\{ \gamma_\nu S^f(x_{\rm op},x)\gamma_\rho S^f(x,z) 
\gamma_\kappa S^f(z,y)\gamma_\sigma S^f(y,x_{\rm op}) \}\Big\>_{{\rm SU}(3)}
\nonumber\\ && \times \sum_{x',y',z'} G_{\rho\rho'}(x,x') G_{\sigma\sigma'}(y,y') G_{\kappa\kappa'}(z,z')\;
\la{eq:FnuA}\\ && \Big( S_0(x_{\rm snk},x') \gamma_{\rho'} S_0(x',z') \gamma_{\kappa'} S_0(z',y')\gamma_{\sigma'} S_0(y',x_{\rm src})
\nonumber\\ && + S_0(x_{\rm snk},z') \gamma_{\kappa'} S_0(z',x') \gamma_{\rho'} S_0(x',y')\gamma_{\sigma'} S_0(y',x_{\rm src})
\nonumber\\ &&  + \textrm{four other permutations}\Big).
\nonumber
\ea
Here $S^f(x,y)$ is the propagator of quark flavour $f$, while
\be
S_0(x,y)=\int \frac{d^4p}{(2\pi)^4} \frac{\rme^{ip(x-y)}}{ip_\mu\gamma_\mu+m}
\ee
is the free muon propagator, and $G_{\mu\mu'}(x,y)$ is the free photon propagator. While
${\cal F}_\nu(x,y,z,x',y',z')$ is relatively straightforward to compute
for fixed values of its arguments, the exact evaluation of the six spacetime sums 
is computationally far too costly. Noting the identity (in Feynman gauge)
\be
\int d^4z \int d^4z' \;G_{\kappa\kappa'}(z,z') f_\kappa(z) g_{\kappa'}(z') 
= \int \frac{d^4k}{(2\pi)^4} \frac{\tilde f_\kappa(-k)\;\tilde g_\kappa(k)}{k^2}
\ee
for test functions $f(x),g(x)$ and using the shorthand notation $\int_k \equiv \int \frac{d^4k}{(2\pi)^4}$, 
we can rewrite Eq.\ (\ref{eq:M4}) as 
\ba
{\cal M}^{(4)}_\nu(x_{\rm src},x_{\rm op},x_{\rm snk}) 
&=&  (-ie)^3 (ie)^3\sum_{f=u,d,s} {\cal Q}_f^4\; \int_{k,p,\ell} \frac{1}{k^2\,p^2\,\ell^2}
\nonumber \\ &&  \bigg(\sum_{x,y,z} \rme^{i(kz+\ell y+px)}\,\Big\<{\rm Tr}\{ \gamma_\nu S^f(x_{\rm op},x)\gamma_\rho S^f(x,z) 
\gamma_\kappa S^f(z,y)\gamma_\sigma S^f(y,x_{\rm op}) \}\Big\>_{{\rm SU}(3)}\bigg)
\nonumber\\ && \times 
\bigg(\sum_{x',y',z'}\rme^{-i(kz'+\ell y'+px')}
\Big( S_0(x_{\rm snk},x') \gamma_{\rho} S_0(x',z') \gamma_{\kappa} S_0(z',y')\gamma_{\sigma} S_0(y',x_{\rm src})
\nonumber\\ && + S_0(x_{\rm snk},z') \gamma_{\kappa} S_0(z',x') \gamma_{\rho} S_0(x',y')\gamma_{\sigma} S_0(y',x_{\rm src})
+ \dots\Big)\bigg).
\la{eq:M4b}
\ea
For fixed momenta  $p,\ell$, the integrand can be computed simultaneously for all $k$ with O($V$) operations.
However, an expensive sampling of the momenta $p,\ell$ then remains.
An exact evaluation is however not required, 
since the average $\<\dots\>_U$ over the gluon degrees of freedom is performed stochastically in any case.

One way to generate ${\cal M}_\nu(x_{\rm src},x_{\rm op},x_{\rm snk})$ and reduce the 
computational burden is thus to treat two of the photon propagators stochastically.
In the method proposed in~\cite{Hayakawa:2005eq}, one considers a difference of 
correlation functions,
\ba\la{eq:2005}
&&\hspace{-0.5cm} {\cal M}^{(4)}_\nu(x_{\rm src},x_{\rm op},x_{\rm snk}) =  e^2\!\!\! \sum_{f=u,d,s} {\cal Q}_f^2 \int_k \frac{1}{k^2}
\\ &&\cdot\;
\bigg[\Big\< \sum_{z} \rme^{ikz}\,{\rm Tr}\Big\{ \gamma_\nu  S^f(x_{\rm op},z)  \gamma_\kappa  S^f(z,x_{\rm op}) \Big\}
  \sum_{z'}\rme^{-ikz'}   S(x_{\rm snk},z')  \gamma_{\kappa} S(z',x_{\rm src})  \Big\>_{{\rm SU}(3)\times {\rm U}(1)}
\nonumber
\\ && -\;\Big\< \sum_{z} \rme^{ikz}\, {\rm Tr}\{ \gamma_\nu  S^f(x_{\rm op},z)  \gamma_\kappa  S^f(z,x_{\rm op}) \}\Big\>_{{\rm SU}(3)\times {\rm U}(1)}
\cdot \sum_{z'}\rme^{-ikz'} \Big\< S(x_{\rm snk},z')  \gamma_{\kappa}
S(z',x_{\rm src})  \Big\>_{{\rm U}(1)} \nonumber\\
&&\quad +\;{\rm O}(\alpha^3)\bigg].
\nonumber
\ea
Here $S(x,y)$ is the muon propagator in a background U(1) field and
$S^f(x,y)$ the quark propagator of flavour $q$ in a background
${\rm SU}(3)\times {\rm U}(1)$ field. The subtraction implies that only the ``1-photon irreducible'' graphs are kept,
i.e.\ more than one photon must be exchanged between the muon and the QCD degrees of freedom. 
By charge conjugation invariance, this number must be odd\footnote{Including the external vertex at $x_{\rm op}$,
the number of insertions of the QCD electromagnetic current is then even.}; hence the first contribution involves
the exchange of three photons and corresponds to the desired HLbL contribution 
to the amplitude ${\cal M}_\nu(x_{\rm src},x_{\rm op},x_{\rm snk})$.
The computational cost for one configuration of 
${\rm SU}(3)\times {\rm U}(1)$ gauge fields has been reduced to O($V$).
At this point it remains to be determined how many samples of the fields are necessary to 
achieve a given accuracy on the result.
Note that inside the square bracket of Eq.\ (\ref{eq:2005}) an O$(e^2)$ contribution cancels,
leaving over an O$(e^4)$ contribution. The latter, including the explicit
$e^2$ factor outside the square bracket, represents the desired set
of diagrams of the connected HLbL contribution (O$(e^6)$).
Proof-of-principle results obtained with this method have been presented in Ref.~\cite{Blum:2014oka}
at pion masses of 330\,MeV and larger.

A different stochastic treatment of the U(1) field was proposed in~\cite{Blum:2015gfa},
which avoids the large cancellation appearing in Eq.\ (\ref{eq:2005}). It involves introducing
two stochastic U(1) fields $A_\mu(x)$ and $B_\mu(x)$, such that 
\be\la{eq:AmuAnuStoch}
\left\<A_\mu(x)\,A_{\mu'}(y)\right\>_A =  \left\<B_\mu(x)\,B_{\mu'}(y)\right\>_B =  G_{\mu\mu'}(x,y).
\ee
Using this equation to replace $G_{\rho\rho'}(x,x')$ and $G_{\sigma\sigma'}(y,y')$ in Eq.\ (\ref{eq:FnuA})
by their stochastic estimates in terms of $A$ and $B$  fields respectively, one obtains
\ba\la{eq:M42015}
 {\cal M}^{(4)}_\nu(x_{\rm src},x_{\rm op},x_{\rm snk})
&=&
e^6\sum_{f=u,d,s} {\cal Q}_f^4\;  \int_{k} \frac{1}{k^2}\; \bigg\< \sum_{x,y,z} A_\rho(x) B_\sigma(y)\,\rme^{ikz}
\nonumber \\ && \Big\<{\rm Tr}\Big\{ \gamma_\nu S^f(x_{\rm op},x)\gamma_\rho S^f(x,z) 
\gamma_\kappa S^f(z,y)\gamma_\sigma S^f(y,x_{\rm op}) \Big\}\Big\>_{{\rm SU}(3)}
\nonumber\\ && \times \sum_{x',y',z'}   A_{\rho'}(x')  B_{\sigma'}(y')\,\rme^{-ikz'}
\\ && \Big( S_0(x_{\rm snk},x') \gamma_{\rho'} S_0(x',z') \gamma_{\kappa} S_0(z',y')\gamma_{\sigma'} S_0(y',x_{\rm src})
\nonumber\\ && + S_0(x_{\rm snk},z') \gamma_{\kappa} S_0(z',x') \gamma_{\rho'} S_0(x',y')\gamma_{\sigma'} S_0(y',x_{\rm src})
+ \dots\Big) \bigg\>_{A,B}.
\nonumber
\ea
The operation just performed allows one to factorise the sums over $x,y$ from those over $x',y'$, so that 
the computational load for one instance of the $A,B$ fields becomes manageable. A concrete prescription 
for the generation of the stochastic U(1) field satisfying Eq.\ (\ref{eq:AmuAnuStoch}) is given in~\cite{Blum:2015gfa}.

\subsection{Sampling the positions of QED vertices\la{sec:stochvert}}

As the reader may have noted, the methods discussed
so far allow for the determination of the full form factors $\hat F_1(q^2)$
and $\hat F_2(q^2)$, when in fact for $(g-2)_\mu$ all that is needed is
$\hat F_2(0)$. It is then natural to ask whether the computational cost
can be reduced if one is interested only in one value of the momentum
transfer.

We therefore return to Eq.\ (\ref{eq:M4b}). 
From the definition (\ref{eq:calMdef}), an important observation is that if $x_{\rm src}$ and $x_{\rm snk}$
are projected on definite spatial momenta $\vec p$ and $\vec p'$, and  $|t_{\rm snk}-t_{\rm src}|\to\infty$,
the dependence of ${\cal M}_\nu(x_{\rm src},x_{\rm op},x_{\rm snk})$ 
on the insertion point $x_{\rm op}$ of the electromagnetic current is completely
determined by four-momentum conservation, since the correlation function is then saturated by the muon.
Following~\cite{Blum:2015gfa}, we re-use expression (\ref{eq:FnuA}) and define
\ba
{\cal F}_\nu(\vec q,x,y,z,x_{\rm op}) &=& \lim_{\substack{t_{\rm src}\to-\infty \\ t_{\rm snk}\to+\infty}}
\rme^{E_{\vec p}(t_{\rm op}-t_{\rm src})+E_{\vec p^{\,\prime}}(t_{\rm snk}-t_{\rm op})}\!\!
\sum_{x_{\rm snk},x_{\rm src}}\!\! \rme^{-i\vec q\cdot(x_{\rm src}+x_{\rm snk})/2} \;\cdot
\\ \nonumber && \qquad \qquad \qquad \qquad \qquad \cdot\;{\cal F}_\nu(x,y,z,x_{\rm op},x_{\rm snk},x_{\rm op}),
\\ 
{\cal M}_\nu(\vec q) &=& \rme^{i\vec q\cdot {\vec x}_{\rm op}} \sum_{x,y,z} {\cal F}_\nu(\vec q,x,y,z,x_{\rm op}).
\ea
The quantity ${\cal M}_\nu(\vec q)$ does not depend on $x_{\rm op}$, 
and $\rme^{i\vec q\cdot {\vec x}_{\rm op}} {\cal F}_\nu(\vec q,x,y,z,x_{\rm op})$
is invariant under a common shift to the position-space vectors ($x,y,z,x_{\rm op}$).
Based on these observations, one can choose the average of $x$ and $y$ to coincide with the origin and write
\be\la{eq:calMc}
{\cal M}_\nu(\vec q) =  \sum_{r,z,x_{\rm op}} \rme^{i\vec q\cdot\vec x_{\rm op}}\;
{\cal F}_\nu(\vec q,r/2,-r/2,z,x_{\rm op}).
\ee
Explicitly, after a few rearrangements the matrix element can be brought into the form 
\ba
&& {\cal M}_\nu(\vec q) =    e^6  \sum_{f=u,d,s} {\cal Q}_f^4\; \sum_{r}
{\cal G}_{\rho\sigma\kappa}(r,\vec p,\vec q)
\nonumber \\ &&  \sum_{z,x_{\rm op}} \rme^{i\vec q\cdot\vec x_{\rm op}+(E_{\vec p}-E_{\vec p^{\,\prime}})t_{\rm op}} \;
\Big\<{\rm Tr}\Big\{ \gamma_\nu S^f(x_{\rm op},r/2)\gamma_\rho S^f(r/2,z) 
\gamma_\kappa S^f(z,-r/2)\gamma_\sigma S^f(-r/2,x_{\rm op}) \Big\}\Big\>_{{\rm SU}(3)},
\nonumber\\ 
&&{\cal G}_{\rho\sigma\kappa}(r,\vec p,\vec q)
=  \sum_{x',y',z'} G_{\kappa\kappa'}(z,z')\; G_{\rho\rho'}(r/2,x') \; G_{\sigma\sigma'}(-r/2,y') 
\\ && \lim_{\substack{t_{\rm src}\to-\infty \\ t_{\rm snk}\to+\infty}}
\sum_{\vec x_{\rm snk},\vec x_{\rm src}} \rme^{i\vec p\cdot \vec x_{\rm src}-E_{\vec p}t_{\rm src} 
 -i\vec p^{\,\prime}\cdot \vec x_{\rm snk} + E_{\vec p^{\,\prime}}t_{\rm snk}}\;
\Big( S_0(x_{\rm snk},x') \gamma_{\rho'} S_0(x',z') \gamma_{\kappa'} S_0(z',y')\gamma_{\sigma'} S_0(y',x_{\rm src})
\nonumber\\ && 
\qquad \qquad \qquad \qquad \qquad \qquad \quad 
+ S_0(x_{\rm snk},z') \gamma_{\kappa'} S_0(z',x') \gamma_{\rho'} S_0(x',y')\gamma_{\sigma'} S_0(y',x_{\rm src})+\dots\Big).
\nonumber
\ea
We note that in the latter form, the second line can, for fixed $(r,\vec p,\vec q)$,  be evaluated for all $z$ and $x_{\rm op}$ with O($V$) operations,
by setting point sources at $r/2 $ and at $-r/2$. The object ${\cal G}_{\rho\sigma\kappa}(r,\vec p,\vec q)$,
which contains only photon and muon propagators, can also be evaluated with O($V$) operations.
This QED part can of course be simplified further, but we will not go into further details. 
The important point is that, after the limit $\vec q\to0$ is taken, the sum over the four-vector $r$ remains to be done.
The authors of~\cite{Blum:2015gfa} have developed a stochastic integration technique to perform this task,
sampling the short distances more finely that the longer distances. 

\begin{figure}[tb]
\begin{center}
\leavevmode
\includegraphics[width=0.78\textwidth]{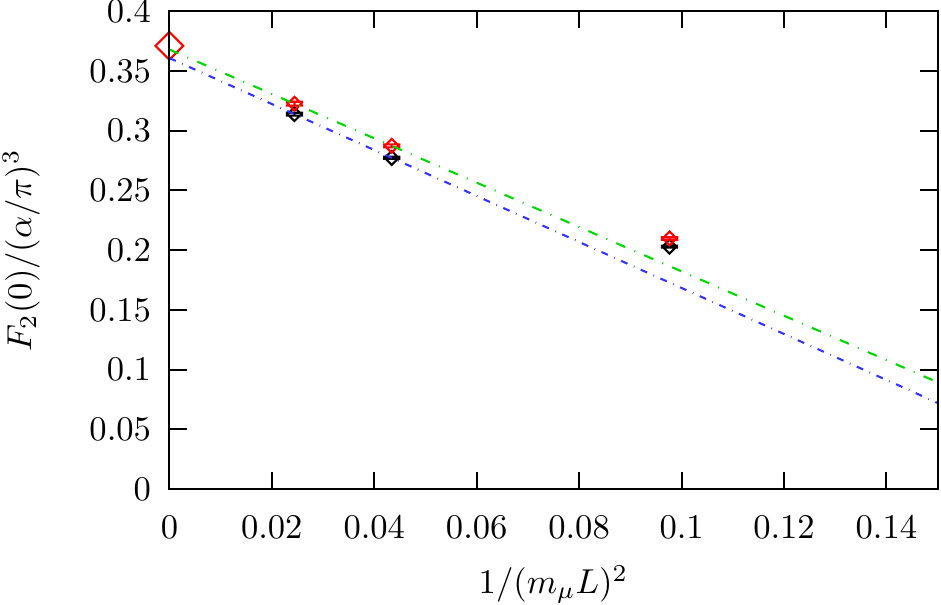}
\vspace{-0.2cm}
\caption{Study of finite-size effects for lattice calculations of $\amuhlbl$~\cite{Blum:2015gfa}:
extrapolation to infinite volume of the free-fermion-loop contribution to $a_\mu^{\rm LbL}$
computed on a torus of dimension $L\times L\times L$ using the method described in Section \ref{sec:stochvert}.
\label{fig:finivol2015gfa}}
\end{center}
\end{figure}

A test of the method for the contribution of a free quark-loop to
$\amuhlbl$ was performed in Ref.~\cite{Blum:2015gfa},
see Fig.\ \ref{fig:finivol2015gfa}.
The known result in the continuum and infinite-volume limits is reproduced with a statistical precision below one percent,
with the dominant systematic uncertainty coming from the linear extrapolation to infinite-volume in the variable $1/L^2$.
This method was found to be more efficient than the use of stochastic U(1) fields, 
for reasons that are largely understood~\cite{Blum:2015gfa}, even though within the latter class 
of methods the treatment based on Eq.\ (\ref{eq:M42015}) represented a significant improvement 
over the older method based on Eq.\ (\ref{eq:2005}).

\subsection{Semi-analytic calculation of the QED kernel\la{sec:semia}}

Position-space methods are often advantageous in lattice QCD, because
the elementary degrees of freedom are treated in position space (quark
fields $\psi(x)$ and gauge variables $U_\mu(x)$), rather than in
momentum space. Moreover, using position-space perturbation theory
for the photons and leptons in infinite volume eliminates the
power-law corrections in the volume that one incurs when treating the
U(1) gauge field on the torus. See~\cite{Lehner:2015bga} for an
alternative idea to avoid the power-law corrections. Below we follow the treatment presented
in~\cite{Green:2015mva,Asmussen:2016lse,Asmussen:2017bup,Asmussen:2018ovy}. A similar
treatment was developed in \cite{Blum:2017cer}, to which we return 
at the end of this subsection.

One may start from Eqs.\ (\ref{eq:Grhosig}) and (\ref{eq:amuhlbl}),
which hold in the infinite-volume, continuum theory. Interchanging
the order of the integrations and expressing the $q_1$ and $q_2$
integrals in terms of position-space propagators, one arrives
at~\cite{Asmussen:2016lse} 
\ba\la{eq:Grs2}
\Gamma_{\rho\sigma}(p,p) &=& -e^6\int_{ x_1, x_2} 
K_{\mu\nu\lambda}( x_1, x_2,p) \;\widehat \Pi_{\rho;\mu\nu\lambda\sigma}( x_1, x_2),
\\
\widehat \Pi_{\rho;\mu\nu\lambda\sigma}( x_1, x_2) &=& 
\int_{x_3} (+ix_3)_\rho\, \Big\<\,J_\mu(x_1)\,J_\nu(x_2)\,J_\sigma(x_3)\, J_\lambda(0)\Big\>,
\ea
with the shorthand notation $\int_x \equiv \int d^4x$ for position-space integrals and 
\ba
\la{eq:Kp}
K_{\mu\nu\lambda}(x_1,x_2,p) &=& 
\gamma_\mu (i p\!\!\!/+ \partial\!\!\!/^{(x_1)} - m) \gamma_\nu (i p\!\!\!/+ \partial\!\!\!/^{(x_1)} + \partial\!\!\!/^{(x_2)}- m) 
 \gamma_\lambda \; {\cal I}(\hat\epsilon, x_1,x_2),   \phantom{\frac{1}{1}}\\ 
{\cal I}(\hat\epsilon,x,y) &=& \int_{q,k} \frac{1}{q^2k^2(q+k)^2} \frac{1}{(p-q)^2+m^2}\frac{1}{(p-q-k)^2+m^2}
\rme^{-i(qx+ky)}.
\la{eq:Ixyp}
\ea
We remind the reader that the unit vector $\hat\epsilon$ parameterises the  momentum of the muon,
$p = i\,m\,\hat\epsilon$.
An important point is that the scalar function ${\cal I}$ requires
infrared regularisation, which however can be dropped after the
derivatives are applied to it to compute $K_{\mu\nu\lambda}(x,y)$.

Combining Eqs.\ (\ref{eq:Grs2}) and (\ref{eq:amuhlbl}),
one arrives at an expression of the form~\cite{Green:2015mva}
\be\la{eq:F2hat_noaver}
\hat F_2(0) = \frac{me^6}{3}\int_{x,y}  {\cal L}_{[\rho,\sigma];\mu\nu\lambda}(\hat\epsilon,x,y) \;
  i\widehat\Pi_{\rho;\mu\nu\lambda\sigma}(x,y).
\ee
Exploiting the fact that $\hat F_2(0)$ is a Lorentz scalar, we may average the right-hand side over
the direction of the muon's momentum, so that~\cite{Green:2015mva} 
\be\la{eq:F2hat_aver}
\hat F_2(0) = \frac{m e^6}{3}\int_{x,y}
\bar {\cal L}_{[\rho,\sigma];\mu\nu\lambda}(x,y)\; i\widehat\Pi_{\rho;\mu\nu\lambda\sigma}(x,y),
\quad \bar {\cal L}_{[\rho,\sigma];\mu\nu\lambda}(x,y) 
= \Big\< {\cal L}_{[\rho,\sigma];\mu\nu\lambda}(\hat\epsilon,x,y)\Big\>_{\hat\epsilon}.
\quad\quad\quad
\ee
The calculation of the kernel proceeds as follows. 
The kernel is written as 
\ba
\bar {\cal L}_{[\rho,\sigma];\mu\nu\lambda}(x,y) 
&=& \sum_{A={\rm I,II,III}} 
{\cal G}^A_{\delta[\rho\sigma]\mu\alpha\nu\beta\lambda} 
T^{(A)}_{\alpha\beta\delta}(x,y).
\ea
The ${\cal G}^A_{\delta[\rho\sigma]\mu\alpha\nu\beta\lambda}$
are sums of products of Kronecker deltas resulting from traces of Dirac matrices.
The rank-three tensor $T^{(A)}_{\alpha\beta\delta}(x,y)$ can be written in terms 
of a scalar, a vector and a tensor component of ${\cal I}(\hat\epsilon,x,y)$ viewed as a function
of the unit vector $\hat\epsilon$,
\ba
T^{({\rm I})}_{\alpha\beta\delta}(x,y) &=&   \partial^{(x)}_\alpha (\partial^{(x)}_\beta + \partial^{(y)}_\beta) 
V_\delta(x,y),
\\
T^{({\rm II})}_{\alpha\beta\delta}(x,y) &=& 
m\, \partial^{(x)}_\alpha 
\Big( T_{\beta\delta}(x,y) + \frac{1}{4}\delta_{\beta\delta} S(x,y)\Big)
\la{eq:TII}\\
\la{eq:TIII}
T^{({\rm III})}_{\alpha\beta\delta}(x,y) &=&  m\, (\partial^{(x)}_\beta + \partial^{(y)}_\beta)
\Big( T_{\alpha\delta}(x,y) + \frac{1}{4}\delta_{\alpha\delta} S(x,y)\Big),
\ea
with
\ba
S(x,y) = \Big\<   {\cal I}\Big\>_{\hat\epsilon}, 
\qquad
V_\delta(x,y) = \Big\<\hat\epsilon_\delta   {\cal I} \Big\>_{\hat\epsilon},
\qquad
T_{\beta\delta}(x,y) = 
\Big\< \Big(\hat\epsilon_\delta\hat\epsilon_\beta-{\textstyle\frac{1}{4}}\delta_{\beta\delta}\Big) \; {\cal I}\Big\>_{\hat\epsilon}.
\ea
Only the scalar contribution $S(x,y)$ contains the infrared divergence of ${\cal I}(\hat\epsilon,x,y)$; the divergence
cancels after the derivatives in (\ref{eq:TII}) and (\ref{eq:TIII}) are applied. In the intermediate steps 
of the calculation, it can for instance be regulated by introducing a photon mass.
The vector and tensor functions are parameterised by, respectively, two and three weight functions,
\ba
S(x,y) &\!\!=\!\!& \bar g^{(0)}(|x|, x\cdot  y, |y|),  \phantom{\frac{1}{1}}
\\
V_\delta(x,y)
&\!\!=\!\!& x_\delta  \bar{\mathfrak{g}}^{(1)}(|x|,x\cdot y,|y|)
+ y_\delta  \bar{\mathfrak{g}}^{(2)}(|x|,x\cdot y,|y|),
\\
 T_{\alpha\beta}(x,y) 
 &\!\!=\!\!& \left(x_\alpha x_\beta - {\textstyle\frac{1}{4}}x^2\,\delta_{\alpha\beta}\right)\; \bar{\mathfrak{l}}^{(1)}
+ \left(y_\alpha y_\beta - {\textstyle\frac{1}{4}}y^2\,\delta_{\alpha\beta}\right)\; \bar{\mathfrak{l}}^{(2)}
+ \left(x_\alpha y_\beta + y_\alpha x_\beta  - {\textstyle\frac{1}{2}}x\cdot y\,\delta_{\alpha\beta}\right)\; \bar{\mathfrak{l}}^{(3)}.\qquad 
\ea
In total, the QED kernel $\bar{\cal L}_{[\rho,\sigma];\mu\nu\lambda}(x,y) $ is thus parameterised by six weight functions
$\bar g^{(0)}$, $\bar{\mathfrak{g}}^{(1)}$, $\bar{\mathfrak{g}}^{(2)}$, $\bar{\mathfrak{l}}^{(1)}$, $\bar{\mathfrak{l}}^{(2)}$
and $\bar{\mathfrak{l}}^{(3)}$,  which are functions of $(x^2,y^2,x\cdot y)$.
The averaging over $\hat\epsilon$ is performed analytically~\cite{Asmussen:2016lse} using the Gegenbauer polynomial technique,
which is the four-dimensional analogue of applying Legendre polynomials to three-dimensional problems; see for instance~\cite{Knecht:2001qf}.
In the final step, the calculation of the form factors involves a two-dimensional numerical integration
of a function defined by an infinite series.
Through the use of Lorentz covariance, 
the calculation of the kernel thus involves a manageable amount of computation and storage.

In the course of an actual lattice calculation, the six form factors
can be read in and combined into ${\cal
  L}_{[\rho,\sigma];\mu\nu\lambda}(x,y)$ ``on the fly'' for fixed
$y$~\cite{Asmussen:2016lse}. In analytic calculations, once all
indices of ${\cal L}_{[\rho,\sigma];\mu\nu\lambda}(x,y)$ are
contracted with $\widehat\Pi_{\rho;\mu\nu\lambda\sigma}(x,y)$, the
eight-dimensional integral over $(x,y)$ reduces to a three-dimensional
integral over $(x^2,y^2,\hat x\cdot\hat y)$. However, in practical
lattice calculations, the sum over $x$ can be carried out explicitly
using the sequential-propagator technique with O($V$) operations and
no significant increase in computational cost. After the $x$ integral
is carried out, by Lorentz invariance of $\amuhlbl$, the $y$ integral
collapses to a one-dimensional integral, $\int d^4y \to 2\pi^2
\int_0^\infty d|y|\;|y|^3$. For that reason, one may expect that the
number of $y$ points at which the integrand needs to be evaluated is
manageable~\cite{Green:2015mva}, perhaps of order twenty.

\begin{figure}[tb]
\begin{center}
\leavevmode
\includegraphics[width=0.49\textwidth]{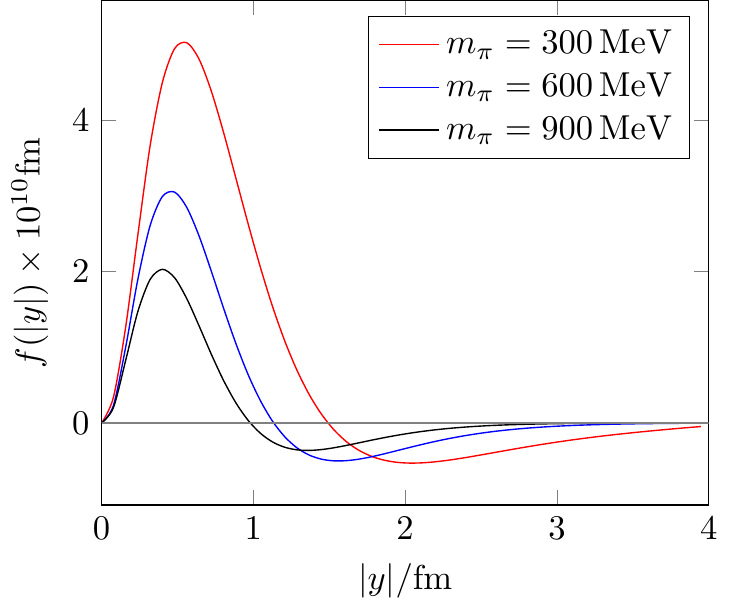}
\includegraphics[width=0.49\textwidth]{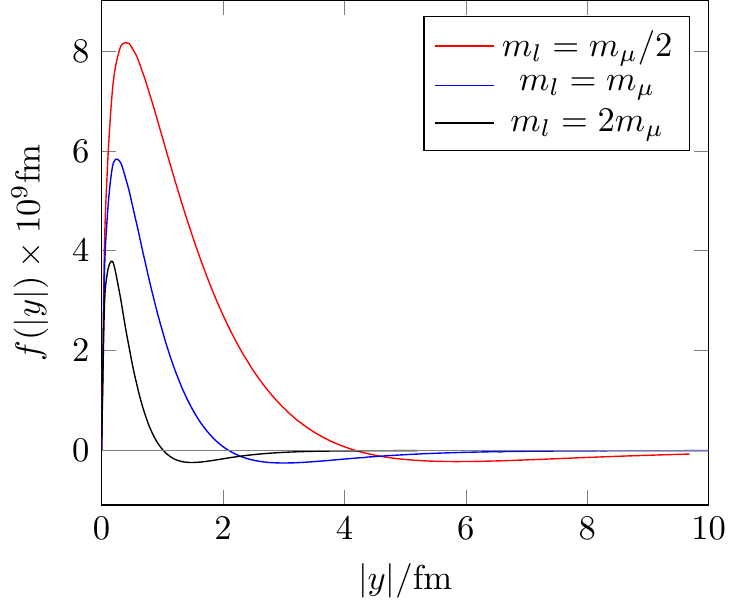}
\vspace{-0.2cm}
\caption{The integrand to obtain the $\pi^0$ pole (left) and the lepton loop (right) contribution to $\amuhlbl$ in infinite-volume position space
based on Eq.\ (\ref{eq:F2hat_aver}). The vector-dominance model was used for the pion transition form factor.
The known results are reproduced at the percent level. Figures from~\cite{Asmussen:2018ovy}.
\label{fig:IntegrandPi0VMD}}
\end{center}
\end{figure}

To demonstrate the validity of the position-space approach described
above, the $\pi^0$  contribution to $\amuhlbl$ was
computed~\cite{Asmussen:2016lse,Asmussen:2017bup} 
 in infinite volume and shown to reproduce the result obtained using momentum-space
methods; see  Fig.\ \ref{fig:IntegrandPi0VMD}.
Similarly, the analytically known contribution of a free quark was reproduced
\cite{Asmussen:2017bup} at the one-percent level. In the latter case, a closed analytic
expression was obtained for $\widehat\Pi_{\rho;\mu\nu\lambda\sigma}(x,y)$.

As noted at the beginning of this subsection, a closely related
approach has been implemented and tested on the lattice using a free
quark loop~\cite{Blum:2017cer}. In the latter publication, however,
the muon rest frame was chosen and the kernel was not parameterised by
Lorentz-invariant scalar functions. The idea to perform either $x$ or
$y$-independent subtractions on the QED kernel in an expression of the
type (\ref{eq:F2hat_noaver}) was introduced and tested. In infinite
volume, such subtractions do not modify the final result for
$\amuhlbl$, because current conservation implies for instance that
$\int d^4x \;\widehat\Pi_{\rho;\mu\nu\lambda\sigma}(x,y)$ vanishes.
Subtraction terms were used to define a new kernel that vanishes at
the contact points $x=0$ and $y=0$. With the latter kernel, the
continuum limit of the free quark loop contribution was found to be
under better control in the tests performed
in~\cite{Blum:2017cer}. Subtractions of this type can be implemented
straightforwardly in the Lorentz-covariant formulation
\eq{eq:F2hat_aver} as well.

\subsection{Lattice QCD results on $\amuhlbl$ \label{sec:HLbL_latresu}}

Lattice QCD results on hadronic light-by-light scattering in
$(g-2)_\mu$ are still scarce. Only one group has published results on
the direct calculation of $\amuhlbl$ on the
lattice~\cite{Blum:2014oka,Blum:2015gfa,Blum:2016lnc}. The first two
publications concern only the fully connected contribution. In the
third publication, first results of the same group concerning the diagrams of
topology (2,2) were obtained. An update was presented at the
Lattice 2016 conference~\cite{Jin:2016rmu}.


We summarise the main result obtained in Ref.~\cite{Blum:2015gfa},
beginning with the fully connected contribution to $\amuhlbl$. It was
obtained using a method based on \eq{eq:calMc}, where the
positions of two quark-photon vertices are summed exactly at short
vertex separations and sampled stochastically at larger separations.

The calculation presented in~\cite{Blum:2015gfa} is based on an $N_f=2+1$ domain-wall-fermion (DWF) lattice ensemble.
It uses a M\"obius variant of the domain-wall fermion operator for the valence quarks, matched to the 
domain-wall action used in the generation of the ensemble. The muon propagator is also computed using 
the DWF action. The lattice size is $32^3\times 64$, the lattice spacing $a=0.144\,$fm and the pion mass $171\,$MeV.
The number of gauge configurations used is 23.
This yields the estimate
\be
({\amuhlbl})_{\rm con}\equiv ({\amuhlbl})^{(4)} = (132.1\pm 6.8)\cdot 10^{-11},
\ee
where the error indicated is purely statistical.

An update was presented at the Lattice 2016
conference~\cite{Jin:2016rmu} and published in\,\cite{Blum:2016lnc}.
On a $48^3\times96$ lattice with a lattice spacing of 0.114\,fm and a
pion mass of 139\,MeV, the result is
\be\la{eq:48Ic}
({\amuhlbl})^{(4)} = (116.0\pm9.6)\cdot 10^{-11}.
\ee
An evaluation of the (2,2) disconnected diagrams (see Fig. \ref{fig:Wtopo}) on the
same lattice ensemble as the connected contribution (\ref{eq:48Ic})
yielded the negative contribution
\be\la{eq:48Ic_disc}
({\amuhlbl})^{(2,2)} = (-62.5\pm 8.0)\cdot 10^{-11}.
\ee
The sign of this contribution was expected on the basis of 
large-$N_c$ arguments reviewed in section \ref{sec:flavoursym},
$N_c$ being the number of colours. In that section,
we will also comment on the magnitude of the results (\ref{eq:48Ic}) and
(\ref{eq:48Ic_disc}). The sum of the
two contributions,
\be\la{eq:48Iall}
  ({\amuhlbl})^{(4)+(2,2)}=(53.5\pm13.5)\cdot 10^{-11},
\ee
is substantially smaller than the ``Glasgow consensus''. However, as
pointed out in \cite{Blum:2016lnc}, finite-volume and discretisation
effects may be large and must be studied in more detail before
\eq{eq:48Iall} can be regarded as a result with fully controlled
systematic errors.

\subsection{The HLbL forward scattering amplitude}

Besides the direct calculation of $\amuhlbl$, two studies pertinent to
this quantity have been carried out on the lattice. The first,
reviewed in this subsection, concerns the calculation of the HLbL
scattering amplitude \emph{per se}~\cite{Green:2015sra}. The second,
reviewed in the next subsection, is the calculation of the pion
transition form factor ${\cal
  F}_{\pi_0\gamma^*\gamma^*}(Q_1^2,Q_2^2)$~\cite{Gerardin:2016cqj}. As
a motivation, in phenomenological and/or dispersive approaches to
$\amuhlbl$, the QCD amplitude has been approximated by the exchange of
mesonic resonances~\cite{Jegerlehner:2009ry}. How well the HLbL
scattering amplitude itself can be described by such an approximation
can be tested using lattice calculations. Furthermore, some of the
experimentally least well constrained parameters can be estimated in
this way. Secondly, the neutral pion pole contribution is thought to
be the single largest contribution to $\amuhlbl$ (see
\fig{fig:PShlbl}). It is determined by the transition form factor (see
e.g.\ \cite{Knecht:2001qf}) and no experimental data exists as of
today for the doubly virtual case~\cite{Nyffeler:2016gnb}.

In~\cite{Green:2015sra}, the HLbL scattering amplitude was computed in
$N_f=2$ lattice QCD. More precisely, the fully connected contribution
was computed using the sequential-propagator methods described in
Section \ref{sec:seqprop}. The amplitude is a function of three
momenta $(q_1,q_2,q_3)$, the fourth one being fixed by momentum
conservation. Using sequential and ``double-sequential'' propagators,
it is possible to obtain the amplitude for all values of $q_3$ at
fixed $(q_1,q_2)$ with O($V$) operations.

The emphasis was placed on the forward scattering amplitudes: the
advantage of studying the latter is that these amplitudes can be
related to the $\gamma^*\gamma^*\to{\rm hadrons}$ cross section via
dispersive sum rules~\cite{Pascalutsa:2012pr}. There are eight such
invariant amplitudes, which are functions of three kinematic
variables: the photon virtualities $q_1^2$ and $q_2^2$ and $\nu =
q_1\cdot q_2$. The dispersive sum rule for one of the amplitudes reads
\be\la{eq:dr} {\cal M}_{\rm TT}(q_1^2,q_2^2,\nu) - {\cal M}_{\rm
  TT}(q_1^2,q_2^2,0) = \frac{2\nu^2}{\pi} \int_{\nu_0}^\infty
d\nu'\frac{ \sqrt{ \nu'{}^2 - q_1^2 q_2^2
}}{\nu'(\nu'{}^2-\nu^2-i\epsilon)}(\sigma_0+\sigma_2)(\nu'),
\ee
where $\sigma_0$ and $\sigma_2$ are the total cross sections
$\gamma^*(q_1^{\,2})\gamma^*(q_2^{\,2})\to {\rm hadrons}$ with total
helicity~0 and~2, respectively. It can be
shown~\cite{Pascalutsa:2012pr} that ${\cal M}_{\rm TT}$ vanishes at
$\nu=0$ if either of the photons is real.

\begin{figure}[tb]
\begin{center}
\centerline{\includegraphics[width=0.6\textwidth]{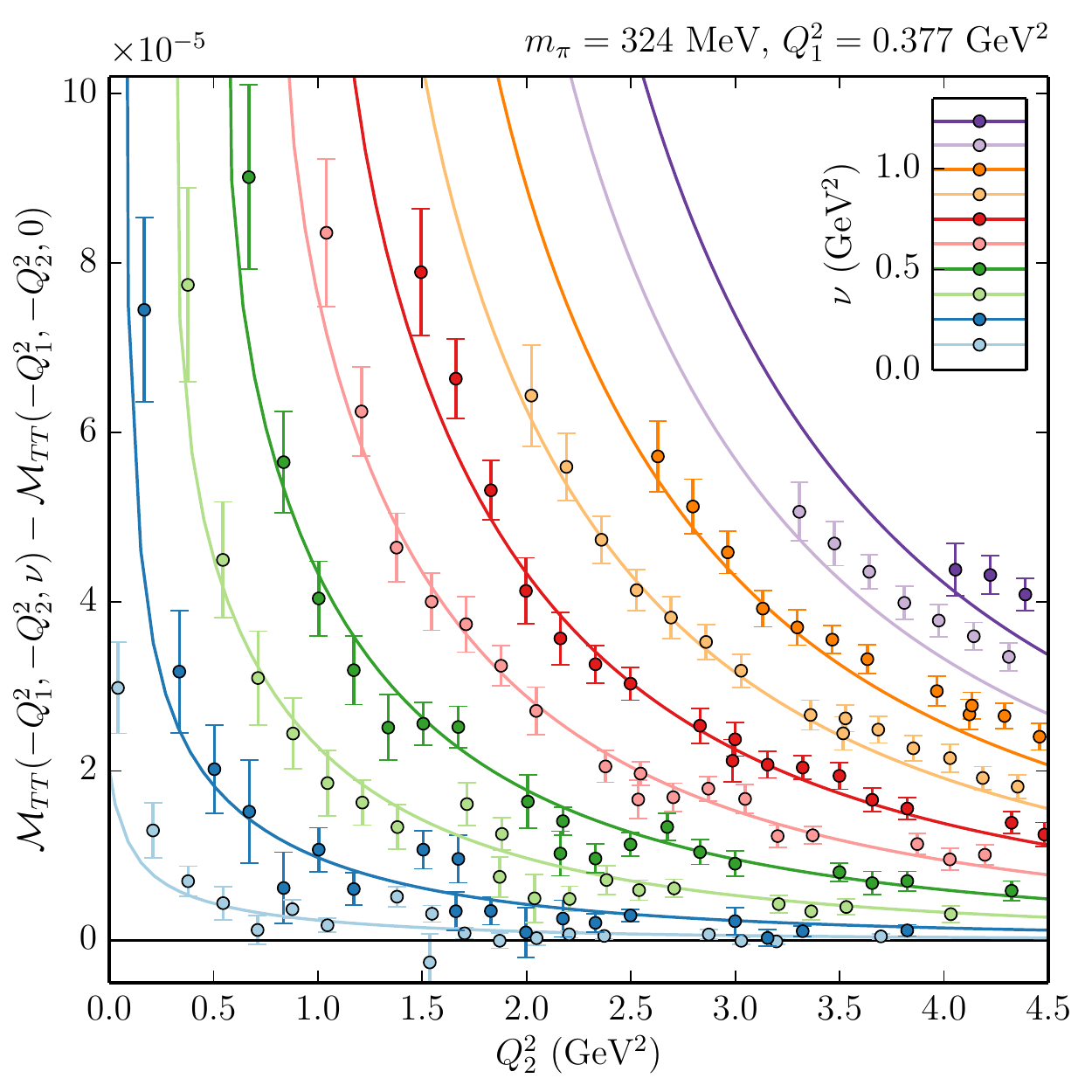}} 
\caption{The forward HLbL scattering amplitude ${\cal M}_{\rm TT}$ computed in $N_f=2$ lattice QCD 
and compared to a model for the $\gamma^*\gamma^*\to{\rm hadrons}$
cross section via the dispersive sum rule (\ref{eq:dr}).
Figure from~\cite{Green:2015sra}.
\label{fig:HLbL3amplit}}
\end{center}
\end{figure}

Figure \ref{fig:HLbL3amplit} shows the amplitude ${\cal M}_{\rm TT}$
for a fixed photon virtuality $Q_1^2=-q_1^2=0.377{\rm GeV}^2$, as a
function of the second virtuality, for various values of the variable
$\nu$. A model for the cross section, known to provide a good
description of experimental data for real-photon scattering,
$\gamma\gamma\to{\rm hadrons}$ and generalised to spacelike photons,
is displayed as well and found to describe the data quite well.

This study has recently been extended to include all eight forward
amplitudes~\cite{Gerardin:2017ryf}. By fitting a resonance-exchange
model for the $\gamma^*(q_1^{\,2})\gamma^*(q_2^{\,2})\to {\rm
  hadrons}$ cross sections simultaneously to the eight amplitudes, the
virtuality dependence of the transition form factors of the scalar,
axial-vector and tensor mesons could be constrained. The simultaneous
analysis of the amplitudes is beneficial, since the resonances
contribute with different weight factors and even different signs to
the various amplitudes, thus enabling a much more constrained
analysis. The successful simultaneous description of the eight amplitudes by the same type 
of model used in estimating $\amuhlbl$ makes it less likely that 
the model estimate is grossly wrong.


If a resonance-exchange model, plus the pion-loop contribution, is
found to describe the forward amplitude, it is relatively
straightforward to extend it to general kinematics. Therefore, a
possible strategy is to take the hadronic model with its parameters
fitted to the forward amplitudes determined by the lattice
calculation, and to compute $\amuhlbl$; here we have especially the
parameters describing the transition form factors in mind. While
still a model-dependent calculation, this procedure would be
constrained by ab initio information from the lattice.

\subsection{The pion transition form factor and $(\ahlbl)^{\pi^0}$}

The transition form factor (TFF) of the neutral pion was computed in
$N_f=2$ lattice QCD~\cite{Gerardin:2016cqj} in the kinematic range
relevant to $\amuhlbl$, with $0< Q_{1,2}^2<1.5\,{\rm GeV}^2$. A
chiral and continuum extrapolation to the physical point was
performed. Three models, in increasing order of complexity, were used
in an attempt to describe the data. While the simplest
``vector-dominance model'' (VMD) fails to describe the data, fits with
the ``lowest meson dominance'' (LMD) model
\cite{Moussallam:1994xp,Knecht:1999gb} and the more refined LMD+V
model \cite{Knecht:2001xc} were found to work.
The LMD+V model accommodates the correct leading asymptotic behavior
of the form factor both in the single-virtual case and in the doubly
virtual case, ${\cal F}_{\pi^0\gamma^*\gamma^*}(Q^2,Q^2)$ and was
therefore chosen to obtain the final result quoted below.

Given the TFF, $(\ahlbl)^{\pi^0}$ can be obtained via a three-dimensional 
integral over the variables $|Q_1|$, $|Q_2|$ and $(Q_1\cdot Q_2)/(|Q_1||Q_2|)$ of two factors of the 
TFF at different arguments and a  kernel known analytically~\cite{Jegerlehner:2009ry}.
Inserting the LMD+V parameterisation of the lattice result for the
transition form factor into the formula, the authors obtain~\cite{Gerardin:2016cqj}
\be
(\ahlbl)^{\pi^0} = (65\pm 8.3)\cdot 10^{-11},
\ee
in good agreement with previous estimates, which are reviewed
in~\cite{Nyffeler:2016gnb}. The novel element of the calculation is
the availability of direct information on the doubly virtual
transition form factor.

Improved calculations in $N_f=2+1$ QCD with increased statistics are
underway. The increased precision should allow for a parameterisation
of the form factor through a systematically improvable family of
functional forms, enabling an informed estimate of the systematic
error. A dispersive determination of the pion transition form factor
has recently appeared~\cite{Hoferichter:2018dmo}, thus allowing for
valuable cross-checks among the different frameworks.

\subsection{Sources of systematic errors in lattice calculations of $\amuhlbl$}

\subsubsection{Finite-volume effects}

When computing $\amuhlbl$ using lattice techniques, precisely which
formulation is used has a big impact on the systematic effects of the
calculation. One important source of systematic error, as in the case
of the hadronic vacuum polarisation, are finite-volume effects. Since
the photon is massless, an important question is how QED is treated.
If it is treated in finite-volume, as
in~\cite{Hayakawa:2005eq,Blum:2015gfa}, power-law corrections on the
result are bound to occur. A quantitative study of these effects was
carried out in~\cite{Blum:2015gfa}, considering the free-fermion loop
contribution to light-by-light scattering in $(g-2)_\mu$.

In the strategy laid out in~\cite{Green:2015mva,Asmussen:2016lse},
position-space perturbation theory is used to derive an expression for
$\ahlbl\equiv \hat{F}_2(0)$ with QED treated in infinite volume, in
order to avoid power-law corrections. Only the QCD four-point function
(see Eq.\ \ref{eq:PihatDef}) is evaluated in finite volume. Of course
this does not necessarily mean that the finite-size effects are then
numerically small for the typical volumes used in lattice QCD. A first
hint that the finite-size effects might indeed be significant is
provided by the calculation of the $\pi^0$ pole contribution using the
position-space method shown in Fig.\ \ref{fig:IntegrandPi0VMD}. For
instance, even with a $\pi^0$ mass of 600\,MeV, an integration up to
at least 2\,fm is necessary to control the result to within
$10\%$. On a torus, the upper bound of integration in the
variable $|y|$ is $L/2$ in the least favourable direction, which
suggests that $L$ would have to be at least 4\,fm, in spite of the
large pion mass.

On the other hand, since the pion pole contribution is computable both
in finite and in infinite volume, the finite-size effect due to this
contribution can be corrected for, provided the pion transition form
factor is known. This possibility represents a further motivation for
computing the pion transition form factor.

\subsubsection{Flavour symmetry and disconnected diagrams in $\amuhlbl$ \label{sec:flavoursym}}

The calculation of all Wick-contraction topologies is demanding.  In
many instances, disconnected diagram contributions have been found to
make numerically small contributions to hadronic matrix elements.
Quark loops generated by a single vector current have been empirically
found to be particularly suppressed; on the other hand, it is well
known that the disconnected diagram is responsible for the difference
between the pion and the $\eta'$ correlator, and is therefore crucial
at long distances.

The importance of the disconnected diagrams in the HLbL amplitude has been pointed out
in~\cite{Bijnens:2015jqa,Bijnens:2016hgx}, showing that the pion, $\eta$  and
$\eta'$ pole contributions would have the wrong weight factors if only
the connected diagrams were included. This result was re-derived in~\cite{Gerardin:2017ryf},
where it was also shown that if the HLbL amplitude is dominated by the pole-exchange
of an iso-vector resonance, isospin symmetry induces relations between
different Wick-diagram topologies.

The arguments, based on the large-$N_c$ motivated idea that an isolated
vector current insertion in a fermion loop gives a suppressed
contribution, lead to the conclusion that the (2,2) disconnected class
of diagrams in \fig{fig:Wtopo} contains all of the contributions from
flavour-singlet meson poles, while the mesons in the adjoint
representation of the flavour symmetry group contribute with a
negative weight factor; the latter is $(-25/9)$ in the SU(2)$_{\rm
  flavour}$ case and $(-2)$ in the SU(3)$_{\rm flavour}$ case.  The
generic large-$N_c$ expectations would further lead to the stronger
conclusion that, in each $J^{PC}$ sector, the non-singlet resonances
cancel the contribution of the flavour-singlet resonances. One
channel, however, where the degeneracy is badly broken is the
pseudoscalar sector, since the pion is much lighter than the $\eta'$
meson. In particular, at low energies in the two-flavour theory one
expects the (2,2) disconnected class of diagrams for a generic HLbL
amplitude ${\cal A}$ to be given to a good approximation by
\be\la{eq:pi0etap} {\cal A}^{(2+2)} \approx -\frac{25}{9} {\cal
  A}^{(\pi^0)} + {\cal A}^{(\eta')}\;.
\ee
In the case of the forward scattering amplitudes, this prediction was
found to be in agreement with the lattice
data~\cite{Gerardin:2016cqj}, albeit with a large uncertainty.
Similarly, one predicts~\cite{Gerardin:2016cqj}
\be\la{eq:amu2p2} (\ahlbl)^{(2,2)} \approx
  \left\{ 
  \begin{array}{l@{~~~~}l}
   -{\displaystyle\frac{25}{9}} (\ahlbl)^{\pi^0} + (\ahlbl)^{\eta'} =
   -(162\pm27)\cdot 10^{-11} & m_s=\infty, \phantom{\Big|} \\[1.0ex]
   -2\left( (\ahlbl)^{\pi^0} + (\ahlbl)^{\eta} \right) +
    (\ahlbl)^{\eta'} = -(142\pm19)\cdot 10^{-11} & 
  m_s=m_{ud}. \phantom{\Big|}
\end{array} \right.,
\ee
where one might expect the value in the real world to lie in between
these two cases.  Taking in addition the result
$\ahlbl=(102\pm39)\cdot 10^{-11}$ from a model
calculation~\cite{Jegerlehner:2015stw} leads to the following estimate
for the fully connected class of diagrams,
\be\la{eq:amu4}
(\ahlbl)^{(4)}_{\rm model}
 \approx \left\{  \begin{array}{l@{~~~~}l}
(264\pm51) \cdot 10^{-11} & m_s=\infty, \phantom{\Big|}
\\
(244\pm46) \cdot 10^{-11} & m_s=m_{ud}. \phantom{\Big|}
\end{array}
\right.
\ee
%
Since the lattice results for $({\amuhlbl})^{(4)}$ and
$({\amuhlbl})^{(2,2)}$ reviewed in Section \ref{sec:HLbL_latresu} (see
Eqs. (\ref{eq:48Ic}) and (\ref{eq:48Ic_disc})) are significantly
smaller in magnitude than these estimates, the authors
of~\cite{Gerardin:2016cqj} conclude that either these lattice results
are severely underestimated, which could be due to discretisation and
finite-volume effects, or the hadronic model based on resonance
exchanges is not viable. Alternatively, the large-$N_c$ inspired
approximations that are used to estimate (\ref{eq:amu2p2}) and
(\ref{eq:amu4}) are inadequate. The deviation might also be a
combination of the above. New and improved lattice calculations will
probably settle the question soon.

\section{Concluding remarks\la{sec:concl}}

The realisation that the Standard Model does not provide a complete
description of nature has sparked a worldwide search for new particles
and forces that are collectively referred to as ``New Physics'' or
``Physics beyond the Standard Model'' (BSM). Precision observables
offer great potential for BSM physics searches, given that new
particles have not been discovered at the LHC in the expected region
so far. The observed discrepancy of $3.5$ standard deviations between
the theoretical and experimental determinations of the muon anomalous
magnetic moment constitutes the most intriguing hint for a deviation
from the Standard Model, provided that the current estimates and
associated uncertainties of hadronic contributions can be
trusted. While lattice QCD provides the appropriate framework for the
calculation of the hadronic vacuum polarisation and light-by-light
scattering contributions from first principles, significant technical
challenges must be overcome before lattice results can have a decisive
impact on resolving -- or confirming -- the muon anomaly.

In this review we have outlined the enormous progress that has been
achieved in computing both $\ahvp$ and $\ahlbl$ on the lattice. The
precision goals for these two quantities are quite different: while a
reliable determination of $\ahlbl$ with a total uncertainty of
$10-15$\% would already have a major impact, lattice calculations of
$\ahvp$ must reach sub-percent precision, in order to be competitive
with the dispersive approach. This requires reliable determinations of
finite-volume effects and isospin-breaking corrections, as well as
computing the contributions from disconnected diagrams and from the
long-distance regime of the vector correlator. The compilation of
results for $\ahvp$ presented in Section~\ref{sec:results} shows that
the errors of current lattice estimates must be further reduced by a
factor $\sim5$. As calculations of $\ahvp$ turn into a flagship
project for lattice QCD, with many collaborations attempting to reduce
the overall error to a competitive level, there are good prospects
that this is achievable within the next few years.

The case of hadronic light-by-light scattering is a lot more
complicated, and we have discussed several complementary strategies
that are being pursued to determine $\ahlbl$ directly or reduce the
model dependence considerably. First results from a direct calculation
of $\ahlbl$ have been published. However, unlike the case of $\ahvp$,
a complete error budget is not yet available. This task is made more
complicated not only because finite-volume effects could be more
severe for $\ahlbl$, but also since there is a much larger class of
disconnected diagrams to be considered. Lattice QCD also provides
crucial input for semi-phenomenological approaches, by either
replacing experimental input or testing the reliability of hadronic
models. For instance, lattice calculations of the transition form
factor for $\pi^0\to\gamma^\ast\gamma^\ast$ allow for a reliable
determination of the expected dominant contribution to $\ahlbl$ from
the pion pole. While the results agree with hadronic models, they do
not suffer from the arbitrary model-dependent error estimates. Another
example is the lattice calculation of a class of forward
light-by-light scattering amplitudes, which can be related to
phenomenological models via dispersion relations. The combined
information from a number of complementary approaches should allow for
a much more reliable and largely model-independent determination of
$\ahlbl$ in the near future.

In order to profit from the new generation of experiments (E989 at
Fermilab and E34 at J-PARC) designed to measure $a_\mu$ with much
enhanced precision, it is clear that theoretical uncertainties must be
substantially reduced in the long run. The stated aim of the recently
formed ``$g-2$ Theory Initiative''\footnote{See {\tt
    https://indico.fnal.gov/event/13795/} and {\tt
    https://wwwth.kph.uni-mainz.de/g-2/}} is to provide the best
theoretical predictions for the hadronic contributions to the muon
$g-2$, with lattice QCD being a cornerstone in this endeavour.

\subsection*{Acknowledgements}

It is a pleasure to thank the members of the Mainz ($g-2$) project for
the fruitful collaboration. In particular, we thank our collaborators
A.~G\'erardin, G.~von Hippel, A.~Nyffeler, V.~Pascalutsa and
H.~Spiesberger for many stimulating discussions over the past few
years. We are grateful to our colleagues within the ``$g-2$ Theory
Initiative'' for interesting discussions and insights. This work was
partially supported by DFG via the Collaborative Research Centre ``The
low-energy frontier of the Standard Model'' (SFB 1044), the
Rhineland-Palatinate Research Initiative, and by the European Research
Council (ERC) under the European Union's Horizon 2020 research and
innovation programme through grant agreement No.\ 771971-SIMDAMA.

\newpage

\begin{appendix} 
\section{Basic concepts of lattice QCD \la{app:lattice}}

In this appendix we give a brief and self-contained introduction to
lattice QCD. We refrain from providing a general and detailed
treatment which can be found in many textbooks
\cite{Creutz:1984mg,Rothe:1992nt,Montvay:1994cy,Smit:2002ug,
  DeGrand:2006zz,Gattringer:2010zz} and review articles
\cite{Wittig:2008zz}. Instead we shall focus on those aspects of the
lattice method that are most relevant for determinations of the
hadronic contributions to the muon $g-2$.

\subsection{Euclidean path integral and expectation values}

Lattice QCD is a rigorous, non-perturbative treatment of the strong
interaction that starts from the expression of physical observables in
terms of the Euclidean path integral, which is given by
\be
   Z=\int D[U]D[\psibar,\psi]\,
   {\rm e}^{-S_{\rm G}[U]-S_{\rm F}[U,\psibar,\psi]},
\ee
where $S_{\rm G}$ and $S_{\rm F}$ denote the Euclidean gluon and quark
action, respectively, and the integration is performed over all gauge
and fermionic fields. After introducing a Euclidean lattice
$\Lambda_{\rm E}$ as the finite set of space-time points $x_\mu=n_\mu
a$ that are integer multiples of the lattice spacing~$a$ and
considering a finite space-time volume of size $L^3{\cdot}T$, the path
integral is mathematically well defined and finite for suitable gauge
invariant discretisations of the gluon and quark action. Thus, the
lattice spacing acts as an ultraviolet regulator which preserves gauge
invariance at all stages during the evaluation of $Z$.

Gauge fields are represented on the lattice by the link variable
$U_\mu(x)$ which is an element of the gauge group SU(3). In contrast
to QCD formulated in terms of the continuum gauge potential $A_\mu(x)$
the integration over the gauge degrees of freedom is performed over
the compact group manifold, and thus the typical gauge fixing
procedure via the Faddeev-Popov ansatz can be avoided. The simplest
discretisation of the gauge action is the Wilson plaquette action
\cite{Wilson:1974sk}
\be
   S_{\rm G}[U] = \frac{6}{g_0^2}\sum_{x\in\Lambda_{\rm
       E}}\sum_{\mu<\nu} \left(1-{\textstyle\frac{1}{3}}{\rm
       Re\,Tr}\,P_{\mu\nu}(x) \right),
\ee
where $P_{\mu\nu}(x)$ denotes the ``plaquette'', i.e. the product of
link variables around an elementary square in the plane defined by
$\mu$ and $\nu$.

A generic expression for the fermionic part of the action is
\be
  S_{\rm F}[U,\psibar,\psi]=a^4\sum_{x\in\Lambda_{\rm E}}
    \sum_{f=u,d,s,\ldots} \psibar_f(x)
    \left((D_{\rm lat}[U]+m_f)\psi_f\right)(x),
\ee
where $D_{\rm lat}[U]$ denotes the massless discretised Dirac
operator, and $m_f$ is the mass of quark flavour~$f$. Since the quark
action is bilinear in the (Grassmannian) fields $\psibar_f$ and
$\psi_f$ one can perform the integration over the quark fields
analytically, which yields
\be
 Z=\int\prod_{x\in\Lambda_{\rm E}}\prod_{\mu=0}^3
 dU_\mu(x)\prod_{f=u,d,s,\ldots}\, {\rm e}^{-S_{\rm G}[U]}
 \det(D_{\rm lat}[U]+m_f).
\ee
The path integral now contains a finite number of integrations over
the group manifold, while the fermionic part is encoded in the quark
determinant. The expectation value of an observable $\Omega$ can
then be defined as
\be
  \left\<\Omega\right\> = \frac{1}{Z} \int\prod_{x\in\Lambda_{\rm
      E}}\prod_{\mu=0}^3 dU_\mu(x) \;\Omega \prod_{f=u,d,s,\ldots}
  \, {\rm e}^{-S_{\rm G}[U]} \det(D_{\rm lat}[U]+m_f).
\ee
The numerical evaluation of $\<\Omega\>$ proceeds by Monte Carlo
integration. An ensemble of gauge configurations  is
generated via importance sampling along a Markov chain, and the factor
$W$, defined by
\be
   W= \prod_{f=u,d,s,\ldots} \det(D_{\rm lat}[U]+m_f)\, {\rm
     e}^{-S_{\rm G}[U]},
\ee
constitutes the statistical weight of an individual gauge
configuration. The most widely used simulation algorithm for QCD with
dynamical quarks is the Hybrid Monte Carlo algorithm, originally
defined in \cite{Duane:1987de}, which has since undergone numerous
improvements \cite{Hasenbusch:2001ne,Luscher:2003vf,
  Urbach:2005ji,Clark:2006fx,Luscher:2007es,Luscher:2008tw,
  Marinkovic:2010eg}.

Monte Carlo integration is inevitably limited to ensembles containing
a finite number of gauge configurations, $N_{\rm cfg}$. Provided that
the configurations are sufficiently decorrelated, the gauge average
$\overline{\Omega}$ is a good approximation of the expectation value
$\<\Omega\>$. Finite statistics also implies a residual uncertainty, so
that the result of the integration must be quoted with a statistical
error which ideally scales like $1/\sqrt{N_{\rm cfg}}$.

\subsection{Lattice actions for QCD}

Before performing the stochastic evaluation of $\<\Omega\>$ one must
make a concrete choice of lattice action for the gauge field (such as
the Wilson plaquette action), as well as the lattice Dirac operator
$D_{\rm lat}$. It is important to realise that the discretisation of
the QCD action is not unique: Different choices for $S_{\rm G}$ and
$S_{\rm F}$ may include any number of irrelevant local operators,
provided that they reproduce the Euclidean action in the continuum as
the lattice spacing is formally sent to zero. An exhaustive list of
common discretisations is given in appendix~A.1 of the FLAG report
\cite{Aoki:2016frl}.

When choosing a particular discretisation one has to balance
computational convenience against conceptual superiority. This is
particularly relevant for the choice of quark action and its
implications for the treatment of the well-known fermion doubling
problem \cite{Nielsen:1981hk,Nielsen:1980rz,Nielsen:1981xu} and the
closely related issue of chiral symmetry breaking. Here we summarise
the general types of fermionic discretisations that are used in
current lattice calculations of $\ahvp$ and $\amuhlbl$. For further
details we refer to the original articles and appendix~A.1 of the FLAG
report~\cite{Aoki:2016frl}. 

{\bf Wilson fermions} \cite{Wilson:1974sk} are among the most widely
used discretisations of the quark action. The massless Wilson-Dirac
operator $D_{\rm w}$ is given by
\be\la{eq:wilson}
   D_{\rm w} ={\textstyle\frac{1}{2}}\Big(\gamma_\mu(\nabla_\mu+\nabla_\mu^\ast)
   -ar\nabla_\mu^\ast\nabla_\mu\Big),  
\ee
where $\nabla_\mu$ and $\nabla_\mu^\ast$ denote the forward and
backward discretisations of the covariant derivative, and the Wilson
parameter $r$ is typically set to one. The addition of the dimension-5
operator proportional to $r$ implies that $D_{\rm w}$ describes a
single fermion species at the expense of explicit chiral symmetry
breaking. The consequences of this are two-fold: First, the leading
discretisation effects in physical observables are linear in the
lattice spacing~$a$, and hence the rate of convergence towards the
continuum limit is slow. Second, the direct lattice transcription of
the electromagnetic current is no longer a conserved quantity. The
local vector current in the discretised theory must therefore be
renormalised by a multiplicative factor $\zv$ before the Ward identity
is satisfied. Still, flavour symmetry remains intact in this
formulation. Furthermore, a conserved vector current can be derived
from the Wilson action via the usual Noether procedure. More details
are provided in \ref{app:vector}.

The leading discretisation effects of $\rmO(a)$ can be removed via the
Symanzik improvement programme \cite{Symanzik:1983dc,Symanzik:1983gh},
which is achieved by adding a suitable counterterm to the Wilson-Dirac
operator. This results in the so-called Sheikholeslami-Wohlert or
``Clover'' action\,\cite{Sheikholeslami:1985ij}, i.e.
\be
   D_{\rm sw} = D_{\rm w}+\frac{ia}{4}
   c_{\rm{sw}}\sigma_{\mu\nu}\widehat{F}_{\mu\nu},
\ee   
where $\widehat{F}$ is a lattice transcription of the gluon field
strength tensor, $\sigma_{\mu\nu}=\frac{i}{2}[\gamma_\mu,\gamma_\nu]$ 
and the coefficient $c_{\rm{sw}}$ must be tuned
appropriately to achieve the complete cancellation of $\rmO(a)$
lattice artefacts. The currents and other local composite operators
must be improved as well by adding appropriate counter\-terms
\cite{Luscher:1996sc,Luscher:1996ug,Luscher:1996jn}.

{\bf{Twisted mass Wilson fermions:}} The removal of the leading
cutoff effects can also be accomplished by adding a chirally twisted
mass term to the Wilson action \cite{Frezzotti:2000nk}. In two-flavour
QCD the corresponding operator describing ``twisted mass QCD'' (tmQCD)
is
\be
   D_{\rm tm}^{(m)}=D_{\rm w}+m_f+i\mu_f\gamma_5\tau^3,
\ee
where $\mu_f$ is the twisted mass parameter of flavour~$f$, and the
superscript ``$(m)$'' on the operator indicates that the massive
operator is considered. The ratio $\mu_{\rm R}/m_{\rm R}$ of the
renormalised twisted and standard quark masses defines the twist
angle, i.e.
\be
   \tan\alpha_{\rm R}=\frac{\mu_{\rm R}}{m_{\rm R}}.
\ee
In Ref.\,\cite{Frezzotti:2003ni} it was shown that the leading cutoff
effects of $\rmO(a)$ are cancelled without the addition of the
Sheikholeslami-Wohlert term, by tuning the bare parameters such that
$\alpha_{\rm R}=\pi/2$ (``maximal twist'').

The presence of the twisted mass parameter introduces a tunable
infrared scale. This is helpful for reducing the numerical effort in
the simulation of the theory in the regime of small quark masses,
where the occurrence of small eigenvalues of the Dirac operator may
lead to a small acceptance rate in the HMC algorithm. 
However, one finds that isospin symmetry is broken in twisted mass QCD.

{\bf{Staggered fermions}} \cite{Kogut:1974ag,Susskind:1976jm} leave a
subgroup of chiral symmetry unbroken but achieve only a partial
lifting of the 16-fold degeneracy that is encountered if $r$ is set to
zero in \eq{eq:wilson}, resulting in four fermionic doubler states,
which are commonly referred to as ``tastes''. To remove this residual
degeneracy one takes fractional powers of the quark determinant, a
procedure known as ``rooting''. Since the taste symmetry is broken by
the interaction with gluons, the rooting procedure produces unitarity
violations. The validity of the rooting procedure has been contested
in Refs.\,\cite{Creutz:2007yg,Creutz:2007yr}. On the other hand,
arguments based on a renormalisation-group approach
\cite{Shamir:2004zc,Shamir:2006nj} suggest that rooted staggered
quarks reproduce the correct, unitary theory in the continuum limit.

The staggered formulation is numerically inexpensive compared to
Wilson-type fermions, since the application of the staggered Dirac
operator requires fewer floating-point operations. The leading lattice
artefacts are of order~$a^2$ without the necessity for twisting or the
addition of counterterms. Furthermore, staggered fermions do not
suffer from accidentally small eigenvalues that slow down the
simulation. These practical advantages must be balanced against the
rooting issue. Violations of the taste symmetry and the overall
influence of cutoff effects can be reduced with the help of the
Symanzik improvement programme \cite{Lepage:1998vj}. The resulting
variants of the staggered action are referred to as the ``Asqtad''
\cite{Orginos:1999cr} and ``HISQ'' \cite{Follana:2006rc} actions.

{\bf{Ginsparg Wilson fermions}}
\cite{Ginsparg:1981bj,Neuberger:1997fp,Neuberger:1998wv,
  Hasenfratz:1998ri,Luscher:1998pqa} preserve chiral symmetry whilst
removing all doublers from the fermion spectrum. The associated Dirac
operator $D$ satisfies
\be\la{eq:GWrel}
   \gamma_5 D+D\gamma_5=aD\gamma_5 D,
\ee
which is known as the Ginsparg-Wilson relation
\cite{Ginsparg:1981bj}. It represents a modification of the usual
requirement that a chirally invariant Dirac operator in the continuum
must anticommute with $\gamma_5$. Explicit constructions of an
operator that satisfies \eq{eq:GWrel} include the ``domain wall''
formulation \cite{Furman:1994ky} in which a $5^{\rm th}$ dimension of
length $N_5$ is introduced while the fermions are coupled to a mass
defect, i.e. a negative mass term. One then finds that modes of
opposite chirality are trapped at the 4-dimensional boundaries. For
any finite values of $N_5$, however, the decoupling of chiral modes is
not exact, leading to residual though exponentially suppressed chiral
symmetry breaking. Domain wall fermions have been employed by the
RBC/UKQCD Collaboration in their calculations of $\ahvp$ and
$\amuhlbl$. In addition to the domain wall construction of
Ginsparg-Wilson fermions, there are other realisations of operators
which satisfy \eq{eq:GWrel}. These include the ``overlap'' or
``Neuberger-Dirac'' operator \cite{Neuberger:1997fp,Neuberger:1998wv},
as well as the truncated fixed point \cite{Hasenfratz:2000xz} and
``chirally improved'' \cite{Gattringer:2000js} actions. None of these
have so far been applied in determinations of the hadronic
contributions to $(g-2)_\mu$.

While Ginsparg-Wilson fermions respect all flavour and chiral
symmetries and reproduce the correct fermion spectrum, they are
numerically much more costly that Wilson or staggered fermions. This
is due to the fact that the operator is defined on a 5-dimensional
lattice in the case of domain wall fermions. If Ginsparg-Wilson
fermions are instead realised via the overlap operator, one is faced
with the problem of evaluating the sign function of a large sparse
matrix. 

\subsection{Vector currents and renormalisation \la{app:vector}}

Both the hadronic vacuum polarisation and light-by-light scattering
contributions to $a_\mu$ are expressed in terms of the electromagnetic
current
\be
  J_\mu(x) = \sum_{f=u,d,s,\ldots}
  {\cal{Q}}_f\psibar_f(x)\gamma_\mu\psi_f(x), 
\ee
where ${\cal{Q}}_f$ denotes the electric charge of quark flavour
$f$. In the continuum, the Ward identities ensure that the current is
conserved, i.e.
\be
   \partial_\mu J_\mu(x)=0.
\ee
However, if the theory is regularised by introducing any of the
discretisations discussed above, one finds that the counterpart of
$J_\mu(x)$ is not the symmetry current which can be derived using
Noether's theorem. The introduction of a lattice cutoff modifies the
theory in the ultraviolet, and these short-distance effects must, in
general, be absorbed into a renormalisation factor. Below we describe
several variants of vector currents that are used in current
calculations of $\ahvp$ and $\amuhlbl$.

Omitting the electric charge factors one defines the local vector
current in the lattice regularised theory for quark flavour~$f$ by
\be
   V_\mu^f(x) = \psibar_f(x)\gamma_\mu\psi_f(x).
\ee
In general, $V_\mu^f(x)$ is renormalised multiplicatively by a factor
$\zv$ which depends on the bare gauge coupling $g_0$. While the
renormalisation factor $\zv$ can be computed in lattice perturbation
theory, it is well known that the perturbative expansion in powers of
the bare coupling $g_0^2$ has poor convergence properties
\cite{Lepage:1992xa}. Several methods have been developed that allow
for the non-perturbative determination of $\zv$ and can also be
applied to the renormalisation of many other local operators. In a
first step one defines a scheme that allows for imposing a
non-perturbative renormalisation condition for the operator under
consideration. In the second step this condition is evaluated in a
numerical simulation. The most widely used schemes are based on the
Schr\"odinger functional
\cite{Luscher:1992an,Sint:1993un,Jansen:1995ck,Luscher:1996jn} and the
regularisation-independent momentum subtraction (RI-MOM) scheme
\cite{Martinelli:1994ty} and its variants \cite{Sturm:2009kb}. A
simple renormalisation condition for the vector current, which can
also be evaluated with good statistical precision, demands that the
forward matrix element of the current between pseudoscalar mesons be
equal to one, which yields (ignoring quark-mass dependent O($a$)
effects)
\be
   \frac{1}{\zv} =
   \frac{\left\<P,\vec q\left|V_0^f\right|P,\vec q\right\>}
        {\left\<P,\vec q|P,\vec q\right\>},
\ee
where $|P,\vec q\>$ denotes a pseudoscalar meson state consisting of
quarks with flavour $f$ and momentum $\vec q$.

Below we discuss variants of the vector current for different
discretisations. If the fermionic part of the QCD action is
discretised using the {\bf Wilson quark action}, the leading
discretisation effects of $\rmO(a)$ can be cancelled by adding the
Sheikholeslami-Wohlert term to the action. While this ensures that
spectral quantities such as hadron masses approach the continuum limit
with a rate proportional to $a^2$, this is not true for operator matrix
elements involving the current. 
In isospin-symmetric QCD, the general form of the renormalised isovector
vector current, which is consistent with $\rmO(a)$ improvement reads
\cite{Luscher:1996sc,Luscher:1996jn,Bhattacharya:2005rb}
\ba\la{eq:Vimp}
   (V_\mu^u(x)-V_\mu^d(x))_{\rm R}&=&\zv(\tilde g_0)
 (1+ \bar b_{\rm V}\; a {\rm Tr}(M) + b_{\rm V}\;a m_{ud})\times
\\ && 
\times   \left\{V_\mu^u(x)-V_\mu^d(x)+ac_{\rm V}\partial_\nu (T_{\mu\nu}^u(x)-T_{\mu\nu}^d(x)) \right\},
\nonumber
\ea
where $b_{\rm V},\, \bar b_{\rm V},\, c_{\rm V}$ are improvement
coefficients that depend on the gauge coupling. In this expression,
$\tilde g_0$ denotes the modified gauge coupling consistent with
O($a$) improvement \cite{Luscher:1996sc}, $M$ is the bare subtracted
quark-mass matrix, with elements $M_{11}=M_{22}=m_{ud}$ corresponding
to the light flavours~$u,d$, and
\be
   T_{\mu\nu}^f(x)=i   \psibar_f(x)\sigma_{\mu\nu} \psi_f(x)
\ee
is the tensor current. While $\zv$ and the improvement coefficient $b_{\rm V}$
have been calculated in perturbation theory \cite{Sint:1997jx}, a
non-perturbative determination is desirable.
The coefficient $c_{\rm V}$ has been addressed in~\cite{Guagnelli:1997db, Bhattacharya:1999uq, Harris:2015vfa, Heitger:2017njs};  
the coefficients $b_{\rm V}$ and $\bar b_{\rm V}$ were computed by different methods in~\cite{Korcyl:2016ugy,Fritzsch:2018zym}.
Recently, a high-accuracy determination of the renormalisation 
factors of the vector and axial-vector currents in the massless theory was achieved~\cite{DallaBrida:2018tpn} by using 
the chirally rotated Schr\"odinger functional framework~\cite{Sint:2010eh,Brida:2016rmy}.
The pre\-sence of the terms
proportional to $b_{\rm V}$ and $\bar b_{\rm V}$ implies that the renormalisation factor
contains a mass-dependent piece. The improvement of the isoscalar part 
of the electromagnetic current involves an additional improvement 
coefficient $f_{\rm V}$, which contributes to an O($a$) 
mass-dependent counterterm proportional to the 
flavour-singlet current~\cite{Bhattacharya:2005rb}.

As an alternative, one can use the vector current derived from the
Wilson action via Noether's theorem, i.e.
\be\la{eq:Vcons}
   \hat{V}_\mu^f(x)=\frac{1}{2}\left\{
     \psibar_f(x+a\hat\mu) (1+\gamma_\mu)U_\mu(x)^\dagger \psi_f(x)
    -\psibar_f(x) (1-\gamma_\mu)U_\mu(x) \psi_f(x+a\hat\mu) \right\},
\ee
which is also referred to as the ``point-split'' vector current, as it
contains fields located at neighbouring lattice sites. Since
$\hat{V}_\mu^f$ is conserved by construction, the multiplicative
renormalisation factor is trivial, $Z_{\rm\hat{V}}=1$. As it stands,
the point-split current in \eq{eq:Vcons} is, however, not improved. An
$\rmO(a)$ improved, renormalised version of $\hat{V}_\mu(x)$ 
requires an additive counterterm given by the divergence of the tensor current,
analogously to the term present in 
 \eq{eq:Vimp}. Non-perturbative determinations of the improvement
coefficient $c_{\rm V}$ have been described in
\cite{Guagnelli:1997db,Harris:2015vfa,Heitger:2017njs}.

{\bf Domain wall fermions:} The lattice Dirac operator
describing domain wall fermions reads
\be
   D_{\rm dwf} = \left(D_{\rm w}
   -M\right)_{xy}\delta_{st}+\delta_{xy}D_{st}^{(5)}, 
\ee
where $s,t$ label coordinates along the $5^{\rm th}$ dimension of
length $N_5$, $D_{\rm st}^{(5)}$ is the corresponding coupling term,
and the parameter $M$ denotes the domain wall height. Quark fields
that are defined on the 4-dimensional subspace are obtained through
the projection \cite{Furman:1994ky}
\be
   q(x) = P_{+}\psi(x,N_5-1)+P_{-}\psi(x,0),\quad
   P_{\pm}={\textstyle\frac{1}{2}}\left(1\pm\gamma_5\right), 
\ee
where the fermion field $\psi(x,s)$ is defined on the full
5-dimensional manifold. The local vector current is defined as
\be
   V_\mu(x) = \overline{q}(x)\gamma_\mu q(x),
\ee
while the expression for the conserved current is quite similar to
\eq{eq:Vcons}, i.e.
\be
   \hat{V}_\mu(x) = \frac{1}{2}\sum_{s=1}^{N_5} \left\{
   \psibar(x+a\hat\mu,s)(1+\gamma_\mu)U_\mu(x)^\dagger \psi(x,s)
   \right. 
    \left.-\;\psibar(x,s)
   (1-\gamma_\mu)U_\mu(x) \psi(x+a\hat\mu,s) 
   \right\}.
\ee
Since the decoupling of chiral modes is only exact for $N_5\to\infty$,
the renormalisation factor of $\hat{V}_\mu(x)$ differs from unity by
terms proportional to the residual additive renormalisation of the
quark mass, i.e. $Z_{\rm\hat{V}}=1+\rmO(am_{\rm res})$.

\subsection{Vector correlator and polarisation tensor on the lattice
  \la{app:PolTens}} 

Using for simplicity the definitions of the conserved vector currents in the previous section,
one obtains the expression for the vacuum polarisation tensor of \eq{eq:PolTens} on a space-time lattice as
\be\label{eq:Pimunudef}
  \Pi_{\mu\nu}({Q})=  a\,\delta_{\mu\nu}\, \sum_f {\cal{Q}}_f^2\, \<T^f_\mu(0)\> + a^4\sum_{f, f^\prime}\,{\cal{Q}}_f
     {\cal{Q}}_{f^\prime}  \sum_x\,{\rm e}^{iQ(x+\frac{a}{2}(\hat\mu-\hat\nu))}\,
  \left\langle \hat V_\mu^{f}(x) \hat V_\nu^{f'}(0)\right\rangle ,
\ee
where ${\cal{Q}}_f, {\cal{Q}}_{f^\prime}$ denote the electric charges
of quark flavours $f$ and $f^\prime$.
  The role of the tadpole terms $\<T^f_\mu(0)\>$ is to remove 
a quadratic divergence~\cite{Gockeler:2003cw}\footnote{In the case of Wilson lattice QCD, 
$T^f_\mu(x) = \frac{1}{2}\left\{
     \psibar_f(x+a\hat\mu) (1+\gamma_\mu)U_\mu(x)^\dagger \psi_f(x)
    +\psibar_f(x) (1-\gamma_\mu)U_\mu(x) \psi_f(x+a\hat\mu) \right\} $.},
but we also note that this term drops out in the subtraction performed 
in \eq{eq:Pimunudefsub} below.\footnote{Several groups
  \cite{Boyle:2011hu,DellaMorte:2017dyu} have considered ``mixed''
  correlators involving the local and conserved currents,
  e.g. $\left\langle V_\mu^f(x)
  \hat{V}_\nu^{f^\prime}(0)\right\rangle$. This requires 
  using the relevant renormalisation factor in \eq{eq:Pimunudef} and the other
  correlators defined here.  In addition, no tadpole term is required in this case.} 
The polarisation tensor given in \eq{eq:Pimunudef} is transverse in the following sense:
\be
\sum_{\mu=0}^3 \hat{Q}_\mu\; \Pi_{\mu\nu}({Q}) = \sum_{\nu=0}^3 \hat{Q}_\nu\; \Pi_{\mu\nu}({Q}) =  0,
\ee
where $\hat{Q}=\frac{2}{a}\sin\left(aQ_\mu/2 \right)$ is the lattice momentum.
Note that on a finite space-time lattice, the
momentum variable $Q_\mu$ assumes integer multiples of $2\pi/L_\mu$,
where $L_\mu$ is the length in direction $\mu$.

It has been noted in \cite{Bernecker:2011gh,Aubin:2015rzx} (see also
\cite{Malak:2015sla,Blum:2015gfa}) that the vacuum polarisation tensor
does not vanish at $Q=0$ in finite volume, $\Pi_{\mu\nu}(0)\neq0$. In
order to reduce finite-volume effects and to suppress the short-distance region,
it is then advantageous to subtract the contribution $\Pi_{\mu\nu}(0)$, which is easily
accomplished via a simple modification of the phase factor in
\eq{eq:Pimunudef}, i.e.
\be\label{eq:Pimunudefsub}
  \Pi_{\mu\nu}({Q})-\Pi_{\mu\nu}({0}) = a^4 \sum_{f,f^\prime}
  {\cal{Q}}_f {\cal{Q}}_{f^\prime} \sum_x\, 
  \left(\rme^{iQ(x+\frac{a}{2}(\hat\mu-\hat\nu))}-1\right) \left\langle
 \hat V_\mu^{f}(x)\hat V_\nu^{f'}(0) \right\rangle.
\ee
The spatially summed vector correlator, $G(x_0)$, is the central
quantity for the determination of $\ahvp$ using the time-momentum
representation (see \eq{eq:Gx0def}). On the lattice the corresponding
expression reads
\be\label{eq:Gdef}
   G(x_0)\delta_{kl} = -a^3 \sum_{f, f'}{\cal{Q}}_f
   {\cal{Q}}_{f^\prime}  \sum_{\vec{x}}  \left\langle
  \hat  V_k^{f}(x)\hat V_l^{f'}(0) \right\rangle.
\ee
In the case of Wilson fermions, the improvement term proportional to
the divergence of the tensor current must implicitly be included in
the vector currents appearing in \eq{eq:Gdef}, if full O($a$)
improvement is to be achieved.
The sum $\sum_{f,f^\prime}\ldots$ in equations (\ref{eq:Pimunudef}),
(\ref{eq:Pimunudefsub}) and (\ref{eq:Gdef}) runs over all quark
flavours included in the electromagnetic currents. However, one is
often interested in the contributions to $\ahvp$ from individual
quark flavours. Noting that the dominant contributions arise from
quark-connected diagrams, one defines
\ba
 & & \Pi_{\mu\nu}^{f}({Q})=   {\cal{Q}}_f^2 \,\Big( 
    a\,\delta_{\mu\nu}\, \<T^f_\mu(0)\> + a^4 
    \sum_x\,   \rme^{iQ(x+\frac{a}{2}(\hat\mu-\hat\nu))}\, \left\langle
  \hat V_\mu^{f}(x)\hat V_\nu^{f}(0) \right\rangle\Big), \label{eq:Pimunuf} \\ 
 & &  G^{f}(x_0)= -\frac{a^3}{3} {\cal{Q}}_f^2 
   \sum_{\vec{x}}  \left\langle
   \hat V_k^{f}(x)\hat V_k^{f}(0)   \right\rangle,\quad f=(ud), s, c, \ldots
  \label{eq:Gfdef}
\ea
In the above expressions it is
understood that the expectation value is restricted to quark-connected
diagrams only. We have assumed that the up and down quarks are mass-degenerate,
$m_u=m_d$ while ${\cal{Q}}_{ud}^2=5/9$. 
By performing the tensor decomposition according to
\eq{eq:PimunuQ} one obtains $\Pi^f(Q^2)$, i.e. the (connected)
contribution of quark flavour $f$ to the vacuum polarisation
function. The corresponding fraction of the anomalous magnetic moment,
$(\ahvp)^f$, is obtained by inserting $\Pi^f(Q^2)$ and $G^f(x_0)$ into
equations (\ref{eq:amublum2}) and (\ref{eq:TMRamu}), respectively.

\subsection{Systematic effects \la{sec:systematics}}

Below we give an overview of the main systematic effects that are
common to all lattice calculations.

{\bf{Discretisation errors:}} Lattice estimates of renormalised,
dimensionless quantities differ from their continuum counterparts by
terms proportional to $a^p$, where $a$ denotes the lattice spacing,
and the positive integer $p$ depends on the details of the
discretisation. In order to obtain the result in the continuum limit,
an extrapolation of data computed at several values of the lattice
spacing must be performed. Obviously, the convergence to the continuum
limit is faster for large values of $p$. While $p=1$ for unimproved
Wilson fermions, one finds $p=2$ for most other fermionic
discretisations, including staggered, domain wall and twisted-mass
Wilson fermions. The Symanzik improvement programme, when applied to
Wilson fermions, also yields $p=2$. Typical values of the lattice
spacing in current simulations are of order $0.1\,\fm$ and smaller.

{\bf{Quark mass dependence:}} Quark masses are ``external'' parameters
of QCD that cannot determined by the theory itself. In lattice
simulations the physical values of the quark masses are {\it a priori}
unknown and must be fixed by comparing lattice results to
experiment. To this end one computes observables at several different
values of the bare quark mass and determines the result corresponding
to the physical situation by an inter- or extrapolation in the quark
mass. The ansatz used to fit the quark mass dependence is often
motivated by chiral effective theory.

Before the mid-2000s, lattice simulations with dynamical quarks had
been restricted to unphysically large values of the up and down-quark
masses, since the numerical effort for producing statistically
decorrelated configurations showed a strong growth as the chiral
regime was approached. Lattice results were therefore subject to a
potentially large systematic uncertainty arising from chiral
extrapolations to the physical point. Thanks to a number of
algorithmic improvements \cite{Hasenbusch:2001ne,Luscher:2003vf,
  Urbach:2005ji,Clark:2006fx,Luscher:2007es,Luscher:2008tw,
  Marinkovic:2010eg} and increasing numerical resources, simulations
at or very near the physical pion mass are now carried out on a
routine basis, so that the uncertainty associated with the quark mass
dependence of observables can be significantly reduced.

{\bf{Finite volume effects:}} Results computed in lattice
simulations are usually affected by the finite extent of the spatial
and temporal dimensions of the lattice. It can be shown, however, that
finite-volume effects in stable hadron masses, decay constants and
spacelike form factors are exponentially suppressed, provided that the
spatial size in unit of the mass of the lightest bound state, the
pion, is sufficiently large. Finite-volume effects thus contain the
universal factor $\exp\{-(m_\pi L)\}$, and one finds empirically that
for many quantities such as hadron masses and decay constants they are
negligible provided that
\be
  m_\pi L \gtrsim 4.
\ee
Obviously, this criterion is not universally applicable and must be
verified on a case-by-case basis. 

Finite-volume effects are not just a nuisance in lattice calculations
but offer a method to gain information on hadronic systems. The
L\"uscher method
\cite{Luscher:1985dn,Luscher:1986pf,Luscher:1990ux,Luscher:1991cf} is
a powerful formalism for the characterisation of resonances in lattice
QCD, by providing an exact relation between scattering phase shifts
and the energy levels of multi-particle states in a finite volume. As
we will see later in this review, this is highly relevant for the
determination of the hadronic vacuum polarisation contribution.

While it is very costly to perform simulations for the same set of
parameters on different volumes to check whether the results show a
significant dependence on the extent of the lattice, one can also
employ effective theories such as Chiral Perturbation Theory to
estimate finite-volume effects\,\cite{Gasser:1986vb, Gasser:1987ah,
  Gasser:1987zq, Colangelo:2005gd}.

{\bf{Critical slowing down:}} The most widely used simulation
algorithm for dynamical fermions, the Hybrid Monte Carlo (HMC)
algorithm becomes rapidly inefficient at producing decorrelated gauge
configurations as the lattice spacing is reduced to
$a\lesssim0.05\,\fm$. This is commonly referred to as ``critical
slowing down''. Gauge configurations can be classified in terms of
their topological properties, characterised by the winding number or
topological charge. Critical slowing down manifests itself in the
observed inability of the HMC algorithm to tunnel between different
sectors of topological charge, which results in extremely long
autocorrelation times \cite{Schaefer:2010hu}. As a result the
statistical errors that are assigned to the results may not be
reliable. A number of proposals have been suggested to alleviate the
problem of ``topology freezing'' and the associated autocorrelation
times, including the use of open boundary conditions
\cite{Luscher:2011kk} or multi-scale Monte Carlo equilibration
\cite{Endres:2015yca}.

\end{appendix} 

\newpage


\end{document}